\documentclass[a4paper,11pt]{article}

\usepackage{jheppub}

\usepackage{hyperref}
\usepackage{graphicx}
\usepackage{amsmath}
\usepackage{amssymb}
\usepackage{xspace}
\usepackage{mathrsfs}
\usepackage{subfigure}
\usepackage{slashed}
\usepackage[normalem]{ulem}
\usepackage{xcolor}
\usepackage{epstopdf}
\usepackage{soul}
\usepackage[utf8]{inputenc}

\DeclareRobustCommand{\eq}[1]{eq.~\eqref{eq:#1}}
\DeclareRobustCommand{\eqs}[2]{eqs.~\eqref{eq:#1} and \eqref{eq:#2}}

\DeclareRobustCommand{\fig}[1]{fig.~\ref{fig:#1}}
\DeclareRobustCommand{\figs}[2]{figs.~\ref{fig:#1} and \ref{fig:#2}}
\DeclareRobustCommand{\sec}[1]{sec.~\ref{sec:#1}}

\DeclareRobustCommand{\subsec}[1]{sec.~\ref{subsec:#1}}

\DeclareRobustCommand{\app}[1]{app.~\ref{app:#1}}
\DeclareRobustCommand{\rcite}[1]{ref.~\cite{#1}}
\DeclareRobustCommand{\rcites}[1]{refs.~\cite{#1}}

\usepackage{marginnote}

\newcommand{\ord}[1]{\mathcal{O}(#1)}

\newcommand{\Mae}[3]{\bigl\langle#1\bigr\rvert#2\bigr\rvert#3\bigr\rangle}

\newcommand{\df}{\mathrm{d}}

\newcommand{\img}{\mathrm{i}}
\newcommand{\ri}{\mathrm{i}}
\newcommand{\re}{\mathrm{e}}

\newcommand{\sdt}{\!\cdot\!}
\newcommand{\tr}{\textrm{tr}}

\newcommand{\eps}{\epsilon}

\newcommand{\cB}{{\mathcal B}}

\newcommand{\cL}{{\mathcal L}}

\newcommand{\bn}{{\bar{n}}}
\newcommand{\bnP}{\overline {\mathcal P}}

\newcommand{\nn}{\nonumber}

\newcommand{\lqcd}{\Lambda_\mathrm{QCD}}

\newcommand{\Ecm}{E_\mathrm{cm}}

\newcommand{\zero}{{(0)}}
\newcommand{\one}{{(1)}}
\newcommand{\two}{{(2)}}

\newcommand{\SCETb}{\ensuremath{{\rm SCET}_{\rm II}}\xspace}

\newcommand{\pT}{\vec{p}_{T}}
\newcommand{\pTsq}{\vec{p}_{T}^{\,2}}
\newcommand{\delpT}{\delta^\two (\pT)}
\newcommand{\LpT}[1]{\mathcal{L}_{#1}(\pT,\mu)}
\newcommand{\operp}{\otimes_\perp}

\newcommand{\mhsq}{\hat{m}^2}

\def\Li{\textrm{Li}}
\def\nn{\nonumber}

\def\df{\textrm{d}}

\def\MS{\overline{\rm MS}}

\allowdisplaybreaks[2]

\preprint{\begin{flushright}
FR-PHENO-2023-03
\end{flushright}}

\title{Two-loop bottom mass effects on the Higgs transverse momentum spectrum in top-induced gluon fusion}

\author[a]{Piotr Pietrulewicz,}
\author[b]{Maximilian Stahlhofen}

\affiliation[a]{No affiliation (formerly DESY, Hamburg)}
\affiliation[b]{Albert-Ludwigs-Universit\"at Freiburg, Physikalisches Institut, D-79104 Freiburg, Germany}

\emailAdd{maximilian.stahlhofen@physik.uni-freiburg.de}

\abstract{
We compute bottom mass ($m_b$) corrections to the transverse momentum ($q_T$) spectrum of Higgs bosons produced by gluon fusion in the regime $q_T \sim m_b \ll m_H$ at leading power in $m_b/m_H$ and $q_T/m_H$, where the gluons couple to the Higgs via a top loop.
To this end we calculate the quark mass dependence of the transverse momentum dependent gluon beam functions (aka gluon TMDPDFs) at two loops in the framework of SCET.
These functions represent the collinear matrix elements in the factorized gluon-fusion cross section for small $q_T$.
We discuss in detail technical subtleties regarding rapidity regulators and zero-bin subtractions in the calculation of the virtual corrections present for massive quarks.
Combined with the known soft function for $m_b \neq 0$ our results allow to determine the resummed Higgs $q_T$ distribution in the top-induced gluon fusion channel at NNLL$^\prime$ (and eventually N$^3$LL) with full dependence on $m_b/q_T$.
We perform a first phenomenological analysis at fixed order, where the new corrections to the massless approximation lead to percent-level effects in the peak region of the Higgs $q_T$ spectrum. Upon resummation they may thus be relevant for state-of-the-art precision  predictions for the LHC.
}

\begin{document}
\maketitle

\section{Introduction}
\label{sec:Intro}

The transverse momentum ($q_T$) spectrum of the Higgs boson is one of the most  important observables at the LHC. High-luminosity measurements by the ATLAS and CMS detectors promise $q_T$-differential Higgs cross section data with relative experimental uncertainties of eventually only a few percent, see e.g.~\rcite{Cepeda:2019klc}.
This precision can be exploited to discover potential new physics effects from modified (effective) Higgs couplings in the spectrum at low and moderate $q_T$~\cite{Grazzini:2016paz,Bishara:2016jga,Soreq:2016rae,Bonner:2016sdg} (i.e.\ $q_T \ll m_H$ and $q_T \lesssim m_H$ with $m_H$ the Higgs mass, respectively).
The shape of the spectrum in the low-$q_T$ region, where the peak of the distribution is located, for instance, is sensitive to modifications of the Higgs-bottom Yukawa coupling~\cite{Grazzini:2016paz}.
To tap the full potential of such new physics analyses the theory uncertainty of the Standard Model (SM) background should be at the same level (or smaller) than the experimental one.

The state-of-the-art theoretical predictions of the Higgs $q_T$ spectrum in gluon fusion, which is by far the dominant Higgs production channel at the LHC, already come close to the few-percent precision goal.
They have reached the fiducial N$^3$LL$^\prime$ + N$^3$LO level%
\footnote{
  In the primed counting of the logarithmic accuracy, N$^n$LL$^\prime$ implies that the fixed-order ingredients in the corresponding factorized cross section are included at N$^n$LO, and not only at N$^{n-1}$LO like at N$^n$LL.
  Throughout this paper fixed orders are counted w.r.t.\ the inclusive Higgs production process, i.e.\ N$^n$LO for the $q_T$ spectrum corresponds to N$^{n-1}$LO for Higgs + 1jet production. We will at times refer to the latter counting as N$^{n-1}$LO$_1$.}%
, i.e.\ they include fixed-order inclusive as well as fiducial corrections and resummation of large logarithms $\propto \ln(q_T^2/m_H^2)$ up to third order in QCD~\cite{Billis:2021ecs,Re:2021con,Becher:2020ugp,Bizon:2018foh,Chen:2018pzu,Bizon:2017rah}.
Recently, even some of the ingredients required to achieve N$^4$LL resummation became available~\cite{Duhr:2022yyp,Moult:2022xzt,Agarwal:2021zft}.
Small-$q_T$ resummation is not only necessary to properly describe the shape of the spectrum, but also improves the prediction for the total fiducial cross section of Higgs production measured by the LHC experiments~\cite{Billis:2021ecs}.
This in turn allows e.g.\ to probe the effective Higgs coupling to gluons.
The currently available high-order resummed results for the Higgs $q_T$ spectrum~\cite{Billis:2021ecs,Re:2021con,Becher:2020ugp,Bizon:2018foh,Chen:2018pzu,Bizon:2017rah} are obtained in the heavy-top limit ($m_t \to \infty$)%
\footnote{The LO top-mass dependence can simply be restored by rescaling the heavy-top limit results with the full Born cross section~\cite{Billis:2021ecs,Re:2021con}, which affects the low-$q_T$ spectrum due to resummation.}%
, while all other SM quarks are treated as massless.

Finite top mass effects become relevant for large $q_T$ ($\gtrsim m_t$). The full top mass dependence of the Higgs $q_T$ distribution is known at NNLO (= NLO$_1$ in Higgs\,+\,1jet production)~\cite{Jones:2018hbb}.
Regarding bottom mass effects there exists quite some literature, see e.g.~\rcites{Bonciani:2022jmb,Caola:2018zye,Lindert:2017pky,Grazzini:2013mca,Melnikov:2016emg,Caola:2016upw,Greiner:2016awe,Bagnaschi:2015bop,Banfi:2013eda,Mantler:2012bj,Keung:2009bs}.
Concentrating on the $q_T \lesssim m_H$ region and working in the heavy-top limit we distinguish bottom mass corrections proportional to (at least one power of) the bottom Yukawa coupling $y_b\sim m_b/m_H$ and those at leading power in $1/m_H$ and thus $\ord{y_b^0}$.
To the best of our knowledge all available literature is concerned with the former type of corrections.

In gluon fusion these arise from a bottom (instead of a top) loop connecting the produced Higgs boson and the two incoming gluons at the amplitude level.
Virtual corrections to the bottom quark mediated gluon-gluon-Higgs ($ggH$) form factor give rise to large (double) logarithms $\propto \ln (m_b^2/m_H^2)$, which partly compensate the $y_b\, m_b/m_H \sim m_b^2/m_H^2$ suppression. Their systematic resummation has been achieved recently~\cite{Liu:2022ajh}.%
\footnote{See \rcite{Liu:2021chn} for partial resummation at even higher powers in $m_b/m_H$.}
For the class of corrections where a real gluon is attached to the bottom loop inducing the $ggH$ interaction it is currently unknown how to consistently resum potentially large logarithms of the type $\ln (m_b^2/q_T^2)$ or $\ln(m_b^2/s)$~\cite{Melnikov:2016emg,Caola:2016upw}.

In the intermediate-$q_T$ region, where (formally) $m_b \ll q_T \lesssim m_H \sim \sqrt{s}$ with $s$ the partonic center of mass energy, the bottom mass corrections proportional to one power of $y_b$ (top-bottom interference) are dominant.
All of them were computed at NLO in \rcite{Grazzini:2013mca} and at NNLO in \rcite{Lindert:2017pky}, where only the leading terms of an expansion in $m_b^2/m_H^2$, $m_b^2/q_T^2$, $m_b^2/s$ are kept in the relevant two-loop amplitudes~\cite{Melnikov:2016qoc}.
Recently, also the full NNLO result for the top-bottom interference contribution became available~\cite{Bonciani:2022jmb}.
In \rcite{Caola:2018zye} different heuristic prescriptions to supplement these results with partial or ambiguous resummation of large logarithms at NNLL were studied.
The bottom mass effects on the Higgs spectrum at moderate $q_T$ were found to be of $\ord{5\%}$ (and negative), while the remaining uncertainties due to unknown higher-order (logarithmic) terms $\propto y_b$ were estimated to be at the one-percent level.
The leading electroweak corrections to the spectrum in the $q_T \lesssim m_H$ range are somewhat smaller in size~\cite{Keung:2009bs}.

For a determination of $y_b$ from the shape of the spectrum at low $q_T$ also the bottom annihilation channel to Higgs production ($\propto y_b^2$) plays an important role.
This contribution is known to NNLL + NNLO~\cite{Harlander:2014hya}
in the ``five-flavor scheme", where bottom quarks are included in the parton distribution functions (PDFs) and which effectively corresponds to the $m_b\to0$ limit with fixed $y_b$.
Beyond this approximation, i.e.\ consistently assuming $\lqcd \ll m_b \ll m_H$, the resummed $q_T$ spectrum in bottom quark annihilation can be calculated in full analogy to the ``primary mass effects" in the Z-boson $q_T$ spectrum (using four-flavor PDFs) following \rcite{Pietrulewicz:2017gxc}.
\looseness=-1

In the present paper we consider bottom mass corrections to the Higgs $q_T$ spectrum at leading power in
$1/m_H \sim 1/s$.
In contrast to the corrections discussed above these are independent of the bottom Yukawa coupling $y_b\sim m_b/m_H$ and insensitive to the structure of the $ggH$ interaction mediated by a top loop, i.e.\ we can safely work in the heavy-top limit.
The $q_T$-dependent contributions of this type can be written as a series of $(m_b/q_T)^{2n}$ terms (with $n\in \mathbb{N}$) and first appear at NNLO in the spectrum.
Hence, they come with an additional factor of the strong coupling $\alpha_s$ compared to the NLO contributions $\propto y_b$.
Near the peak of the Higgs $q_T$ distribution at $q_T \approx 2 m_b \approx 10$ GeV, both types of effects may therefore be of similar size.
The aim of this work is to compute the leading $\ord{y_b^0}$ bottom mass corrections in the regime $\lqcd \ll m_b \sim q_T \ll m_H$ and to provide a first analysis of their numerical impact on the NNLO (= NLO$_1$) spectrum.
\looseness=-1

Cross sections for sufficiently inclusive measurements in high-energy processes that involve largely different (energy) scales can often be shown to factorize to a good approximation.
This means that the physics at the various scales can be described independently by separate factorization functions, which typically simplifies the calculation substantially.
The factorization of the $q_T$-differential cross section of color-singlet production in the presence of a massive quark flavor was worked out in \rcite{Pietrulewicz:2017gxc} in the context of the Drell-Yan process.
For the (four) relevant hierarchies between the hard scale $Q$ set by the invariant mass of the color singlet, its transverse momentum $q_T \ll Q$, and the quark mass, generically denoted by $m$, the appropriate factorization theorems were formulated there using soft-collinear effective theory (SCET)~\cite{Bauer:2000ew, Bauer:2000yr,
Bauer:2001ct, Bauer:2001yt, Bauer:2002nz, Beneke:2002ph}.%
\footnote{In the case where all quarks (except for the infinitely heavy top) are treated as massless the corresponding factorization theorem for $q_T \ll Q$ was first derived in direct QCD~\cite{Collins:1984kg} and later in SCET~\cite{Becher:2010tm, Chiu:2012ir, GarciaEchevarria:2011rb}.}
This factorization framework also applies to Higgs production in gluon fusion with $m=m_b$ and allows to systematically resum all types of large logarithms at leading power in the small scale ratios.
The factorized cross section for the regime $q_T \sim m \ll Q$ takes a special role in this approach, because the factorization theorems for the adjacent regions, i.e.\ $q_T \ll m \ll Q$ and $m \ll q_T \ll Q$, represent its large and small mass limits.
The latter can therefore be derived together with the corresponding power corrections by an expansion of the former in $q_T/m$ and $m/q_T$, respectively.

The only missing (and arguably most complex) ingredients to compute the bottom mass effects on the Higgs $q_T$ spectrum we are interested in are the transverse momentum dependent (TMD) gluon beam functions for $q_T \sim m \ll Q\, (=m_H)$.
These factorization functions describe the initial state radiation collinear to the proton beams in the gluon fusion process at the LHC.%
\footnote{Bottom-quark initiated Higgs production requires the corresponding heavy-quark beam functions computed in \rcite{Pietrulewicz:2017gxc}.
}
It is the main purpose of this work to calculate the quark mass corrections to the gluon TMD beam functions to NNLO, while their massless version is already known to N$^3$LO~\cite{Ebert:2020yqt,Luo:2020epw}.
This will enable the small-$q_T$ resummation in the spectrum with full bottom mass dependence at NNLL$^\prime$ and N$^3$LL level (and leading power in $1/Q$).
The beam functions are independent of the hard scattering process. Our results therefore not only contribute to the Higgs $q_T$ distribution at hadron colliders, but also to many other TMD cross sections.
In particular, all analytic expressions in this paper directly carry over to the transverse momentum spectrum of any color singlet final state produced by gluon fusion.

The paper is structured as follows: In \sec{factorization} we discuss the factorization theorem for the $q_T$-differential color-singlet production cross section and its ingredients in the regime $q_T \sim m \ll Q$.
We also give some details on the renormalization group (RG) evolution of the TMD beam functions.
The two-loop calculation of the quark mass corrections to the gluon beam function is presented in \sec{calculation}.
The renormalized results are derived and summarized in \sec{results}.
In \sec{limits} we cross check our results containing the full dependence on $m/q_T$ with known expressions in the small and large mass limits, $m \ll q_T$ and $m \gg q_T$.
In \sec{numerics} we analyze the numerical impact of the computed NNLO bottom mass corrections to the Higgs $q_T$ spectrum.
We conclude in \sec{conclusion}.
\looseness=-1

\section{Factorization with massive quarks}
\label{sec:factorization}

As the prototype of a TMD observable we consider in this work the $q_T$ spectrum of a color singlet state $X$ with invariant mass $Q$ produced in proton-proton collisions.
Using this process as an example we discuss in the following factorization in SCET with an active heavy quark flavor of mass $m$ and $n_l$ massless quark flavors in the regime where $\lqcd \ll q_T \sim m \ll Q$.
The relevant effective field theory (EFT) modes in this kinematic region are $n_a$-collinear, $n_b$-collinear, and soft.
They are defined by the scaling of their typical momenta:
\begin{align}
n_a \text{-collinear:}
&\quad
p_{n_a}^\mu \sim \Bigl(\frac{q_T^2}{Q},Q,q_T\Bigr) \sim \Bigl(\frac{m^2}{Q},Q,m\Bigr)
\,, \nn \\
n_b \text{-collinear:}
&\quad
p_{n_b}^\mu\sim \Bigl(Q,\frac{q_T^2}{Q},q_T\Bigr) \sim \Bigl(Q,\frac{m^2}{Q},m\Bigr)
\, , \nn\\
\text{soft:}
&\quad
p_{s}^\mu\sim (q_T,q_T,q_T) \sim (m,m,m)
\,,
\label{eq:modes}
\end{align}
using the (light-cone) notation
\begin{align} \label{eq:lc}
  p^\mu = n_a \sdt p\,\frac{n_b^\mu}{2} + n_b \sdt p\,\frac{n_a^\mu}{2} + p_{\perp}^\mu \equiv  (n_a \sdt p,n_b \sdt p,p_{\perp}) \equiv  (p^+,p^-,p_\perp)
\,,\end{align}
with opposite light-like beam directions $n_a$, $n_b$ and $\bar{n}_a \equiv n_b$, $\bar{n}_b \equiv n_a$, $n_a^2= \bar{n}_a^2=0$, $n_a \!\cdot \bar{n}_a = 2$.
In addition to the perturbative modes in \eq{modes} we also define nonperturbative ultra-collinear modes with the momentum scaling $(\lqcd^2/Q,Q,\lqcd)$ and $(Q,\lqcd^2/Q,\lqcd)$ to describe the incoming protons (inside the two beams with radius $\sim 1/\lqcd$) and their constituents.
The associated ultra-collinear fields are part of a SCET with $n_l$ massless quark flavors, where the heavy quark field has been integrated out.
The matching between the two SCET versions with and without massive quarks yields the beam function matching coefficients we compute in this paper.
For a detailed account on the EFT framework for the factorization including other possible kinematic regimes with different hierarchies between the scales $q_T$, $m$, $Q$ as well as the connections among them we refer to \rcite{Pietrulewicz:2017gxc}.

The mode setup in \eq{modes} is of \SCETb type, because soft and collinear degrees of freedom have parametrically the same invariant mass $\sim m^2 \sim q_T^2$ and are only separated in rapidity.
As a consequence quantum corrections to soft and collinear operators will in general generate rapidity divergences, which require renormalization and eventually manifest themselves as large rapidity logarithms $\sim \ln(m/Q)$ or $\sim \ln(q_T/Q)$ in fixed-order predictions of physical observables.
We will employ the MS-type approach to rapidity renormalization devised in \rcites{Chiu:2011qc, Chiu:2012ir}.
It allows to systematically resum the rapidity logarithms by means of rapidity renormalization group equations (RRGEs).
The corresponding rapidity renormalization scale is denoted by $\nu$, whereas $\mu$ represents the standard virtuality-type $\MS$ renormalization scale.

The factorization theorem for the gluon fusion process we are concerned with is given in analogy to the (quark-antiquark initiated) Drell-Yan process studied in \rcite{Pietrulewicz:2017gxc} by
\begin{align}\label{eq:factXsec}
& \frac{\df \sigma}{\df q^2_T \,\df Q^2 \,\df Y} =  H^{\{n_l+1\}}_{gg}(Q,\mu)\, \int \df^2 p_{T,a}\,\df^2 p_{T,b}\, \df^2 p_{T,s} \, \delta(q_T^2-|\vec{p}_{T,a}+\vec{p}_{T,b}+\vec{p}_{T,s}|^2)
 \\
 &\qquad\quad \times
 \biggl[\,\sum_{k \in \{q,\bar{q},g\}} \mathcal{I}_{gk,\mu \nu}\Bigl(\vec{p}_{T,a},m,x_a,\mu,\frac{\nu}{\omega_a}\Bigr) \otimes f^{\{n_l\}}_k (x_a,\mu)\biggr]\, S_{gg}(\vec{p}_{T,s},m,\mu,\nu)
 \nn \\
 &\qquad\quad  \times \biggl[\,\sum_{l \in \{q,\bar{q},g\}}
  \mathcal{I}^{\mu \nu}_{gl}\Bigl(\vec{p}_{T,b},m,x_b,\mu,\frac{\nu}{\omega_b}\Bigr) \otimes f^{\{n_l\}}_l (x_b,\mu)\biggr]  \, \biggl[1+\mathcal{O}\biggl(\frac{q_T}{Q},\frac{m}{Q},\frac{\lqcd^2}{m^2},\frac{\lqcd^2}{q_T^2}\biggr)\biggr] \nn
 \, ,\end{align}
 where
 \begin{align}\label{eq:labelmomenta}
 \omega_a = Q e^Y \, , \quad \omega_b = Q e^{-Y} \, , \quad x_{a,b} = \frac{\omega_{a,b}}{\Ecm}
 \,,\end{align}
and $Y$ is the rapidity of the color-singlet state.
Here and in the following the superscripts $\{n_l+1\}$ and $\{n_l\}$ on the $m$-independent factorization functions indicate whether the associated operators belong to SCET with $n_l+1$ or $n_l$ active quark flavors.
Their renormalization group (RG) evolution (w.r.t.\ $\mu$) is performed in the same flavor scheme above and below their characteristic (matching) scale.
Inside these functions the running QCD coupling $\alpha_s(\mu)$ must be consistently evaluated in the respective $\{n_l+1\}$ or $\{n_l\}$ flavor scheme in order to avoid large logarithms when $\mu$ is of the order of the characteristic scale.
In \eq{factXsec} this applies to the hard function $H^{\{n_l+1\}}_{gg}$ governed by the hard scale $Q$ and the parton distribution functions (PDFs) $f^{\{n_l\}}_k$ governed by the hadronization scale $\lqcd$.
The hard function is process-dependent but independent of the observable (here in particular $q_T$).
At leading power in $m/Q$ it corresponds to the squared coefficient of a SCET current operator resulting from the matching between QCD and SCET carried out with $n_l+1$ massless active quark flavors at $\mu \sim Q$. Explicit expressions of the hard function for gluon fusion Higgs production ($gg \to H$) can be found up to NNLO e.g.\ in \rcite{Berger:2010xi} and at N$^3$LO (in the large top mass limit) in \rcite{Gehrmann:2010ue}.

The $m$-dependent factorization functions in \eq{factXsec} are the TMD gluon beam function kernels $\mathcal{I}^{\mu \nu}_{gk}$ and the gluon-fusion TMD soft function $S_{gg}$.
They can be regarded as matching coefficients at leading power in $\lqcd/m$ and are located at the flavor threshold where the massive quark is integrated out, i.e.\ their matching scale is $\sim m$.
Accordingly, the RG evolution above and below $\mu \sim m$ is performed with $n_l+1$ active flavors (concerning the running of soft and beam functions) and $n_l$ active flavors (concerning the PDF running), respectively.
The renormalized $m$-dependent functions $\mathcal{I}^{\mu \nu}_{gk}$ and $S_{gg}$ can be expressed in terms of $\alpha_s(\mu)$ in either the $\{n_l+1\}$ or the $\{n_l\}$ flavor scheme without introducing large logarithms for $\mu \sim m$.
For this reason we did not assign flavor scheme superscripts to $\mathcal{I}^{\mu \nu}_{gk}$ and $S_{gg}$ in \eq{factXsec}.
For concreteness we will, however, by default use $\alpha_s^{\{n_l+1\}}(\mu)$ in explicit expressions of these functions and indicate this by superscripts as $\mathcal{I}^{\mu \nu \{n_l+1\}}_{gk}$ and $S_{gg}^{\{n_l+1\}}$ when necessary.
The mass-dependent TMD soft function for gluon fusion processes $S_{gg}$ is up to three-loop order related to the corresponding soft function $S_{q\bar{q}}$ in the quark-antiquark channel by Casimir scaling.
At NNLO it can therefore be directly obtained from the result for $S_{q\bar{q}}$ computed in \rcite{Pietrulewicz:2017gxc} for Drell-Yan processes by replacing the quadratic Casimir coefficient  $C_F \to C_A$.
We give the explicit expression in \app{Softfctresults}.

In this work we are mainly concerned with the matching coefficients $\mathcal{I}^{\mu \nu}_{gk}$ of the TMD gluon beam functions $B_g^{\mu \nu \,\{n_l+1\}}$ onto the standard collinear PDFs $f^{\{n_l\}}_k$.
The leading power matching relation was used to formulate \eq{factXsec} and reads (following \rcites{Collins:1981uw, Fleming:2006cd, Stewart:2009yx})
\begin{align}\label{eq:BeamFctMatching}
  B_g^{\mu \nu \,\{n_l+1\}}\Bigl(\vec{p}_{T},m,x,\mu,\frac{\nu}{\omega}\Bigr)
  &\!= \!\!\!\!\sum_{k \in \{q,\bar{q},g\}} \!\!\!\mathcal{I}^{\mu \nu}_{gk}\Bigl(\vec{p}_{T},m,x,\mu,\frac{\nu}{\omega}\Bigr) \otimes_x f^{\{n_l\}}_k (x,\mu) \biggl[1+\mathcal{O}\biggl(\frac{\lqcd^2}{m^2},\frac{\lqcd^2}{\pT^{\,2}}\biggr)\biggr],
\end{align}
with $\mu \sim q_T \sim m$ and $\nu \sim Q$ representing the virtuality and rapidity matching scales, respectively.
Here and in the following we use the symbol $\otimes_z$ for the (Mellin-) convolution
\begin{align}
  f(z)\otimes_z g(z)\equiv \int_z^1\frac{\df x}{x}f(x) \, g\Big(\frac{z}{x}\Big)\,.
\end{align}
The parton indices $q$ and $\bar{q}$ in the sums in \eqs{factXsec}{BeamFctMatching}, stand for all massless quark and corresponding antiquark flavors: $q = u,d,s,\ldots$
We will use $Q$ as the index for the massive quark flavor in the following and $g$ represents the gluon.
The universal perturbative matching kernels $\mathcal{I}^{\mu \nu}_{gk}$ describe the $n_{a,b}$-collinear initial-state radiation characterized by the beam function matching scales for the case of a gluon entering the hard scattering process.

 The resummation of logarithms $\ln(q_T^2/Q^2)$ and $\ln(m^2/Q^2)$ is accomplished by performing the (R)RG evolution of the different factorization functions in \eq{factXsec} from their characteristic scales to common renormalization scales $\mu$, $\nu$.
 The resummation kernels for each factorization function (not shown in \eq{factXsec} for compactness) contain the resummed logarithms and are obtained by solving the corresponding (R)RGEs.
 The $\mu$ evolution of the TMD gluon beam function is  determined by the RGE
 \begin{align}
    \mu\frac{\df}{\df \mu} \,
    B_g^{\mu \nu\,\{n_l+1\}}\Bigl(\vec{p}_{T},m,\mu,\frac{\nu}{\omega}\Bigr) =
     \gamma^{\{n_l+1\}}_{B_g} \!\Bigl(\mu,\frac{\nu}{\omega} \Bigr)\,
     B_g^{\mu \nu\,\{n_l+1\}}\Bigl(\vec{p}_{T},m,\mu,\frac{\nu}{\omega}\Bigr) \,.
     \label{eq:BeamfctRGE}
 \end{align}
Similar RGEs hold for the hard and soft functions in \eq{factXsec}.
RG  consistency  in \eq{BeamFctMatching} implies that the dependence on the matching scale $\mu$ of the coefficients $\mathcal{I}^{\mu \nu}_{gk}$ is  subject to
 \begin{align}
 \mu\frac{\df}{\df \mu} \,\mathcal{I}^{\mu \nu}_{gk} \Bigl(\vec{p}_{T},m,z,\mu,\frac{\nu}{\omega}\Bigr) ={}&
 \gamma^{\{n_l+1\}}_{B_g} \!\Bigl(\mu,\frac{\nu}{\omega} \Bigr)\,
 \mathcal{I}^{\mu \nu}_{gk} \Bigl(\vec{p}_{T},m,z,\mu,\frac{\nu}{\omega}\Bigr)
 \nn\\
 &- \sum_{j \in \{q,\bar q,g\}} \Bigl[\mathcal{I}^{\mu \nu}_{gj} \otimes_z \gamma^{\{n_l\}}_{f,jk}\Bigr] \Bigl(\vec{p}_{T},m,z,\mu,\frac{\nu}{\omega}\Bigr) \, ,
 \label{eq:muRGE}
 \end{align}
where $\gamma^{\{n_l\}}_{f,jk}$ are the PDF anomalous dimensions (splitting functions) collected at one loop in \app{splitting}.
In contrast to the $\mu$ anomalous dimensions, like $ \gamma^{\{n_l+1\}}_{B_g}$ in \eq{BeamfctRGE}, the beam and soft $\nu$ anomalous dimensions depend on the quark mass $m$. The corresponding RRGEs read
 \begin{align} \label{eq:nuRGE}
 \nu \frac{\df}{\df \nu}
 B_g^{\mu \nu\,\{n_l+1\}}\Bigl(\vec{p}_{T},m,\mu,\frac{\nu}{\omega}\Bigr)
 & = \gamma^{\{n_l+1\}}_{\nu,B_g} (\pT,m,\mu)
  \,\otimes_\perp B_g^{\mu\nu\,\{n_l+1\}}\Bigl(\pT,m,\mu,\frac{\nu}{\omega}\Bigr)\, ,
 \\
 \nu \frac{\df}{\df \nu} S_{gg}^{\{n_l+1\}}(\pT,m,\mu,\nu)
 & =  \gamma^{\{n_l+1\}}_{\nu,S_g}(\pT,m,\mu) \, \otimes_\perp S_{gg}^{\{n_l+1\}}(\pT,m,\mu,\nu)\, .
\end{align}
The symbol $\otimes_\perp$ denotes the convolution%
\footnote{This definition differs by a factor of $(2\pi)^2$ from the definition in \rcites{Chiu:2012ir,Luebbert:2016itl}.}
\begin{align}
  \label{eq:ptConv}
  g(\pT)\otimes_\perp f(\pT)\equiv \int\!\!\df^2k_T \,f\bigl(\pT-\vec{k}_T\bigr) \, g\bigl(\vec{k}_T\bigr)
\end{align}
in two-dimensional transverse momentum space.
The $\nu$-independence of the cross section in \eq{factXsec} implies the RG consistency condition
\begin{equation}
  2 \gamma_{\nu,B_g}^{\{n_l+1\}} +  \gamma_{\nu, S_g}^{\{n_l+1\}} = 0\,.
  \label{eq:nuconsistency}
\end{equation}

 The $m$-dependence of the TMD beam function matching coefficients $\mathcal{I}^{\mu \nu}_{gk}$ is currently unknown.
 Their quark mass dependent contributions are the only missing pieces in \eq{factXsec} at NNLO and will be computed in the present work to this order, i.e.\ $\mathcal{O}(\alpha_s^2)$.
 The tensor structure of the $\mathcal{I}^{\mu \nu}_{gk}$ can be decomposed as
\begin{align}
\mathcal{I}^{\mu \nu}_{gk}\Bigl(\vec{p}_{T},m,z,\mu,\frac{\nu}{\omega}\Bigr) = \frac{g^{\mu \nu}_\perp}{2}\,\mathcal{I}_{gk}\Bigl(\pT,m,z,\mu,\frac{\nu}{\omega}\Bigr) + \Bigl(\frac{g^{\mu \nu}_\perp}{2}
+\frac{p_T^\mu p_T^\nu}{\pTsq}\Bigr)\mathcal{J}_{gk} \Bigl(\pT,m,z,\mu,\frac{\nu}{\omega}\Bigr) \, .
\end{align}
Note that for unpolarized proton beams the beam function matching kernels $\mathcal{I}_{gk}$ and $\mathcal{J}_{gk}$ depend on $\pT$ only via $\pTsq$ due to rotation symmetry.
Only $\mathcal{I}_{gk}$ acquires a non-zero contribution from tree-level matching:%
\footnote{Throughout this paper we are frequently using the identity $\delta^{(2)}(\pT)=\delta(\pTsq)/\pi$.}
\begin{align}
  \mathcal{I}_{gk}^{(0)}\Bigl(\pT,m,z,\mu,\frac{\nu}{\omega}\Bigr) =
  \delta_{gk}\, \delta(1-z)\, \delta^{(2)}(\pT)\,
  ,\qquad
  \mathcal{J}_{gk}^{(0)}\Bigl(\pT,m,z,\mu,\frac{\nu}{\omega}\Bigr) =0
  \,.
\end{align}
Here and in the following the superscript $(n)$ with $n=0,1,2,...$
indicates an $n$-th order contribution in the perturbative (loop) expansion, i.e.\ $F^{(n)} \sim \alpha_s^n$ for (fixed-order) functions $F = B_g,\, \mathcal{I}_{gk},\, S_{gg}$, etc.
and $\gamma^{(n)} \sim \alpha_s^{n+1}$ for anomalous dimensions $\gamma = \gamma_{B_g},\, \gamma_{\nu,B_g},\, \gamma_{f, ij}$, etc.
The tensor structure of $\mathcal{J}_{gk}$ is orthogonal to $g^{\mu \nu}_\perp$ (upon contraction of all Lorentz indices).
Thus, for cross sections that are insensitive to the gluon polarizations like \eq{factXsec},
only the two-loop matching coefficients $\mathcal{I}_{gk}^{(2)}$ are required
at NNLO, or  N$^2$LL$^\prime$ and N$^3$LL when including resummation.
Moreover, there are no quark mass corrections to the one-loop coefficient $\mathcal{J}^{(1)}_{gk}$.
Hence, at NNLO  (N$^2$LL$^\prime$, N$^3$LL) accuracy, we only have to consider mass effects on the
``unpolarized" gluon beam function
$B_g \equiv g_{\mu\nu}\, B_{g}^{\mu\nu}$ with matching kernel $\mathcal{I}_{gk} = g_{\mu\nu}\, \mathcal{I}_{gk}^{\mu\nu}$.
We present their calculation in the next section.

\section{Calculation of the two-loop beam functions}
\label{sec:calculation}

The SCET operator matrix element defining the bare unpolarized TMD gluon beam function, which accounts for the effects of the $n$-collinear initial-state radiation ($n=n_a,n_b$), reads
\begin{align} \label{eq:Bgdef}
  &B_g(\vec{p}_{T},m, x) = -\omega\, \theta(\omega)\Mae{p_n(p^-)}{\cB_{n\perp}^{\mu c}(0)
  \bigl[ \delta(\omega - \bnP_n)\,
  \delta^\two(\vec{p}_{T} \!- \vec{\mathcal{P}}_{n\perp}) \,
  \cB_{n\perp\mu}^{c}(0)\bigr]}{p_n(p^-)}
 \,,
\end{align}
where $p_n(p^-)$ denotes the incoming spin-averaged proton with lightlike momentum $p^\mu=p^- n^\mu/2$ and $x\equiv \omega/p^-$. The operator
$\cB_{n\perp}^{\mu c} \equiv 2\, \tr [T^c \cB_{n\perp}^\mu ]$ is the gauge-invariant $n$-collinear gluon field strength in SCET:
\begin{equation}
  \cB_{n\perp}^\mu = \frac{1}{g} \bigl[W_n^\dagger\, \img D_{n\perp}^\mu W_n \bigr]\,,
  \quad
  \img D_{n\perp}^\mu = \mathcal{P}_{n\perp}^\mu + g A_{n\perp}^\mu\,\,,
  \quad
  W_n = \biggl[\,\sum_\mathrm{perms} \exp\Bigl(-\frac{g}{\bnP_n}\,\bn\cdot A_n \Bigr)\biggr] .
\end{equation}
The SCET label momentum operators $\vec{\mathcal{P}}_{n\perp}$ and $\bnP_n\equiv\bar{n}\cdot\mathcal{P}_n$~\cite{Bauer:2001ct} act on the $n$-collinear gluon fields $A_n^\mu$ to their right.
For more details on the involved SCET operators and Wilson lines ($W_n$) we refer to \rcites{Bauer:2001yt,Stewart:2009yx,Stewart:2010qs}.

\begin{figure}[t]
  \begin{center}
    \vspace*{-3 ex}
    \includegraphics[width=0.23 \textwidth]{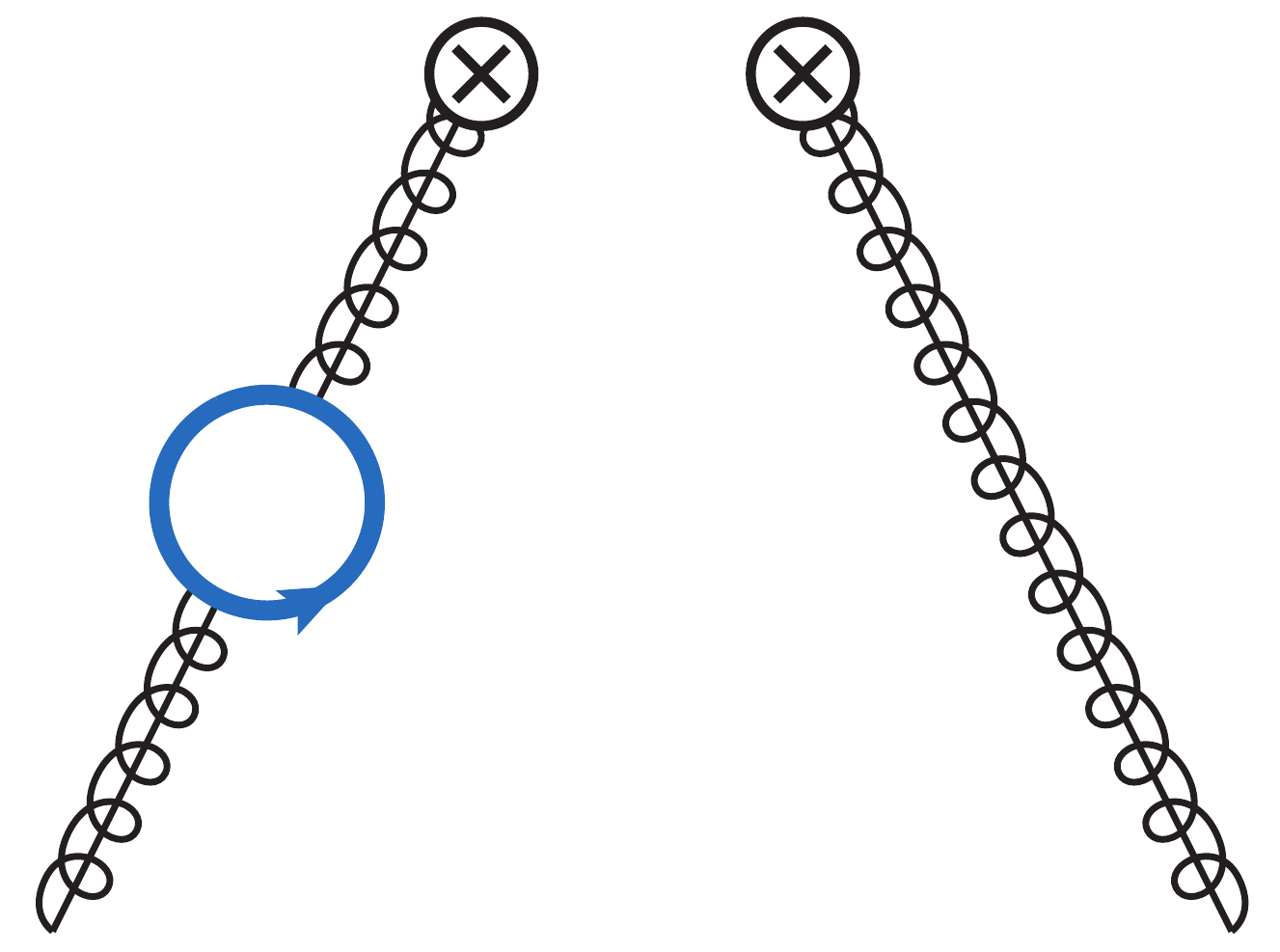}%
    \put(-58,0){(a)} \qquad \qquad
    \includegraphics[width=0.23 \textwidth]{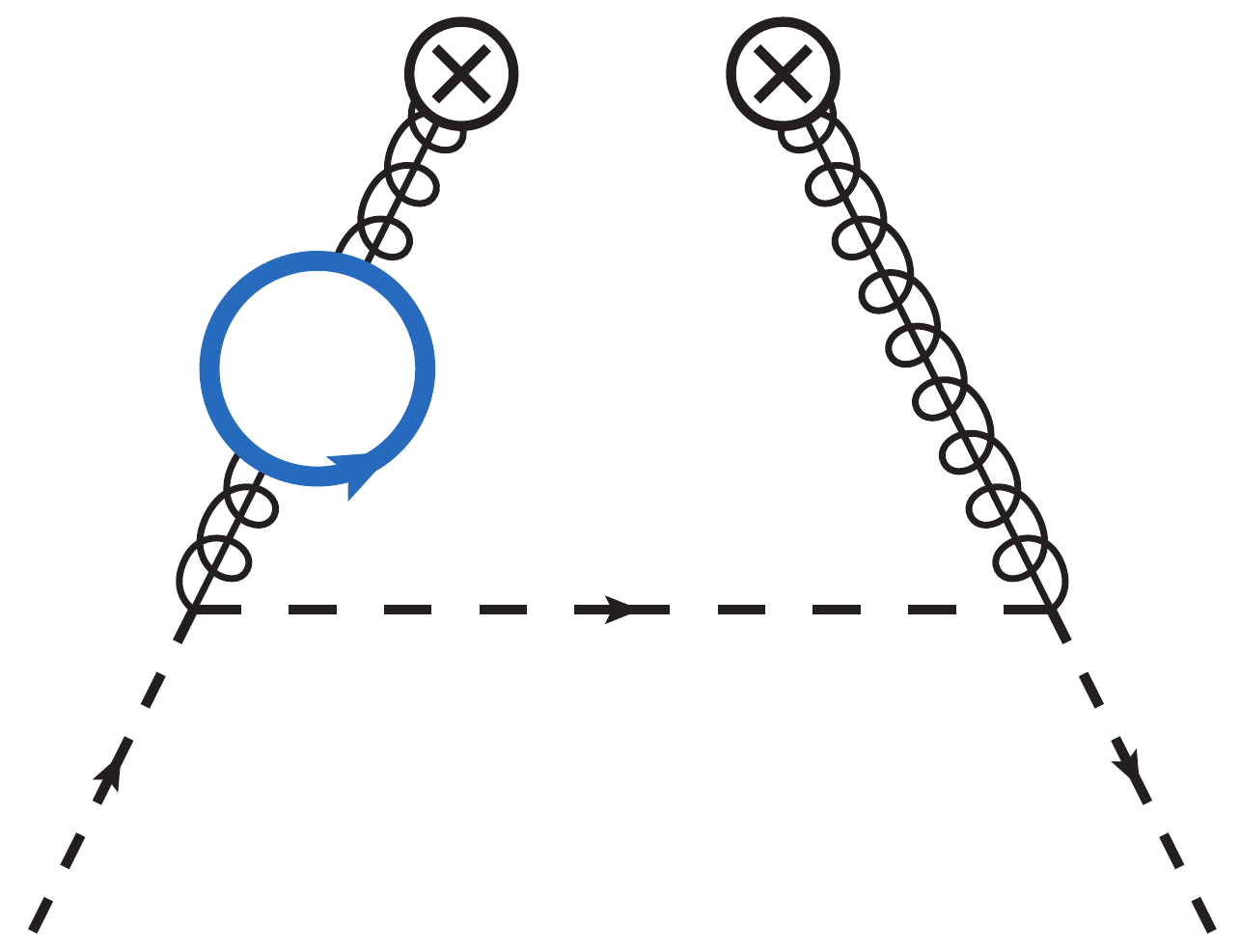}%
    \put(-58,0){(b)}
  \end{center}
  \caption{Lowest order diagrams with massive quark loop (thick blue) contributing to the matching calculation of $\mathcal{I}_{gg}^{(1)}$ (a) and $\mathcal{I}_{gq}^{(2)}$ (b), respectively. The dashed line represents a collinear massless quark $q$, curly lines (with a straight line inside) represent collinear gluons.
  The $\cB_{n\perp}^{\sigma c} \cB_{n\perp\sigma}^{c}$ operator insertion according to \eq{Bgdef} is symbolized by the two crossed circles.
    \label{fig:diagslowestorder}
  }
\end{figure}

The beam function kernels $\mathcal{I}_{gk}$ are in practice computed from a perturbative matching calculation of partonic beam functions $B_{g/j}$ obtained by replacing the incoming proton with parton states (with the same momentum) onto corresponding partonic PDFs $f_{k/j}$~\cite{Stewart:2009yx,Stewart:2010qs}.
The operator between the external states in \eq{Bgdef} is local in time.
We can therefore evaluate the corresponding real-emission Feynman diagrams in \fig{diagslowestorder}b and \fig{realdiags2loopgg} directly as loop diagrams without cutting (or taking a discontinuity or imaginary part).%
\footnote{Of course one may just as well sum the contributions of all possible final-state cuts of each diagram.}
This is in analogy to the SCET calculation of the standard PDFs~\cite{Stewart:2010qs,Berger:2010xi}, but in contrast to that of the virtuality-dependent beam functions~\cite{Stewart:2010qs,Berger:2010xi,Gaunt:2014xga,Gaunt:2014cfa}.
Apart from the vertices for the $\cB_{n\perp}^{\mu c}$ operator insertions~\cite{Berger:2010xi,Gaunt:2014cfa} the usual (time-ordered) QCD Feynman rules can be used~\cite{Bauer:2000yr}.

The bare partonic beam functions are ultraviolet (UV) and infrared (IR) divergent due to the separation of the collinear regions from the hard and ultra-collinear regions in virtuality.
As argued in \rcite{Gaunt:2020xlc}, for the case of the unpolarized gluon beam function UV and IR divergences can be regulated by conventional dimensional regularization upon replacing
\begin{equation}
  \label{eq:deltaReplaceCDR}
  \delta^\two(\pT \!- \vec{\mathcal{P}}_{n\perp}) \,\rightarrow\,
  \frac{(\pTsq)^{-\epsilon}}{\Gamma(1-\epsilon)\pi^{\epsilon}}\;  \delta^{(d-2)}(\pT - \vec{\mathcal{P}}_{n\perp})
\end{equation}
in (the real emission contribution to) \eq{Bgdef} with $d = 4 - 2 \eps$.
For the strong coupling $\alpha_s$ we employ the $\MS$ and for the (bottom) quark mass $m$ the on-shell renormalization scheme.
We adopt the notation%
\begin{equation}
   B_{g/j}^{(n,h)}(\pT,m, x) \equiv
   B_{g/j}^{(n)}(\pT,m, x) -
   B_{g/j}^{(n,l)}(\pT, x)
\end{equation}
for the $n$-loop heavy-flavor correction to the partonic beam function, and analogously for $\mathcal{I}_{gk}^{(n,h)}$, $\gamma_{\nu,B}^{(n,h)}$, etc.
The contributions involving only gluons and the $n_l$ light (i.e.\ massless) flavors are denoted by $B_{g/j}^{(n,l)}$, $\mathcal{I}_{gk}^{(n,l)}$, $\gamma_{\nu,B}^{(n,l)}$, etc.
In both parts we let $\alpha_s$ evolve with $n_l+1$ active flavors, i.e.\ $\alpha_s \equiv \alpha_s^{\{n_l+1\}}(\mu)$ throughout this paper, unless indicated otherwise.

\subsection{Rapidity regulator}
\label{subsec:rapreg}

To regulate the rapidity divergences present in the real emission as well as the purely virtual contributions to $B_g(\vec{p}_{T},m, x)$ we choose the symmetric Wilson line regulator of \rcites{Chiu:2011qc, Chiu:2012ir}.
The same rapidity regulator has been used in the calculation of the NNLO TMD soft function $S_{gg}$ for $m_b \neq 0$~\cite{Pietrulewicz:2017gxc} and the massless NNLO TMD beam functions in \rcite{Luebbert:2016itl} with which we will combine the massive quark contributions to be computed in the present paper.%
\footnote{NNLO results for the massless gluon TMD beam functions obtained with different rapidity regulators are found in \rcites{Catani:2022sgr,Luo:2019bmw,Echevarria:2016scs,Gehrmann:2014yya}, see also \rcite{Catani:2011kr}.}
It may be implemented by modifying the $n$-collinear Wilson lines as%
\footnote{
In general, there are exceptions from this prescription.
Consider, for example, the real quark-antiquark cut and the real gluon cut of diagram \ref{fig:realdiags2loopgg}i, which are separately rapidity divergent (as $x\to1$). In our calculation of $B_g(\vec{p}_{T},m, x)$ the sum of the cuts vanishes exactly, such that the diagram in total does not contribute.
For measurements that put different weights on the one- and two-particle final states like the ones in \rcites{Gangal:2016kuo,Bell:2022nrj,Abreu:2022zgo}, however, the contributions from \fig{realdiags2loopgg}i must cancel with similar terms from diagrams \ref{fig:realdiags2loopgg}g and \ref{fig:realdiags2loopgg}h related by gauge symmetry (just like the corresponding rapidity divergences in the soft function).
This cancellation must be preserved by the rapidity regulator. In practice this means that we have to add (i.e.\ cancel) these particular terms before regulating the involved diagrams by different powers of $|\bn\cdot \mathcal{P}_g|^{-\eta}$.
See \app{realZBs} for a zero-bin calculation where such cancellation is crucial.
\label{footnote:rapreg}
}
\begin{align}
  W_n = \biggl[\,\sum_\mathrm{perms} \exp\Bigl(
  -\frac{g\, w^2}{\bnP_n}\,
  \frac{|\bn\cdot \mathcal{P}_g|^{-\eta}}{\nu^{-\eta}}\,\bn\cdot A_n
  \Bigr)\biggr]\,.
  \label{eq:etaregWL}
\end{align}
Logarithmic rapidity divergences manifest themselves as $1/\eta$ poles in loop (and phase space) integrals. The parameter $w$ obeys the RRGE
$\nu\, \df/\df \nu\, w = - \eta\, w /2$ and is set to one after the derivation of the $\nu$ anomalous dimension.
To obtain the correct anomalous dimensions it is crucial to take the limit $\eta \to 0$ before $\epsilon \to 0$~\cite{Chiu:2012ir}.%
%

Because the relevant rapidity-divergent two-loop diagrams for $B_{g/j}^{(2,h)}$
in fact correspond to one-loop graphs, where either a gluon line is dressed with a massive quark bubble or a triple gluon vertex is replaced by a massive quark triangle subgraph, only single $1/\eta$ poles occur in our calculation.
This is consistent with the requirement that the rapidity divergences of beam and soft functions cancel in the cross section \eq{factXsec} at NNLO.
Moreover, since we have at most a single gluon attached to a Wilson line, the $n$-collinear ``group momentum" operator $\bn \cdot \mathcal{P}_g$~\cite{Chiu:2012ir} can be replaced with the standard label momentum operator $\bnP_n$ for our purposes.

For technical reasons we will furthermore employ the ``$\delta$-regulator" of \rcite{Chiu:2009yx} in the calculation of the virtual diagram in \fig{virtdiags}a. This corresponds to assigning an offshellness to the involved eikonal (Wilson line) propagators.
Although diagram~\ref{fig:virtdiags}a is rapidity-finite, we introduce an auxiliary
$\delta$-regulator to avoid rapidity divergences in (intermediate) expressions generated by an integration by parts (IBP) reduction to master integrals, see \subsec{virtualdiags}.
In the context of IBP reduction the $\delta$-regulator proves more efficient than the $\eta$-regulator, which modifies the power of eikonal propagators to non-integer values, in the sense that it yields a smaller set of master integrals.

Before applying them in our calculation let us point out an interesting peculiarity of the rapidity regulators of the $\eta$- and $\delta$-type.
Consider the following rapidity-finite one-loop integral
\begin{align}
  I_0(a,b) =
  \int\!\! \frac{\df^d k}{(2 \pi)^d}
  \;\frac{1}{(-k^2)^a \, (-k^2+M^2)^b}
  =
  \frac{\ri\, 2^{-d} \pi ^{-d/2}\, \Gamma \bigl(\frac{d}{2}-a\bigr)
  \Gamma \bigl(a+b-\frac{d}{2}\bigr)}{\Gamma(b) \,
  \Gamma \bigl(\frac{d}{2}\bigr)} \, M^{d-2 (a+b)} \,,
  \label{eq:I0Int}
\end{align}
where $M \sim m$ denotes some mass parameter (and we suppress the causal $\ri 0$ prescription).
An integral of this type for example contributes to the unregulated virtual diagram in \fig{virtdiags}b, where it arises after integrating the massive quark bubble (leaving a single-parameter integral, see \subsec{dispersionmethod}) and canceling a factor of $k^- \equiv \bn \cdot k$ in the numerator and the (Wilson line propagator) denominator.
Implementing the $\eta$-regulator in this diagram according to \eq{etaregWL} yields
\begin{align}
  I_\eta(a,b) = \nu^\eta \int\!\! \frac{\df^d k}{(2 \pi)^d}
  \;\frac{|k^-|^{-\eta}}{(-k^2)^a \, (-k^2+M^2)^b}
  = 0 \,.
  \label{eq:Ieta}
\end{align}
Similarly, with the $\delta$-regulator we have%
\footnote{Adding a factor $(k^- +\delta)^{-c}$ in the integrand of \eq{I0Int} results in an additional factor $\delta^{-c}$ after integration.}
\begin{align}
  I_\delta(a,b) &= \int\!\! \frac{\df^d k}{(2 \pi)^d}
  \;\frac{k^-}{(-k^2)^a \, (-k^2+M^2)^b \, (k^- \!+ \delta) } \nn\\
  &= I_0(a,b) -
  \int\!\! \frac{\df^d k}{(2 \pi)^d}
  \;\frac{\delta}{(-k^2)^a \, (-k^2+M^2)^b \, (k^- \!+ \delta) }
  = 0 \,.
  \label{eq:Idelta}
\end{align}
In both cases the limit of vanishing regulator is not continuous:
\begin{align}
  \lim_{\eta \to 0} I_\eta(a,b) = \lim_{\delta \to 0} I_\delta(a,b) = 0 \neq I_0(a,b).
  \label{eq:discrap}
\end{align}
While this does not pose a conceptual problem (as long as the rapidity regulator is correctly implemented such that the combined soft and collinear contributions to an observable reproduce the leading-power full-theory result at fixed order),
it forces us to consistently regulate also rapidity-finite terms which may complicate their calculation in practice.

The discontinuous behavior of the rapidity-regulated integral in \eq{discrap} is caused by the absence of an external minus momentum component (here $p^-$) in the denominator of the  integrand.
For example, the $\eta$-regulated rapidity-finite integral ($p^\mu=p^- n^\mu/2$)
\begin{align}
  J_\eta(a,b) &= \nu^\eta \int\!\! \frac{\df^d k}{(2 \pi)^d}
  \;\frac{|k^-|^{-\eta}}{[-(k-p)^2]^a \, (-k^2+M^2)^b} \nn\\
  &=
  \frac{\ri\, 2^{-d} \pi ^{-d/2}  \Gamma \bigl(\frac{d}{2}-a\bigr)
  \Gamma (a-\eta )
  \Gamma \bigl(a+b-\frac{d}{2}\bigr)}{\Gamma (a) \Gamma (b) \Gamma \bigl(\frac{d}{2}-\eta \bigr)}\,\biggl(\frac{p^-}{\nu} \biggr)^{-\eta } M^{d-2(a+b)}
   \,
   \label{eq:Jeta}
\end{align}
has a smooth $\eta \to 0$ limit%
\footnote{Note that due to the pole structure in the complex $k^+$ plane the integrand has only support for $0<k^-<p^-$. We can thus replace $|k^-|^{-\eta} \to (k^-)^{-\eta}$ in \eq{Jeta} without changing the integral.}%
, and analogously using the $\delta$-regulator:
\begin{align}
  \lim_{\eta \to 0} J_\eta(a,b) = \lim_{\delta \to 0} J_\delta(a,b) = J_0(a,b) = I_0(a,b).
\end{align}

On the other hand, if the scalar loop integral in \eq{I0Int} corresponds to a term of an $n$-collinear SCET diagram like \fig{virtdiags}b the independence of the  large lightcone momentum component $p^-\sim Q$ gives rise to a non-vanishing soft zero-bin (aka soft-bin)~\cite{Manohar:2006nz,Chiu:2009yx}:
Adopting soft scaling of the loop momentum, i.e.\ $k^\mu \sim (m,m,m)$, and expanding the integrand (with $M \sim m$) accordingly leaves the integral unchanged.
Hence, subtracting the zero-bin exactly removes the contribution of \eq{I0Int}  from the diagram.
This cancellation is not affected by the rapidity regulator.
Indeed, we find by explicit calculation that after zero-bin subtraction (see \app{ZBs}) all rapidity-finite contributions to the bare two-loop beam function have a smooth $\eta \to 0$ limit.
We conjecture that this holds true for any rapidity regulator of $\eta$- and $\delta$-type and for any collinear matrix element at any loop order, such that one can always safely drop the rapidity regulator in rapidity-finite terms when consistent zero-bin subtractions are performed.

\subsection{Dispersion relation for massive bubble diagrams}
\label{subsec:dispersionmethod}

The two-loop Feynman diagrams in \fig{diagslowestorder}b, \fig{realdiags2loopgg}e-i, and \fig{virtdiags}b-d contain the one-loop off-shell gluon self-energy consisting of a massive quark bubble as subdiagram.
For their evaluation in general covariant gauge ($\xi=0$ corresponds to Feynman gauge) we conveniently employ the dispersion relation~\cite{Gritschacher:2013pha,Pietrulewicz:2014qza}
\begin{align}
  \label{eq:dispersion}
  &
  \raisebox{-3 ex}{
  \includegraphics[width=0.2 \textwidth]{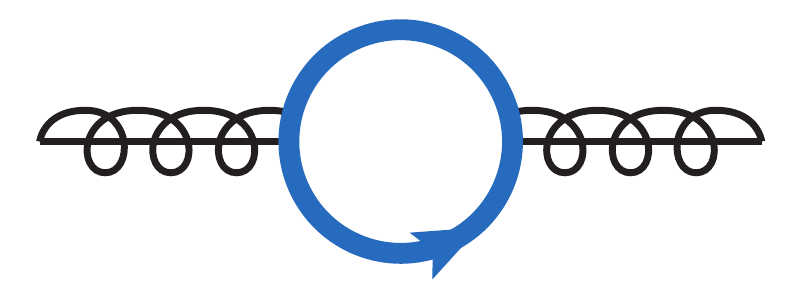}
  \put(-53,13){1PI}
  }
  =
  \frac{-\ri \Bigl(g^{\mu\rho}- \xi \frac{p^\mu p^\rho}{p^2} \Bigr) }{p^2+\ri 0}
  \;\Pi_{\rho\sigma}(p^2,m^2)
  \,\frac{-\ri \Bigl( g^{\sigma\nu} - \xi \frac{p^\sigma p^\nu }{p^2}\Bigr)  }{p^2+\ri 0}
  \\
  &\qquad\qquad
  =\frac{1}{\pi}  \int\limits_{4m^2}^\infty\!\frac{\df {M}^2}{{M}^2}\,
  \frac{-\ri \Bigl(g^{\mu\nu}- \kappa  \frac{p^\mu p^\nu}{p^2} \Bigr)}{p^2-{M}^2+\ri 0}\;\mathrm{Im}\left[\Pi({M}^2,m^2)\right]
  -\frac{-\ri\Bigl(g^{\mu\nu}- \kappa \frac{p^\mu p^\nu}{p^2}\Bigr)}{p^2 +  \ri 0}\;\Pi(0,m^2)
  \,.
 \nn
\end{align}
Note that the gluon propagators to the left and right of the one-particle irreducible (1PI) gluon self-energy bubble are included in \eq{dispersion}, color indices are suppressed, and we have introduced the bookkeeping parameter $\kappa \equiv 1$ in the second line for convenience of presentation, see below.
The first term in the second line represents a weighted integral over a massive gluon propagator (in Landau gauge), the second term is proportional to a massless gluon propagator.
The vacuum polarization function due to a virtual massive quark pair is defined by
\begin{align}
  \Pi_{\mu\nu}^{ab}(p^2,m^2) = -i(p^2 g_{\mu \nu}-p_\mu p_\nu)\Pi(p^2,m^2)\delta^{ab} = \int\df^d x\; \re^{ipx}\langle 0|\mathrm T J_\mu^a(x)J_\nu^b(0)|0\rangle \,,
  \label{eq:Pimunudef}
\end{align}
with the vector current $J^a_\mu \equiv i g {\bar Q} T^a \gamma_\mu Q$.
The relevant one-loop expressions are
\begin{align}
  \Pi^{(1)}(p^2,m^2) &=\frac{\alpha_s T_F}{4\pi} \,
  2^{2 \epsilon +3} \pi ^{\epsilon } \tilde{\mu}^{2 \eps}\, \Gamma(\eps)
  \int_0^1 \df y \,y^{1-\eps}(1-y)^{1-\eps} \biggl(- p^2 + \frac{m^2}{y(1-y)}  \biggr)^{\!\!-\eps},
  \nn\\
  \mathrm{Im}\big[\Pi^{(1)}(p^2,m^2)\big]&=\theta \big(p^2\!-\!4m^2 \big)
  \frac{\alpha_s T_F}{4\pi}  \biggl(\frac{p^2}{\tilde  \mu^2}\biggr)^{\!\!-\varepsilon}\frac{2^{4\varepsilon}\pi^{\frac32+\varepsilon}}{\Gamma(\frac52-\varepsilon)}
  \biggl(\frac{2m^2}{p^2}+1-\varepsilon\biggr)\biggl(1-\frac{4m^2}{p^2}\biggr)^{\!\!\frac12-\varepsilon},
  \nn\\
  \Pi^{(1)}(0,m^2) &=\frac{\alpha_s T_F}{4\pi} \frac{4}{3} (4 \pi)^{\eps} \,
  \Gamma(\eps) \biggl( \frac{m^2}{\tilde{\mu}^2} \biggr)^{\!\!-\eps},
  \label{eq:Pioneloop}
\end{align}
where $\tilde\mu \equiv\mu \, e^{\gamma_E/2}(4\pi)^{-1/2}$ and $\alpha_s \equiv \alpha_s(\mu)$ is the running coupling in the $\MS$ scheme.
Note that the first term in the second line of \eq{dispersion} corresponds to the insertion of the on-shell renormalized vacuum polarization function and is thus UV finite for given $p^\mu$. The UV divergence of the massive quark bubble is contained in the second (massless) term.

Using \eq{dispersion} we can write each of the two-loop diagrams with an off-shell massive quark bubble as sum of two parts.
One part corresponds to the one-loop diagram where the dressed gluon propagator is replaced by a massive gluon propagator with mass $M \ge 2m$, which must be integrated over.
The other part equals the corresponding massless one-loop diagram times a factor $-\Pi^{(1)}(0,m^2)$.
In the virtual diagrams \fig{virtdiags}b-d the latter contribution vanishes because the loop integral is scaleless.
Of course we can also use%
\footnote{We stress again that $\kappa \equiv 1$ is not a gauge parameter or anything alike. Like in \eq{dispersion} the only purpose of $\kappa$ is to label the $p^\mu p^\nu$ part of the dressed propagator in order to trace these terms in the calculations.}
\begin{align}
  \label{eq:bubprop}
  &
  \raisebox{-3 ex}{
  \includegraphics[width=0.2 \textwidth]{Figs/Gluon1loopMassiveSelfBub.pdf}
  \put(-53,13){1PI}}
  = -\frac{-\ri\Bigl(g^{\mu\nu}
  - \kappa \frac{p^\mu p^\nu}{p^2}\Bigr)}{p^2 +  \ri 0}\;\Pi(p^2,m^2)\,.
\end{align}
This leads to one-loop--type integrands including a massive propagator denominator to the power of $\eps$ with mass $m/\sqrt{y(1-y)}$, see first line of \eq{Pioneloop}. The integration over $y$, just like the integration over $M$, is conveniently performed after the loop integration.
In our beam function calculation we used both methods and checked that the results agree for all two-loop diagrams with a massive offshell bubble.

The approach based on the dispersion relation in \eq{dispersion} allows a particularly transparent discussion of the main features of the relevant two-loop diagrams, since their calculation is effectively reduced to a one-loop problem with a massive gluon and integer powers of propagator denominators.
The integration over $M$ does not affect important properties of the original two-loop graph like the presence of a non-vanishing zero-bin, a rapidity divergence, or the gauge-dependence. We will therefore mainly refer to this method in the presentation of our beam function calculation.

In \rcites{Gritschacher:2013pha,Pietrulewicz:2014qza,Pietrulewicz:2017gxc,Hoang:2019fze} it was argued that the terms  ($\propto p^\mu p^\nu$) labeled  by $\kappa$ in \eqs{dispersion}{bubprop} cancel among the two-loop diagrams contributing to gauge-invariant SCET matrix elements such as soft functions or quark jet and beam functions.
The statement also holds for the gluon beam function in \eq{Bgdef}.
For the real-emission diagrams this can be understood from the analogy to the cancellation of the terms linear in the gauge parameter $\xi$ within the (massless) one-loop calculation  of $B_{g/g}^{(1)}$ (or resorting to a Ward identity).
Note that (one/two-particle) real-emission and purely virtual contributions are separately gauge-invariant, i.e.\  independent of $\xi$ (and thus $\kappa$).
It is straightforward to explicitly verify that $\kappa$ drops out separately in the real-emission diagrams \ref{fig:realdiags2loopgg}e and \ref{fig:realdiags2loopgg}f as well as in the sum of diagrams \ref{fig:realdiags2loopgg}g-i already at the integrand level.%
\footnote{After the loop (and before $M$ or $y$) integration all real-emission diagrams are separately $\kappa$-independent.
}

For the purely virtual diagrams in \fig{virtdiags}b-d the analogy to the $\xi$ terms of the corresponding massless one-loop graphs is more subtle, because the latter vanish in dimensional regularization.
However, the gauge-invariant coefficients $\mathcal{I}_{gg}$ can also be obtained from the matching of partonic beam functions and PDFs with offshell external legs or an artificial gluon mass to regulate IR singularities.
In this case also the massless virtual diagrams contribute to $B_{g/g}^{(1)}$ and require zero-bin subtractions.
The $\xi$-independence of the result again suggests that also the $\kappa$ terms from diagrams \fig{virtdiags}b-d must cancel upon zero-bin subtractions.
Indeed, we find by explicit calculation that their total $\kappa$ term before zero-bin subtraction exactly equals the total virtual zero-bin contribution.
Hence, $B_{g/g}^{(2,h)}$ is independent of $\kappa$.
The crucial role of zero-bin subtractions for the gauge invariance of SCET matrix elements involving massive gauge bosons was already pointed out in \rcite{Chiu:2009yx}.

The zero-bin contributions from real-emission graphs vanish, see \app{realZBs}.
We thus conclude that dropping the $\kappa$ terms from the start removes all zero-bin contributions. At the same time this eliminates all terms ($\propto I_0$) that cause the issues with the rapidity regulator discussed in \subsec{rapreg}.
We will therefore mostly exclude the $\kappa$ terms in the following presentation of our beam function calculation.
Instead we will treat them separately and explicitly demonstrate that they exactly cancel the non-vanishing virtual zero-bins in \app{virtualZBs}.

\subsection{Real emission diagrams}

\begin{figure}[t]
  \begin{center}
    \includegraphics[width=0.23 \textwidth]{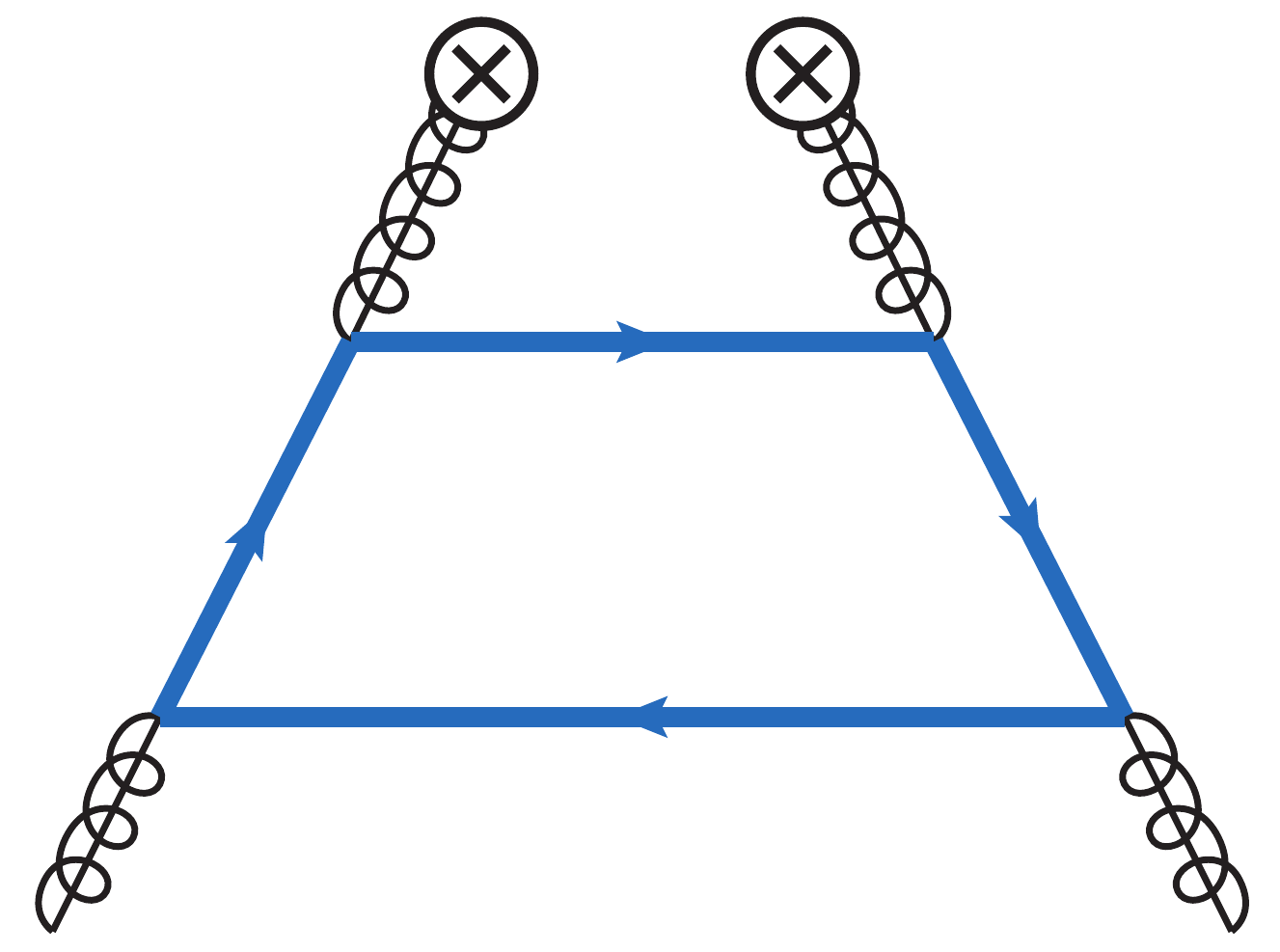}%
    \put(-58,0){(a)} \quad
    \includegraphics[width=0.23 \textwidth]{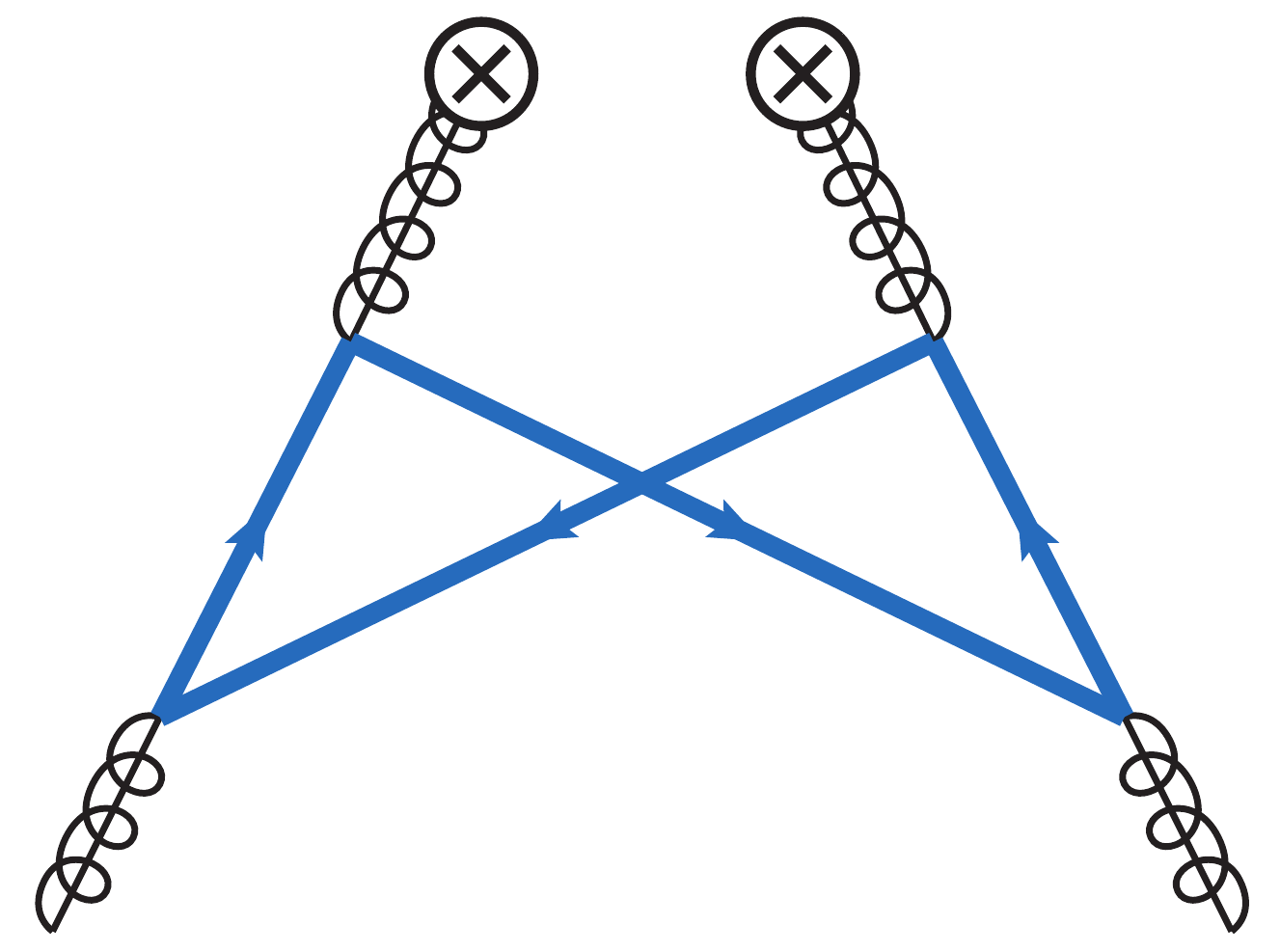}%
    \put(-58,0){(b)} \quad
    \includegraphics[width=0.23 \textwidth]{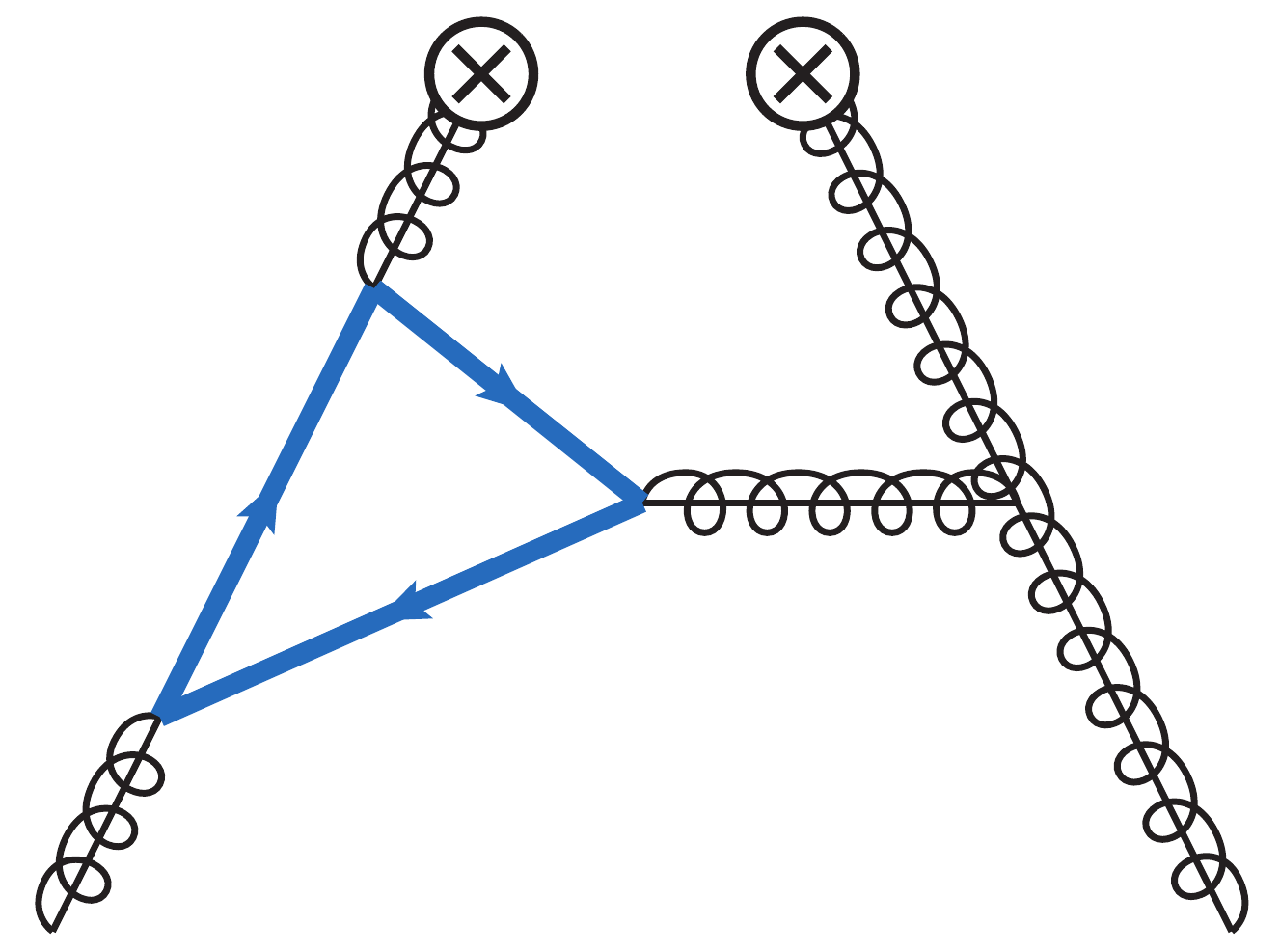}%
    \put(-58,0){(c)} \quad
     \includegraphics[width=0.23 \textwidth]{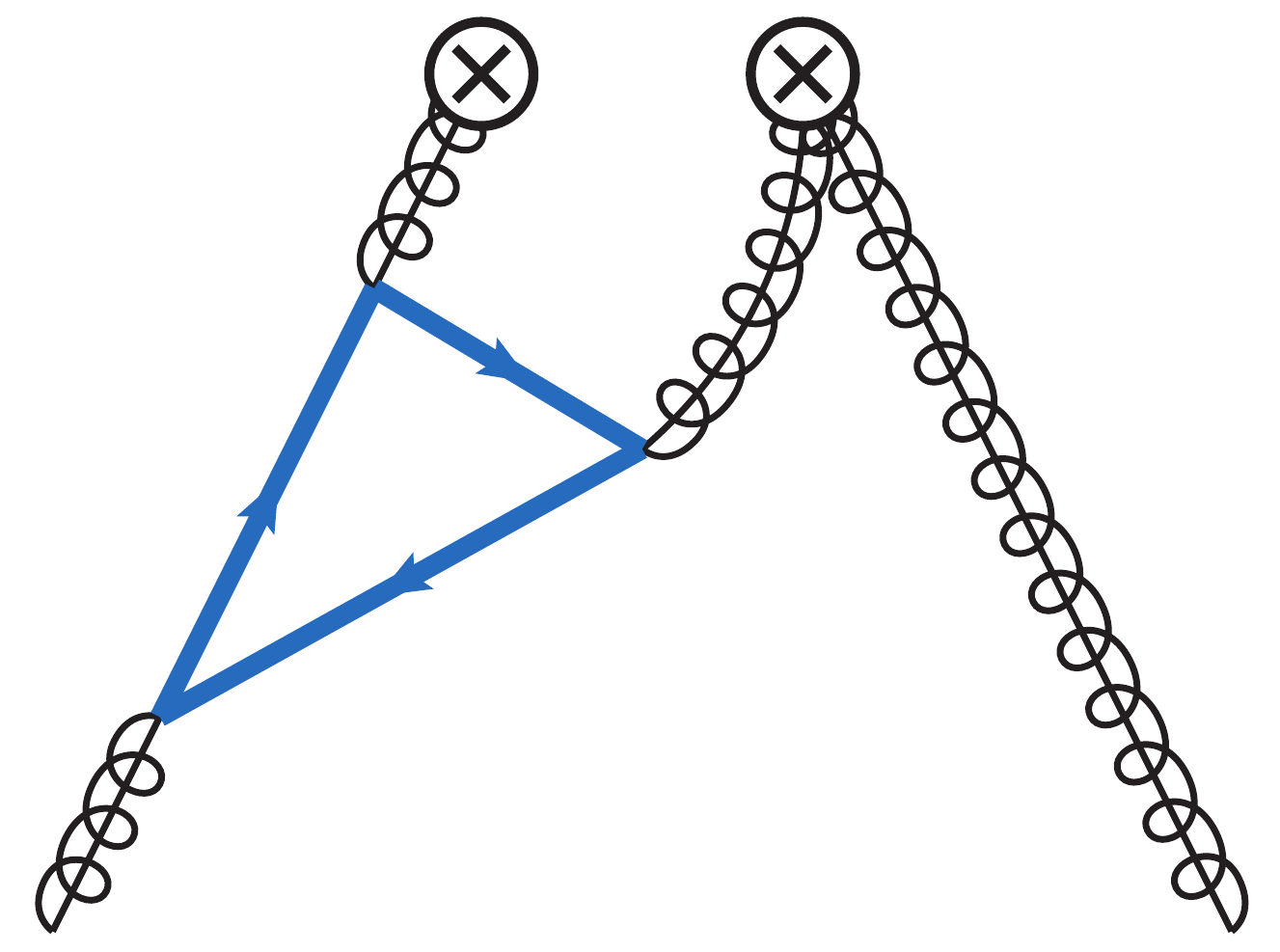}%
    \put(-58,0){(d)}

    \vspace*{3 ex}
     \includegraphics[width=0.23 \textwidth]{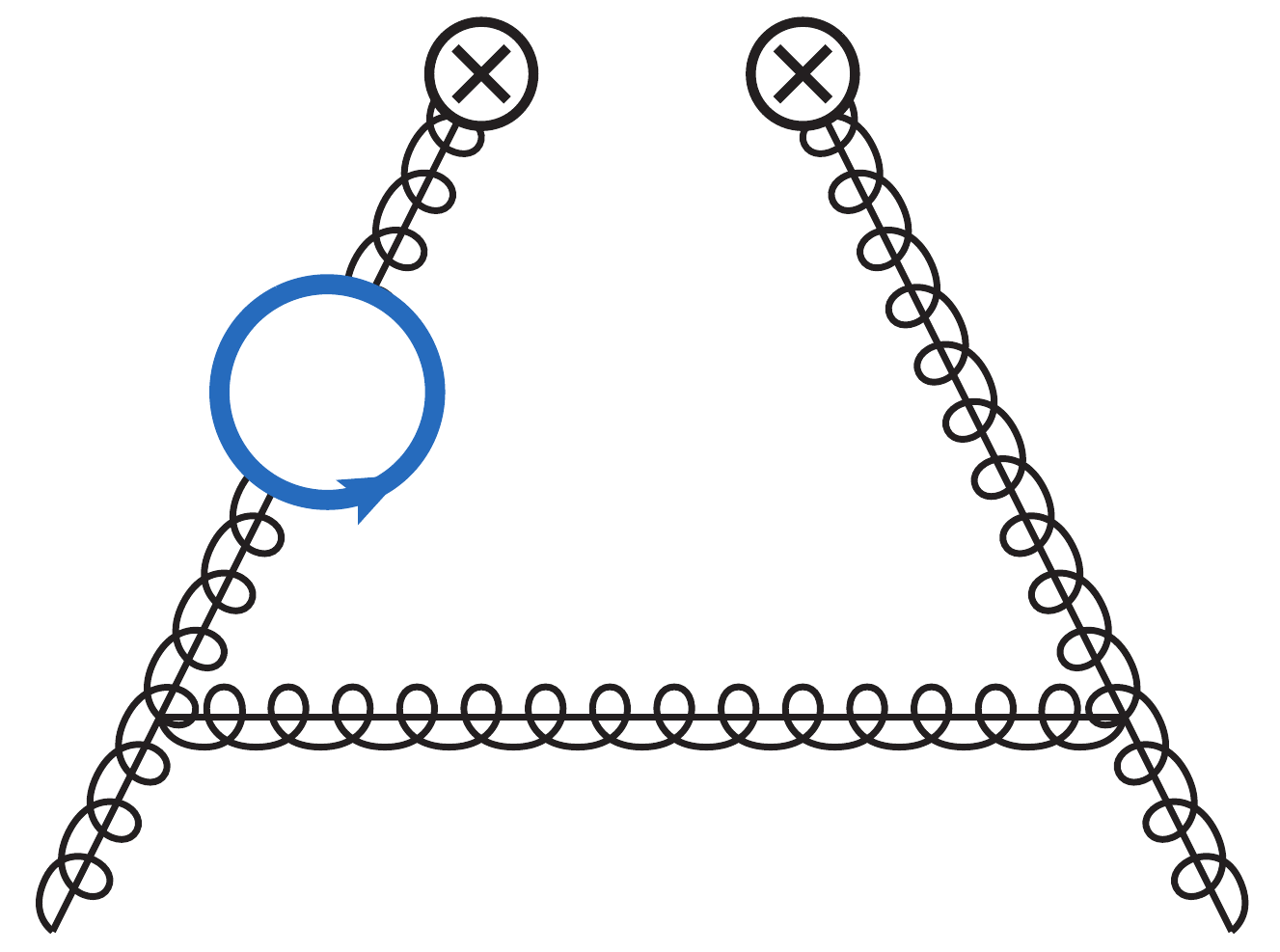}%
    \put(-58,0){(e)} \quad
    \includegraphics[width=0.23 \textwidth]{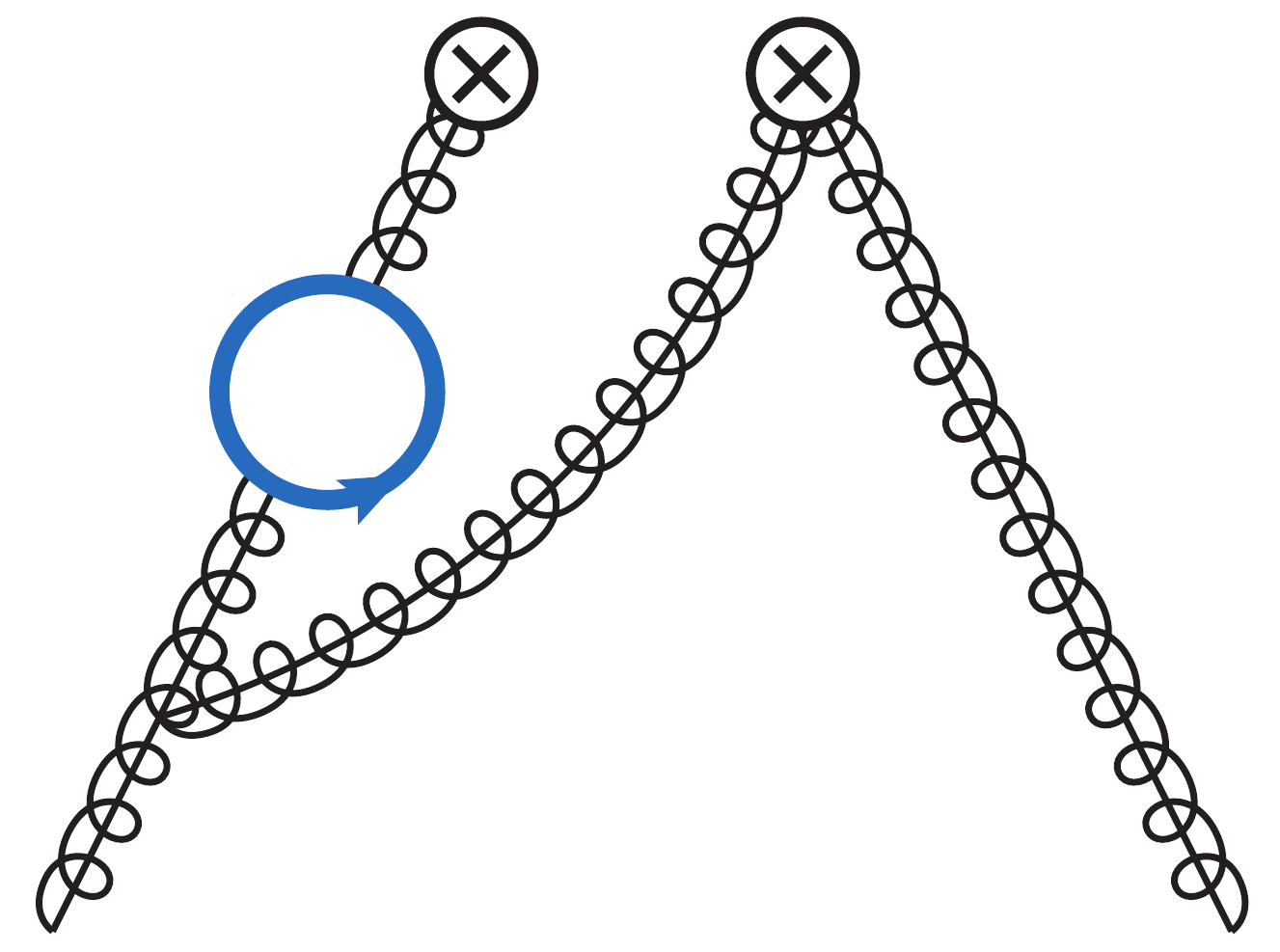}%
    \put(-58,0){(f)} \quad
    \includegraphics[width=0.23 \textwidth]{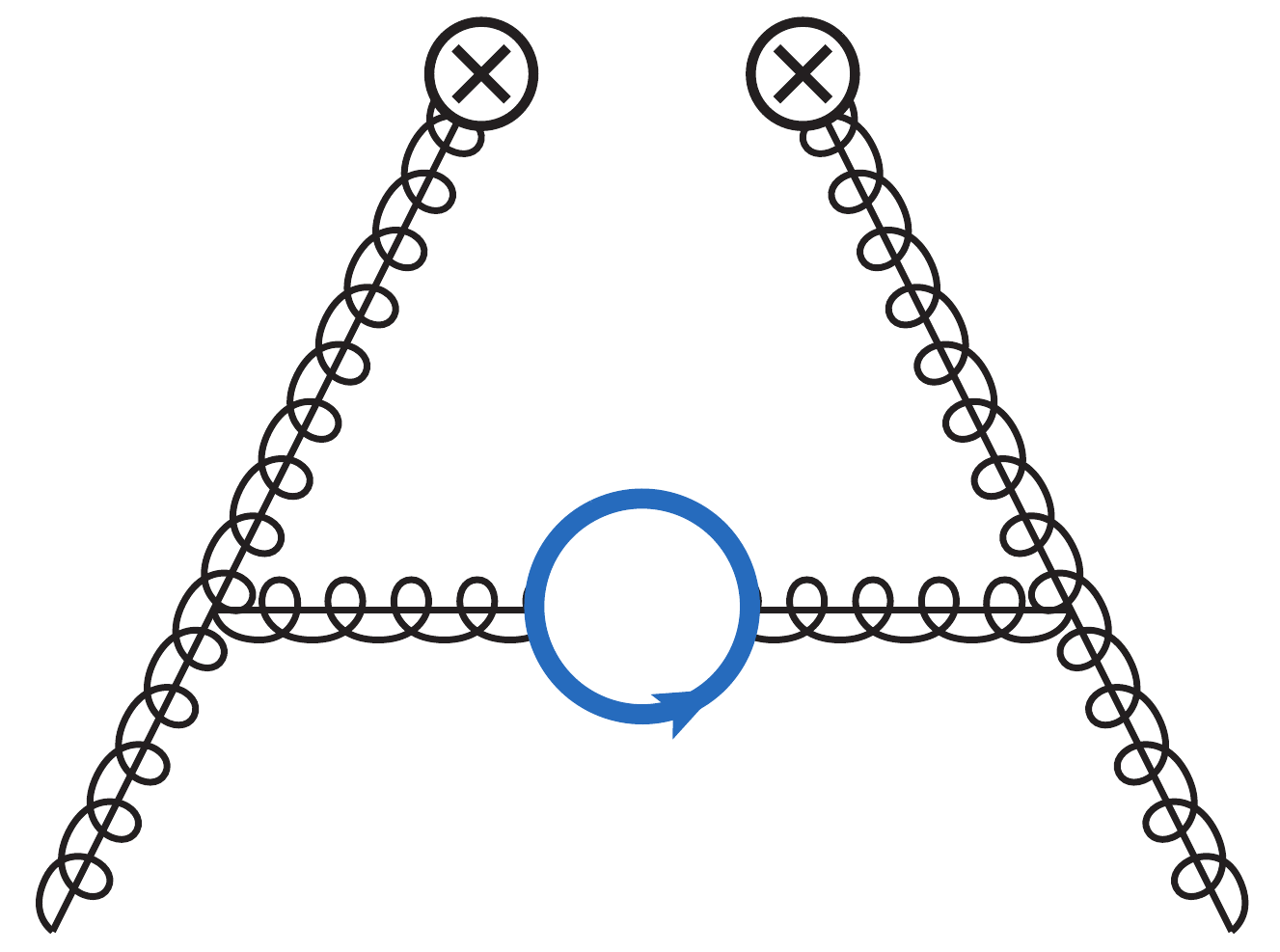}%
    \put(-58,0){(g)} \quad
     \includegraphics[width=0.23 \textwidth]{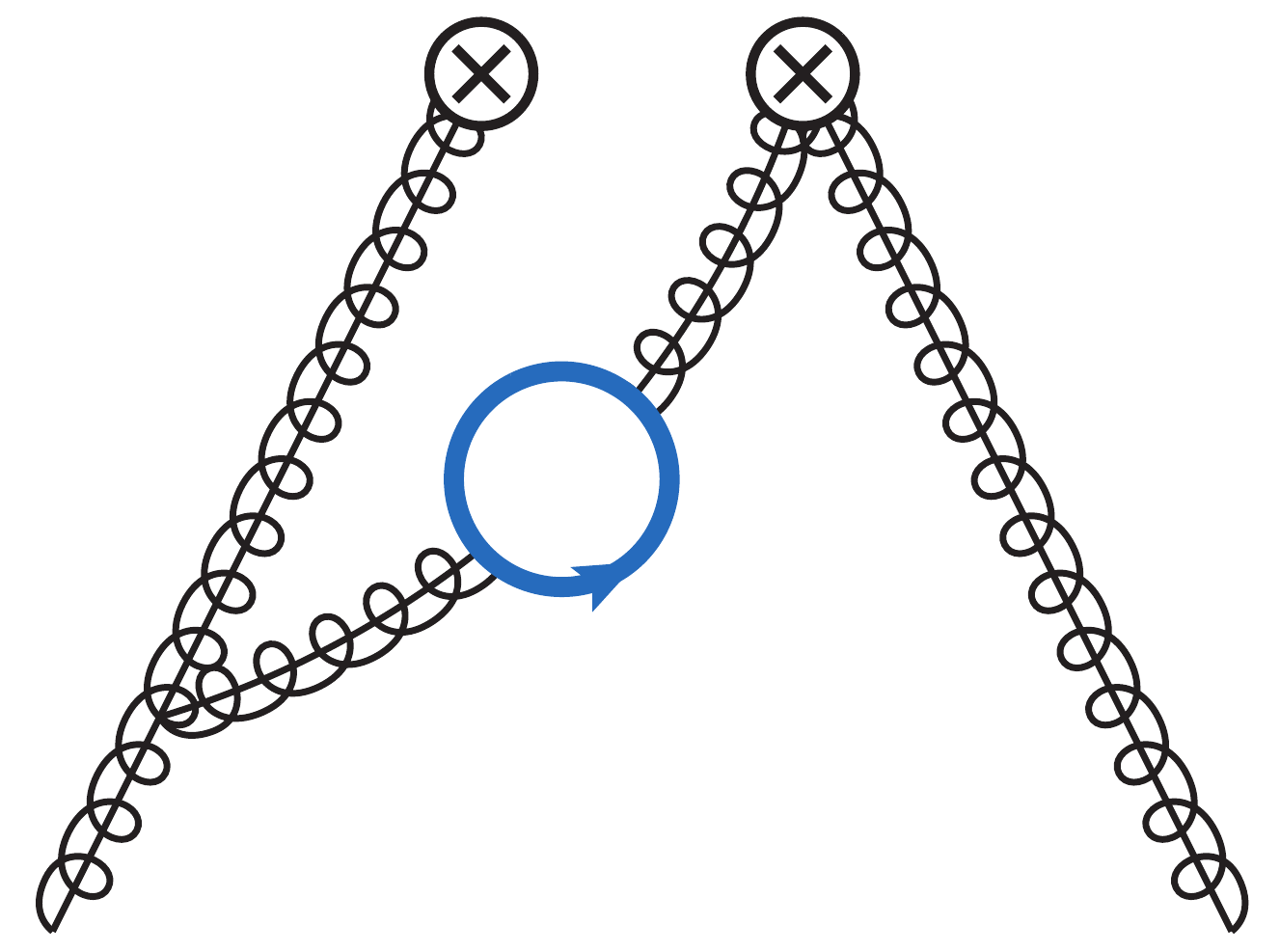}%
    \put(-58,0){(h)}

    \vspace*{3 ex}
     \includegraphics[width=0.23 \textwidth]{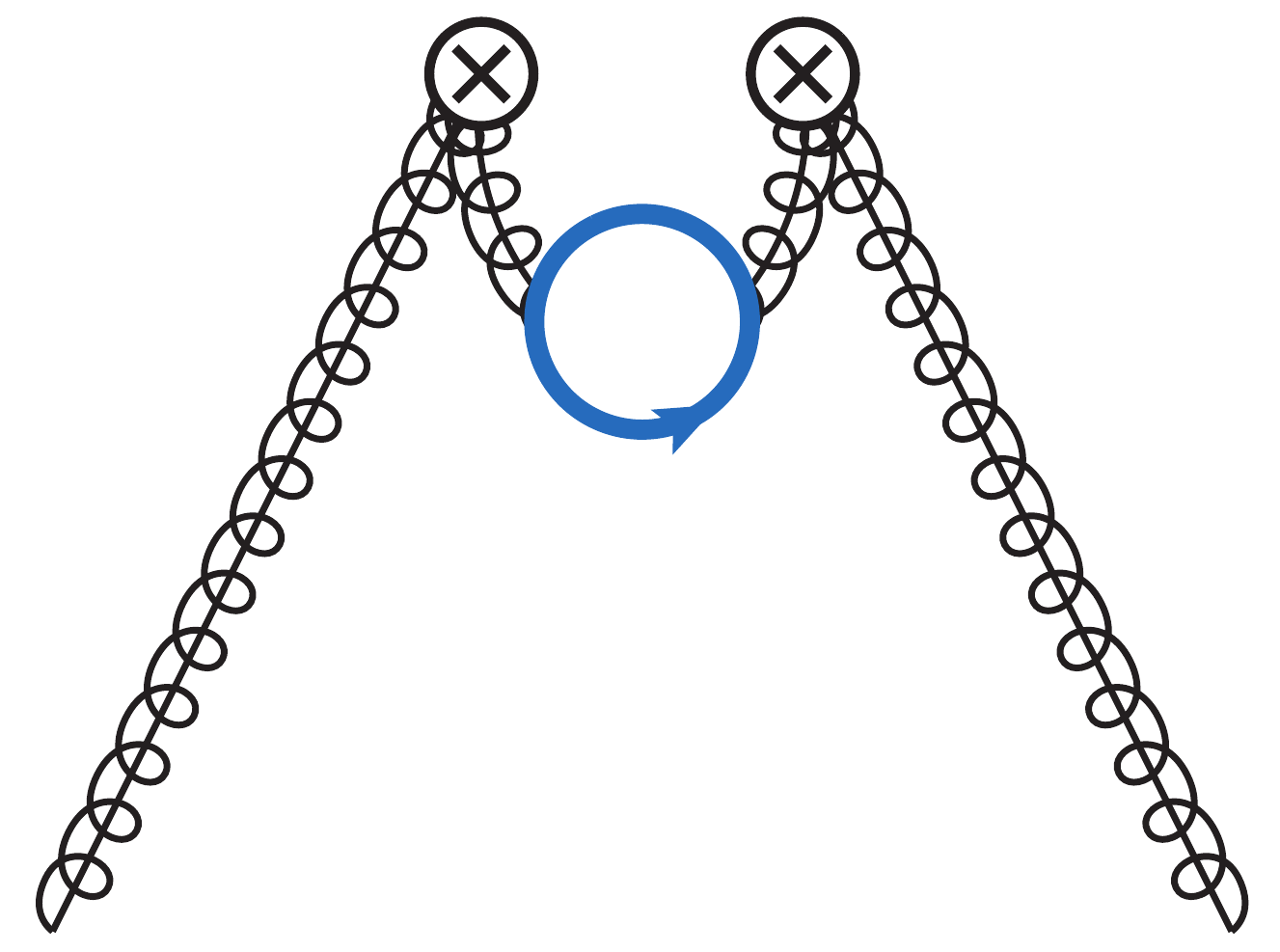}%
    \put(-58,0){(i)} \quad
    \includegraphics[width=0.23 \textwidth]{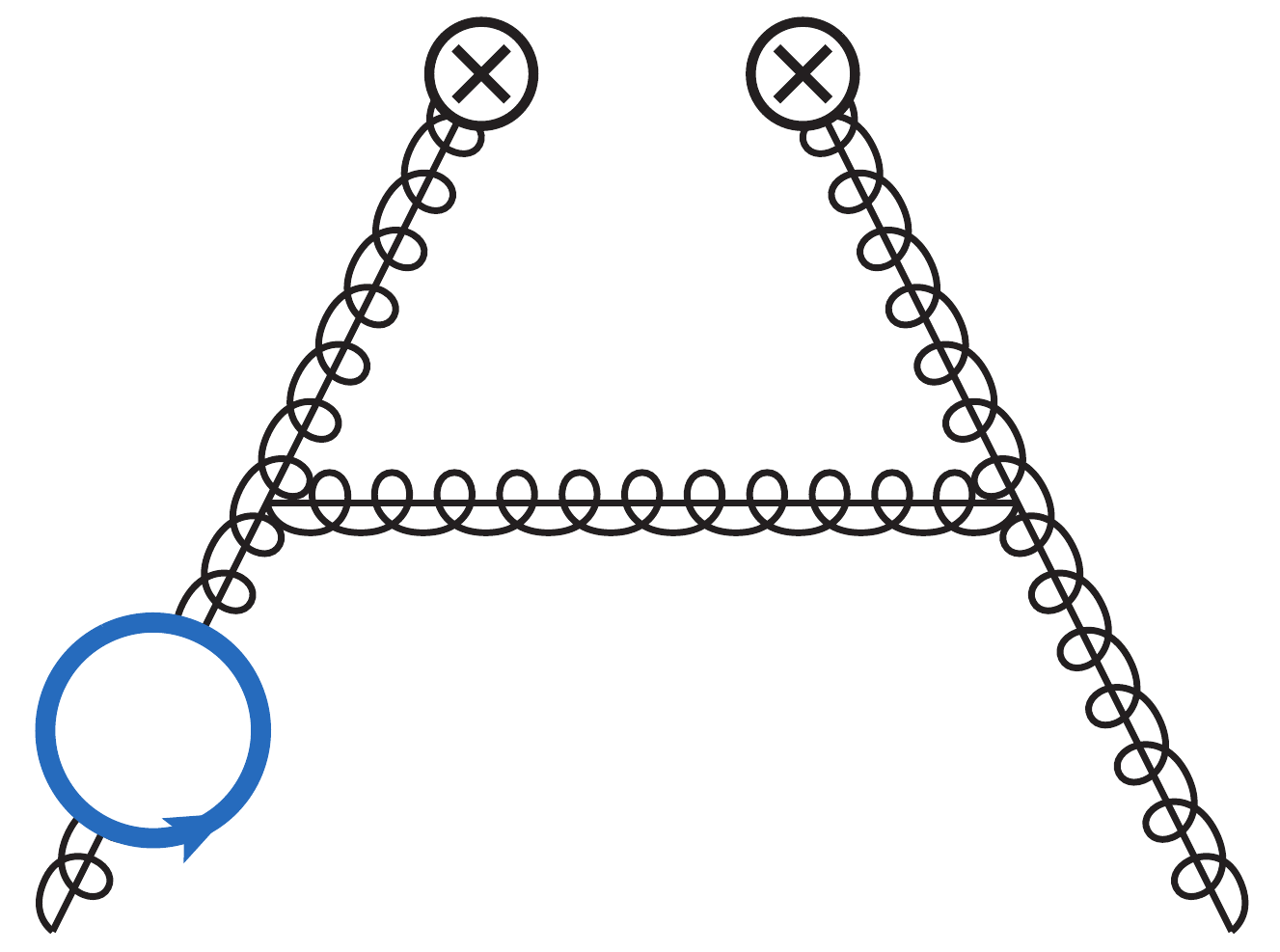}%
    \put(-58,0){(j)} \quad
    \includegraphics[width=0.23 \textwidth]{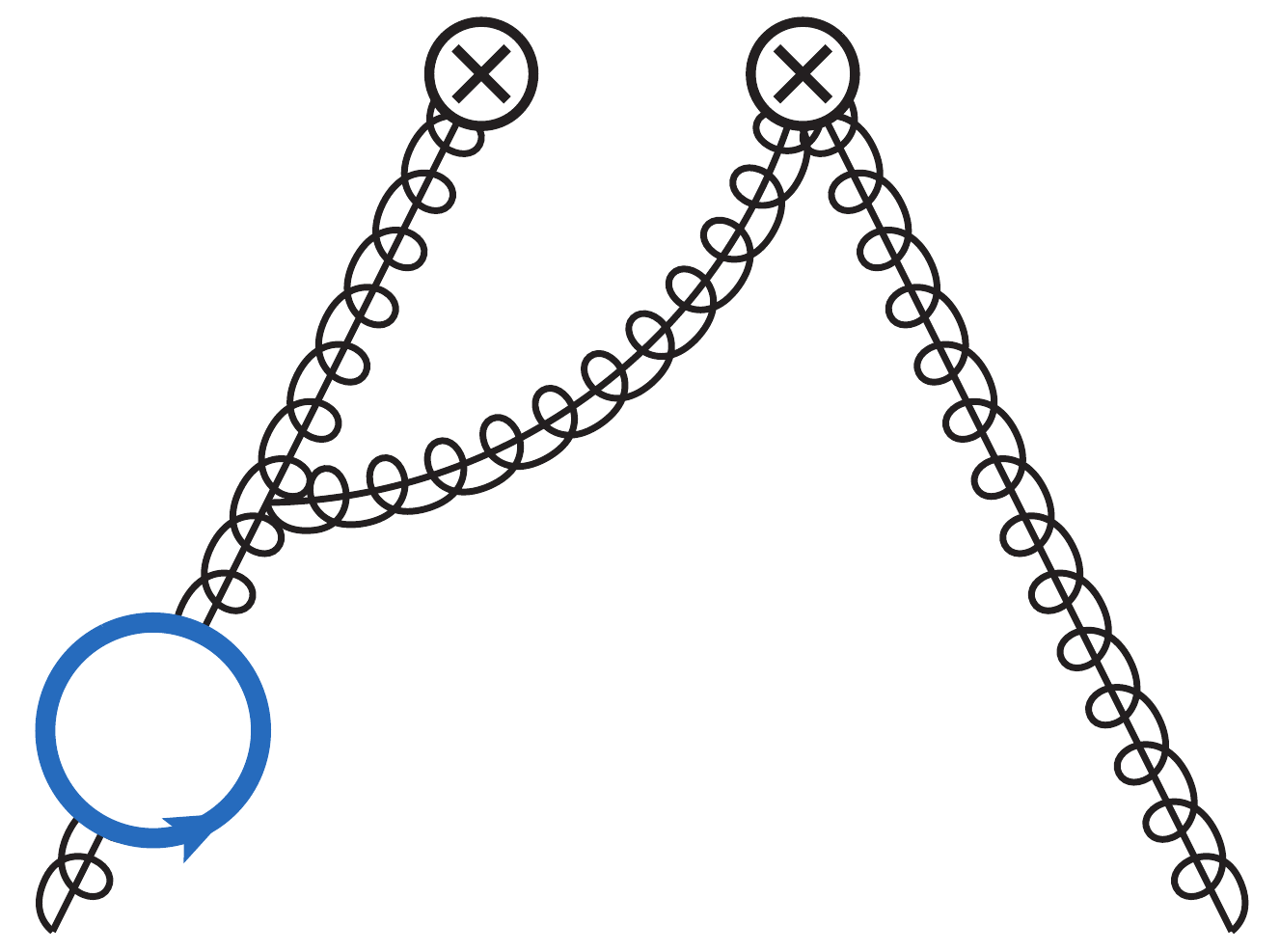}%
    \put(-58,0){(k)}
  \end{center}
  \caption{Real emission diagrams with massive quark loop (thick blue) contributing to the matching calculation of $\mathcal{I}_{gg}^{(2)}$.
  Diagrams j and k represent the one-loop massive wavefunction corrections to the leading order (massless) real contribution.
  The total contribution (sum of all cuts) of diagram~i vanishes.
  Diagrams with a massive quark bubble and a gluon attached to a Wilson line  (d, f, h, k) are rapidity divergent.
  Left-right mirror graphs are not shown, but understood.
    \label{fig:realdiags2loopgg}
  }
\end{figure}

The relevant (one- and two-particle) real emission diagrams for the computation of $B_{g/q}^{(2,h)}$ and $B_{g/g}^{(2,h)}$ are shown in \fig{diagslowestorder}b and \fig{realdiags2loopgg}, respectively.
The evaluation of the graph in \fig{diagslowestorder}b (and its left-right mirror diagram) directly yields
\begin{align}
  B_{g/q}^{(2,h)} ={}&
  \frac{\alpha_s^2 C_F T_F}{3 \pi ^3\, \pTsq}  \,\theta(x) \, P_{gq}(x)
   \biggl[\Bigl(2 (1-x)\mhsq -1\Bigr) c_{1-x}
   \ln \frac{c_{1-x}-1}{c_{1-x}+1}
   +4  (1-x)\mhsq  -  \frac{5}{3}\biggr]
    \nn\\
  &- 2  \Pi^{(1)}(0,m^2) \, B_{g/q}^{(1)} + Z_{\alpha_s}^{(1,h)} B_{g/q}^{(1)}
  + \ord{\eps}
  \,,
  \label{eq:MBgq}
\end{align}
where the splitting function $P_{gq}$ is given in \eq{Pij} and we defined for (later) convenience
\begin{align}
  \hat{m} \equiv \frac{m}{|\pT|}\,,
  \qquad
  c_y = \sqrt{1+4 y\, \mhsq}\,.
  \label{eq:mhatcdef}
\end{align}
According to the dispersion relation \eq{dispersion} the first term in \eq{MBgq} originates from the one-loop diagrams with a massive gluon propagator, while the second term comes from the massless one-loop diagram for $B_{g/q}$.
The third term is due to the conversion of the bare coupling constant to the $\MS$ renormalized
$\alpha_s \equiv \alpha_s^{\{n_l+1\}}  (\mu)$ via the heavy flavor contribution
\begin{equation}
  Z_{\alpha_s}^{(1,h)}
  = \frac{\alpha_s T_F}{4\pi}  \frac{4}{3\eps}
\end{equation}
to the one-loop $\MS$ coupling counterterm.
Note that the $\ord{\eps}$ contribution of the massless one-loop result $B_{g/q}^{(1)}$
given in \eq{Bgq1} contributes to the $\ord{\eps^0}$ part of the second and third term in \eq{MBgq}.
In contrast, the calculation of the finite first term in \eq{MBgq} can be safely performed in $d=4$ dimensions.

As noted above it is easy to see that the contributions $\propto \kappa$ from the  diagrams for $B_{g/g}^{(2,h)}$ with a massive quark bubble cancel.
In particular, using the dispersion relation in \eq{dispersion}, the calculation of diagrams \fig{realdiags2loopgg}g,h resembles the one of the real emission graphs contributing to $B_{q/g}^{(2,h)}$ in \rcite{Pietrulewicz:2017gxc}.
The graphs in \fig{realdiags2loopgg}  with a coupling to a collinear Wilson line are rapidity divergent. (Diagram \ref{fig:realdiags2loopgg}i vanishes upon integration.)
Implementing the $\eta$ rapidity regulator, according to \eq{etaregWL},  for these diagrams amounts to an overall factor of%
\footnote{In the following we will suppress the parameter $w$ and only restore it implicitly when required by the RRG formalism~\cite{Chiu:2012ir}, i.e.\ for the derivation of the $\nu$ anomalous dimension of the beam function in \eq{gammanu1h}.}
\begin{align}
  w^2 \biggl(\frac{1-x}{x} \, \frac{\omega}{\nu} \biggr)^{-\eta }.
\end{align}
This factor regulates the $1/(1-x)$ poles associated with rapidity divergences by translating them to $1/\eta$ poles via the expansion
\begin{align} \label{eq:plus_exp}
\frac{\theta(1-x)}{(1-x)^{1+\eta}} = -\frac{1}{\eta}\, \delta(1-x) + \sum_{n = 0}^\infty \frac{(-\eta)^n}{n!}\, \cL_n(1-x)
= -\frac{1}{\eta}\, \delta(1-x) + \cL_0(1-x) + \ord{\eta}
\end{align}
in terms of the plus distributions
\begin{align} \label{eq:plusdef}
\cL_n(y)
&\equiv \biggl[ \frac{\theta(y) \ln^n y}{y}\biggr]_+
 = \lim_{\eps \to 0} \frac{\df}{\df y}\biggl[ \theta(y- \eps)\frac{\ln^{n+1} y}{n+1} \biggr]
\,.
\end{align}
The rapidity divergences of diagrams \ref{fig:realdiags2loopgg}d-f (and their mirror graphs) cancel exactly.
The rapidity divergence of diagram \ref{fig:realdiags2loopgg}h instead cancels with its soft analog in \fig{softdiags}a within the cross section in \eq{factXsec} at fixed order.

The one-real-gluon cuts of the diagrams in \fig{realdiags2loopgg} (if present) give rise to
terms singular in $\pTsq$, i.e.\ proportional to $\delpT$ or the plus distributions%
\footnote{In \rcites{Chiu:2012ir,Luebbert:2016itl} the notation
$\mathcal{L}_n^T(\pT,\mu) \equiv \frac{(-1)^n}{2} \mathcal{L}_n(\pT,\mu)$ was used.
Useful properties and convolutions of the $\mathcal{L}_n(\pT,\mu)$ are summarized in \rcite{Ebert:2016gcn}.
}
\begin{align}
  \mathcal{L}_n(\pT,\mu) \equiv \frac{1}{\pi \mu^{2}}\,
  \mathcal{L}_n\biggl(\frac{\pTsq}{\mu^{2}}\biggr)
  \,,
  \label{eq:LpTdef}
\end{align}
via the expansion
\begin{align}
  \frac{1}{\pi} \frac{(\mu^2)^{\eps}}{(\pTsq)^{1+\eps}} =
      -\frac{1}{\eps} \delpT + \sum^\infty_{n=0} \frac{(-\eps)^n}{n!} \,
   \mathcal{L}_n(\pT,\mu)
   \,.
\end{align}
The cuts through two massive quark lines  result in regular (non-singular) functions of $\pTsq$.
The reason is that the limit $\pTsq \to 0$ is effectively tied to the limit $m^2 \to \infty$, where massive quarks in the final state are kinematically not allowed.
The massive cuts therefore yield terms proportional to $1/m^2$ rather than to $1/\pTsq$ (in $d=4$ dimensions) and therefore need no regularization in terms of distributions.

The total zero-bin contribution to $B_{g/g}^{(2,h)}$ associated with real emission diagrams is scaleless and vanishes similar to that from the massless diagrams.
Details on the corresponding zero-bin calculation are presented in \app{realZBs}.


\subsection{Virtual diagrams}
\label{subsec:virtualdiags}

\begin{figure}[t]
  \begin{center}
    \vspace*{-2 ex}
    \includegraphics[width=0.23 \textwidth]{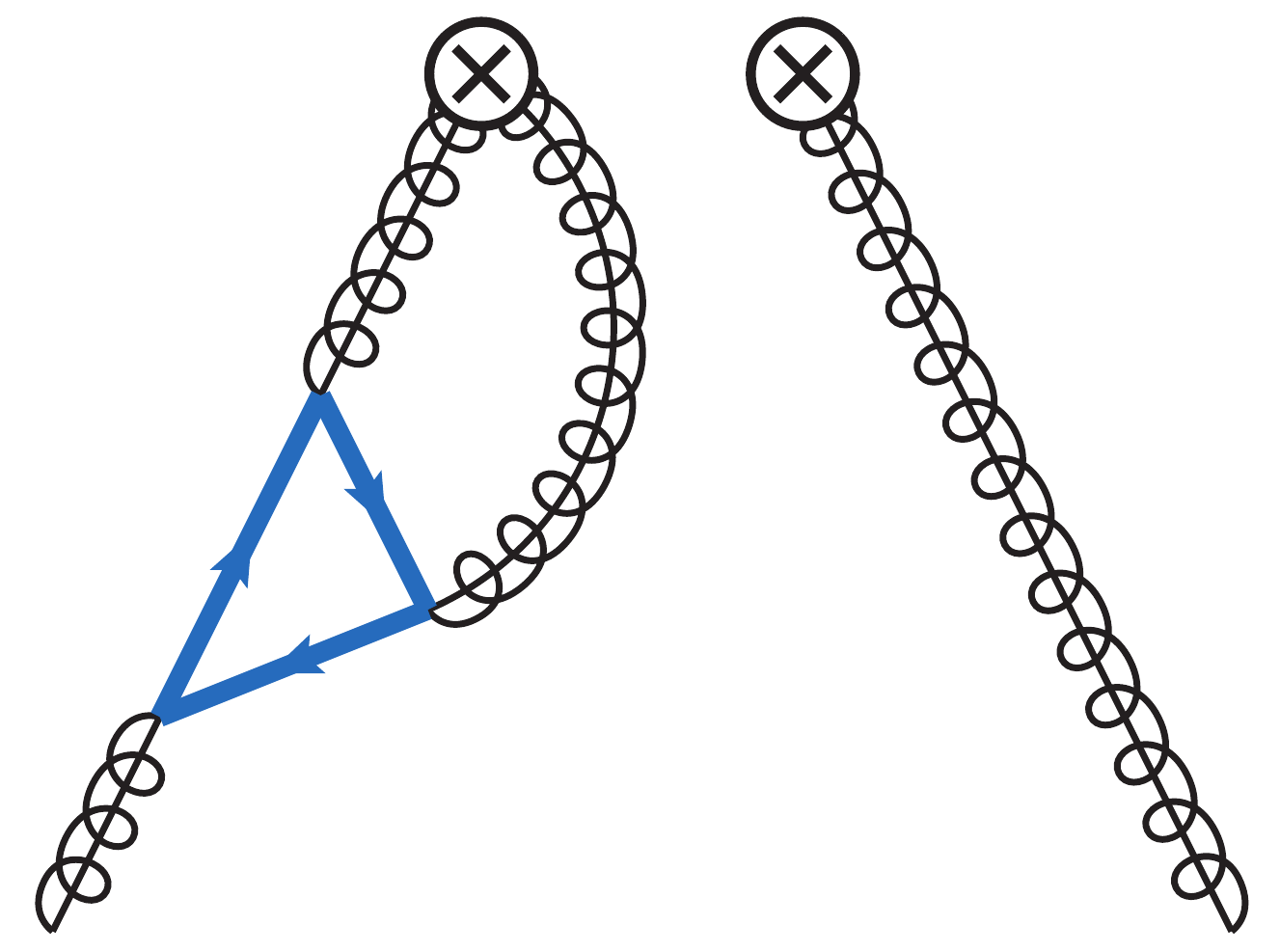}%
    \put(-58,0){(a)} \qquad
    \includegraphics[width=0.23 \textwidth]{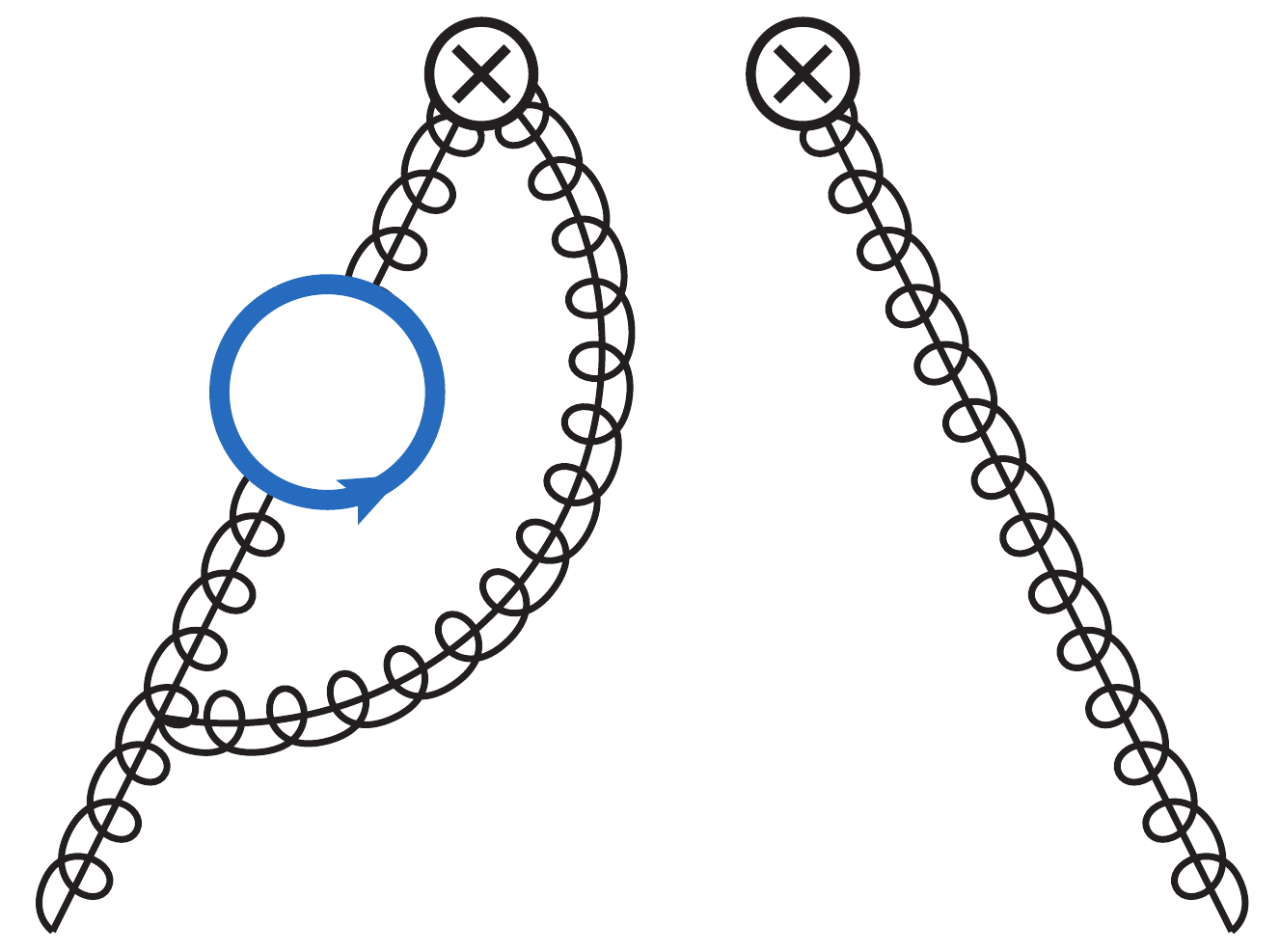}%
    \put(-58,0){(b)} \qquad
    \includegraphics[width=0.23 \textwidth]{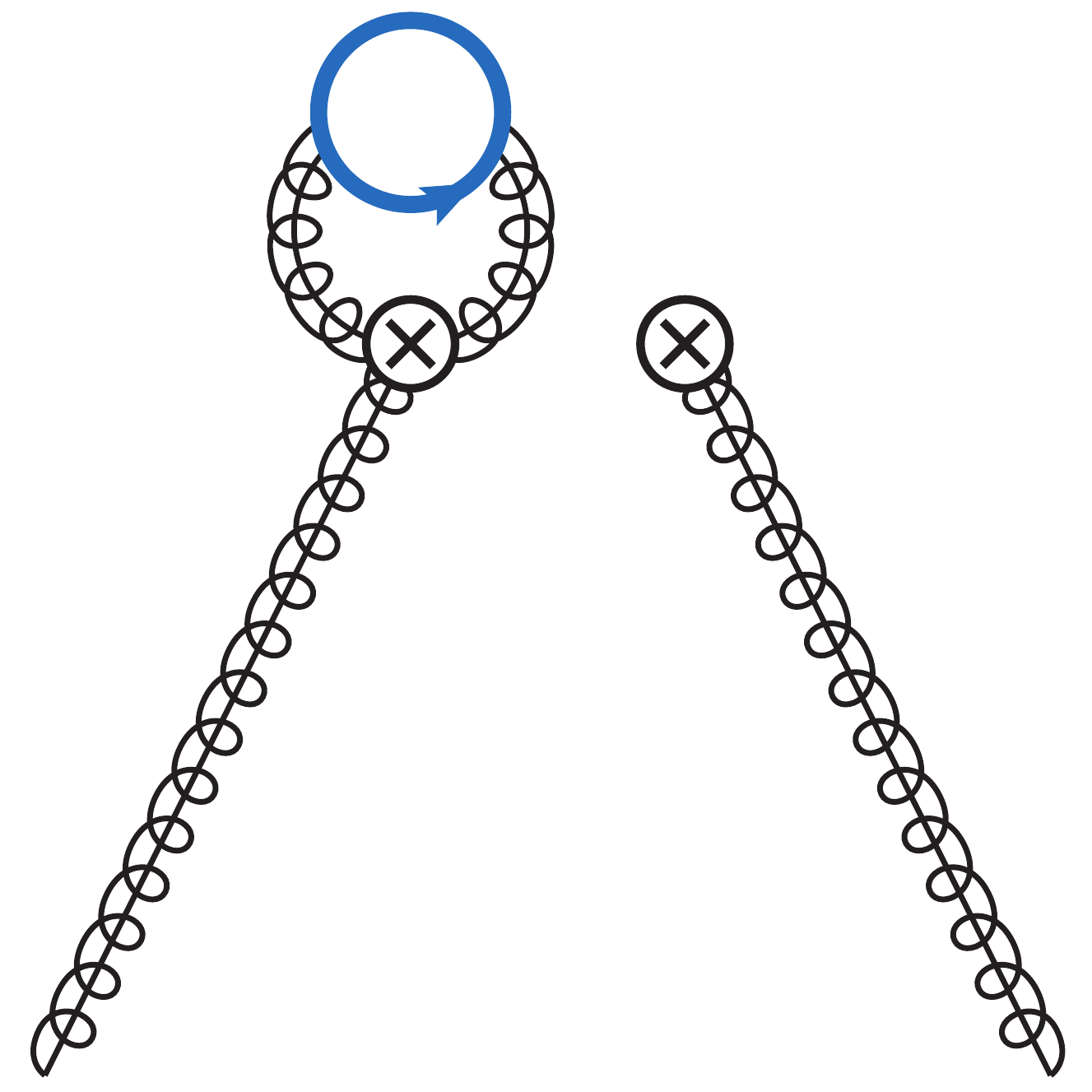}%
    \put(-58,0){(c)}

    \vspace*{3 ex}
    \includegraphics[width=0.23 \textwidth]{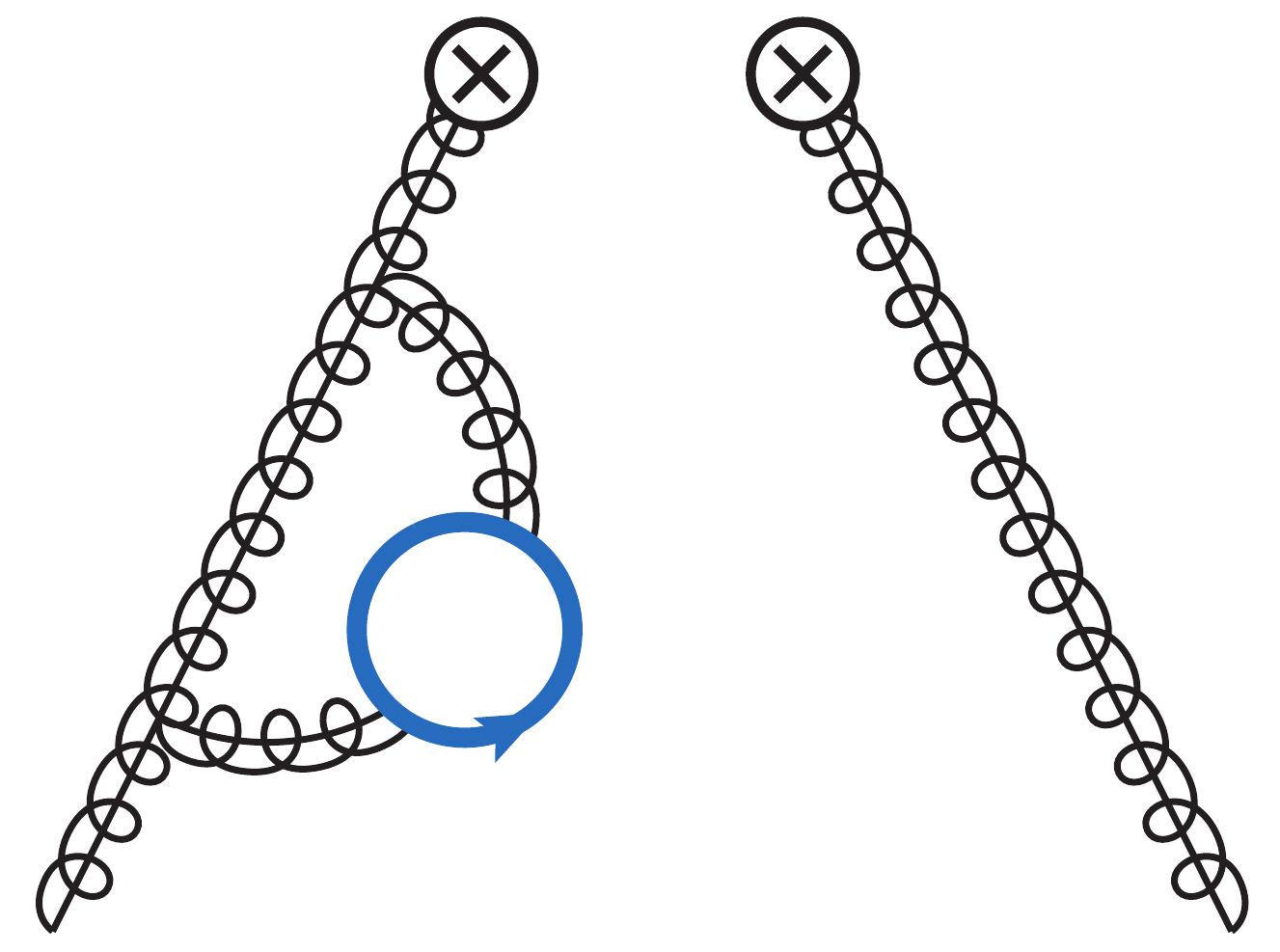}%
    \put(-58,0){(d)} \quad
     \includegraphics[width=0.23 \textwidth]{Figs/BeamFct1loopMassive_gg.pdf}%
    \put(-84.5,38){\tiny{2-}}
    \put(-88.2,33){\tiny{loop}}
    \put(-58,0){(e)}
    \quad
    \includegraphics[width=0.23 \textwidth]{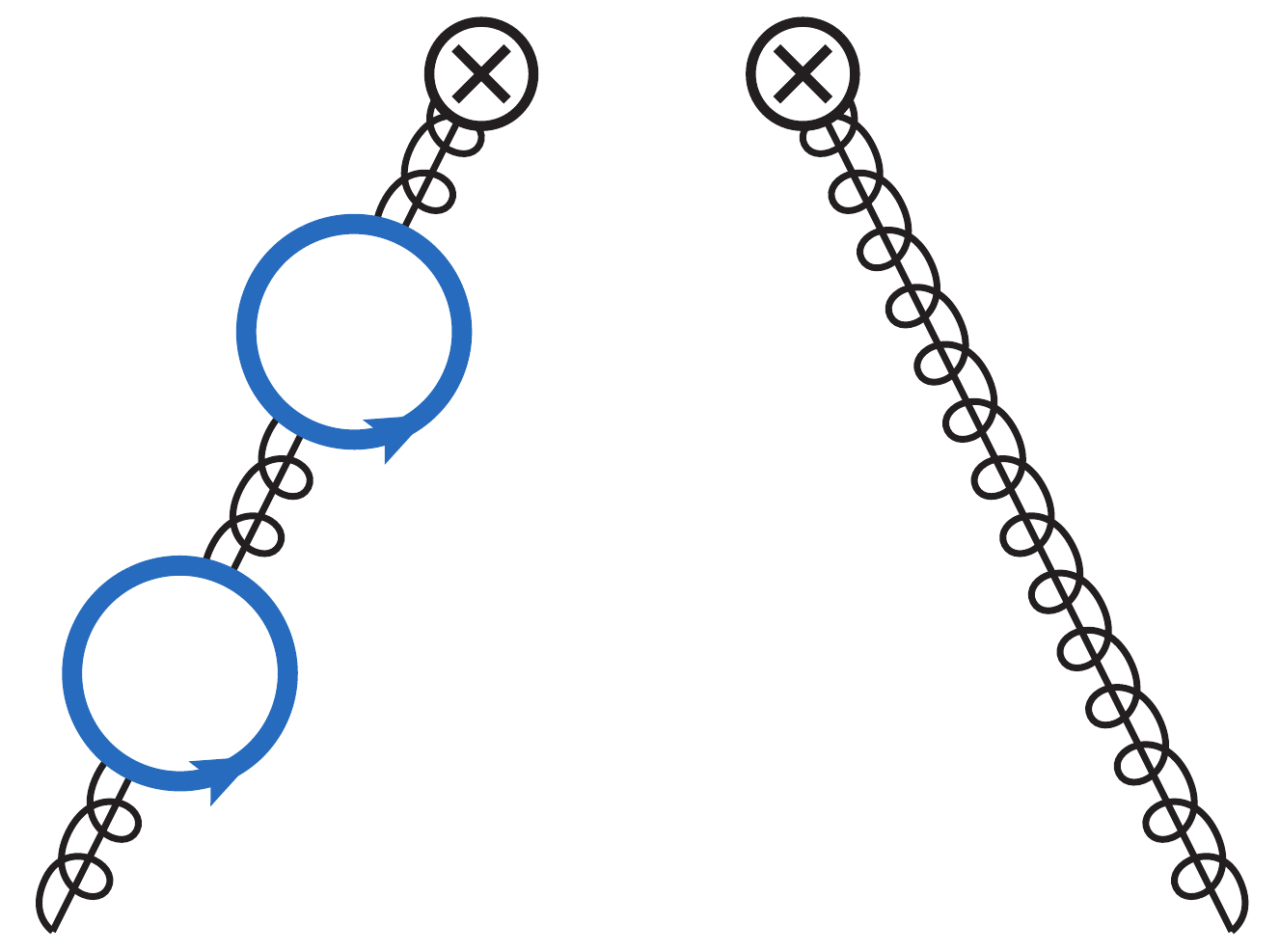}%
    \put(-58,0){(f)} \quad
     \includegraphics[width=0.23 \textwidth]{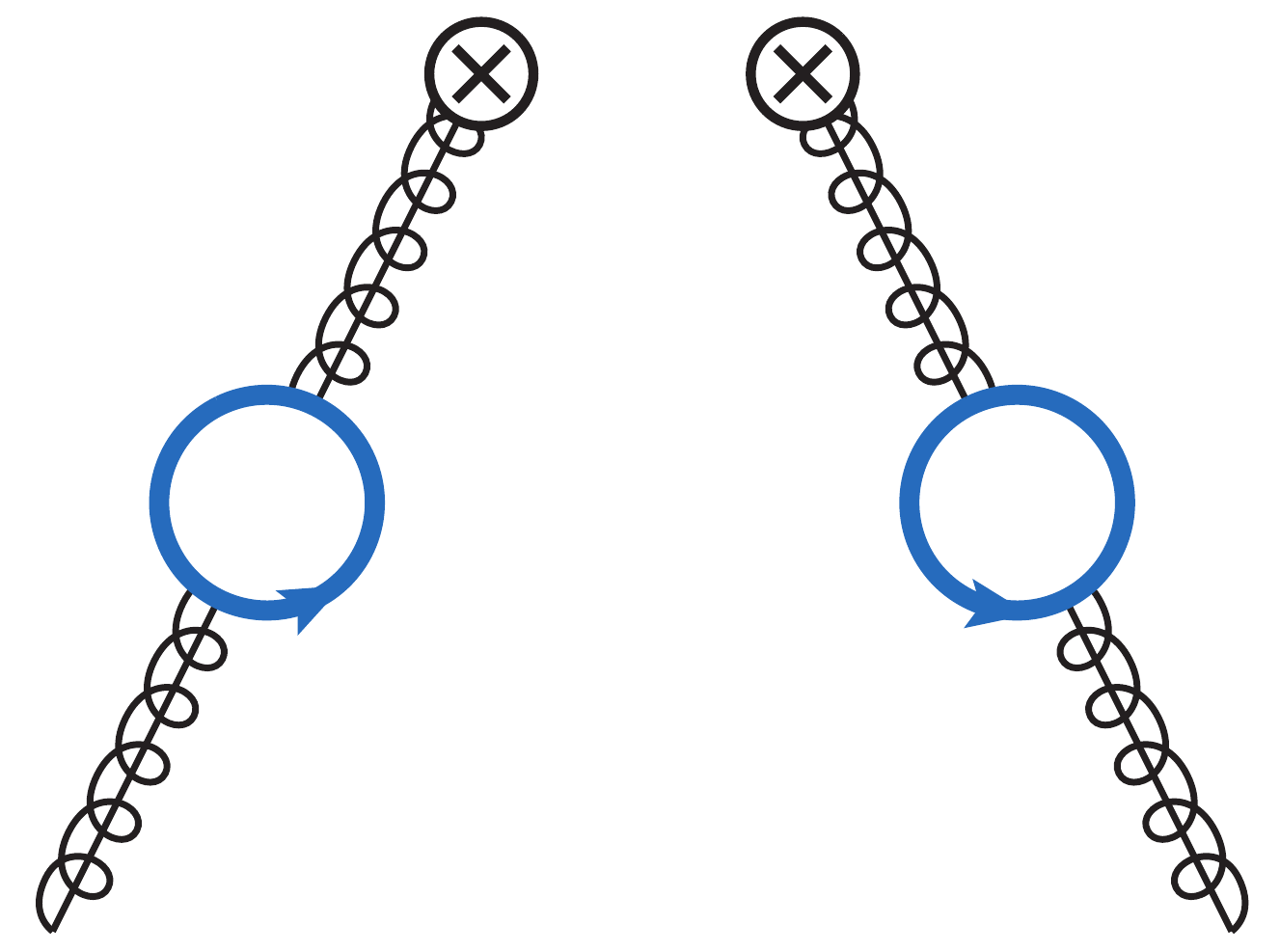}%
    \put(-58,0){(g)}
\end{center}
\caption{Purely virtual two-loop diagrams for the calculation of $\mathcal{I}_{gg}^{(2)}$. Left-right mirror graphs are understood.
  Diagrams~c-g represent wavefunction renormalization corrections.
  Diagram~c corresponds to a Wilson-line self-energy graph (and vanishes after zero-bin subtraction).
  The two-loop bubble in diagram~e symbolizes the sum of all 1PI two-loop vacuum polarization subdiagrams involving the massive quark flavor. Diagram~d is part of diagram~e and contains the complete $\kappa$ term of the two-loop wavefunction contribution. For consistent wave function renormalization the graphs d, e, f, g (as well as their mirror diagrams) have to be multiplied with the factors $1/2$, $1/2$, $3/8$, $1/4$, respectively.
\label{fig:virtdiags}
}
\end{figure}

The one-loop contribution to $B_{g/g}$ due to a massive quark flavor is given by the diagram in
\fig{diagslowestorder}a. It corresponds to a wavefunction-type correction to the tree-level result $B_{g/g}^\zero$ and reads
\begin{align}
  B_{g/g}^{(1,h)} &=
  -\delta(1-x)\,\delta^\two(\pT)\, \Pi^{(1)}(0,m^2)
  \nn\\
  &=
  \frac{\alpha_s T_F}{4\pi}\,
  \delta(1-x)\,\delta^\two(\pT)
  \biggl[
  -\frac{4}{3  \eps}
  +\frac{4 L_m}{3}-\frac{6 L_m^2+\pi ^2 }{9}\,\eps
  +\ord{\eps^2}
  \biggr]\,.
  \label{eq:MBgg1}
\end{align}
Here and in the following we use the shorthand notation
\begin{equation}
  L_m \equiv \ln \biggl(\frac{m^2}{\mu ^2}\biggr)\,.
\end{equation}

The two-loop virtual diagrams contributing to $B_{g/g}^{(2,h)}$ are displayed in \fig{virtdiags}.
As stated above and shown by explicit calculation in \app{virtualZBs} the total virtual zero-bin contribution to $B_{g/g}^{(2,h)}$ exactly equals and thus cancels the terms proportional to the bookkeeping parameter $\kappa$
in the unsubtracted expression for $B_{g/g}^{(2,h)}$.
The calculation of the virtual diagrams is somewhat more involved than the one of the real emission graphs in \fig{realdiags2loopgg}, because the loop integrations are not constrained by the measurement $\delta$-functions in \eq{Bgdef}.
In fact, up to an overall $\delpT$ the virtual contribution equals the one of the massive PDF matching coefficient $\mathcal{M}^{(2)}_{gg}$ computed in direct QCD~\cite{Buza:1996wv}.
Given the complete (one- and two-particle) real-emission contribution we could even extract the virtual part from the known $\mathcal{M}^{(2)}_{gg}$ by taking the small mass ($m \ll \pT$) limit of the TMD beam function.
It is nevertheless instructive to calculate the virtual diagrams (and their zero-bins) in our SCET setup. In \subsec{smallmasslimit} we will then use the small mass limit as a strong cross check of our explicit calculation.
In the following we briefly discuss the evaluation and features of the different virtual diagrams in \fig{virtdiags}:\\

\underline{Diagram~\ref{fig:virtdiags}a}
  has a massive quark triangle subgraph and is arguably the most difficult to compute. As there is no corresponding soft diagram with a massive triangle, all zero-bins are power-suppressed and do not contribute to $B_{g/g}^{(2,h)}$.
  For the same reason the diagram is rapidity-finite and does a priori not require any rapidity regulator.
  To simplify the involved two-loop Feynman integrals we want to use automated integration by parts (IBP) reduction to a minimal set of master integrals, a standard tool of modern multi-loop calculations.
  However, when eikonal (Wilson line) and massive propagators are present at the same time a naive implementation of IBP reduction often fails, see e.g.\ the discussion in \rcite{Hoang:2019fze}.
  For the unregulated rapidity-finite two-loop diagram~\ref{fig:virtdiags}a the IBP program FIRE5~\cite{Smirnov:2014hma} for example leads to a rapidity-divergent expression due to unregulated rapidity-divergent master integrals.
  In the present case there is a pragmatic solution to this problem:
  We introduce an auxiliary rapidity regulator which ensures well-defined integrals in the result and at intermediate steps of the IBP reduction.
  After evaluation of the master integrals the dependence of the IBP reduced  expression of the diagram on the rapidity regulator cancels out. For practical reasons we choose the $\delta$-regulator for this calculation. The IBP reduction with FIRE5~\cite{Smirnov:2014hma} yields four master integrals, which we solved  by direct integration in Feynman parameter representation as functions of $m$ and $\delta/p^-$ (in the limit $\delta \to 0$).
  The final result for diagram~\ref{fig:virtdiags}a is $\delta$-independent, but depends on the gauge parameter $\xi$. The $\xi$-dependent terms exactly cancel the ones of diagram~\ref{fig:virtdiags}b.
  As a check we computed the $\eps$ poles of the diagram (for $\xi=0$) also by direct loop integration without IBP reduction and rapidity regulation.

\underline{Diagram~\ref{fig:virtdiags}b} is rapidity divergent and $\xi$-dependent.
Implementing the $\eta$-regulator according to \eq{etaregWL} it entails rapidity-finite terms ($\propto \kappa$) that are however discontinuous in the $\eta \to 0$ limit as discussed in \subsec{rapreg}. These terms cancel exactly after zero-bin subtraction, see \app{virtualZBs}, leaving a smooth result for the rapidity-finite part as $\eta \to 0$ (which therefore can also be computed without regulator).
Like for all relevant diagrams with a massive quark bubble a particular convenient and transparent way to calculate this diagram is via the dispersion relation as described in \subsec{dispersionmethod}.
Note that for the virtual diagrams the $\Pi(0,m^2)$ term in \eq{dispersion} only gives rise to a vanishing contribution proportional to the corresponding virtual massless one-loop diagram, and can thus be dropped.

\underline{Diagram~\ref{fig:virtdiags}c} is proportional to the bookkeeping parameter $\kappa$ when deploying \eq{dispersion}.
To derive the integrand one must carefully implement the Feynman rule for the triple gluon field strength vertex given in app.~D of \rcite{Gaunt:2014cfa}.
One way to do this is to assign an auxiliary offshellness $\delta$ to the Wilson line propagators and perform the necessary color index contractions.
The resulting Feynman integral then takes the form of $I_\delta(1,1)$ in \eq{Idelta}.
Setting $\delta \to 0$ before the loop integration
we end up with a result for diagram~\ref{fig:virtdiags}c that is proportional to $I_0(1,1)$ in \eq{I0Int} and corresponds to a (soft) Wilson line self-energy  correction, cf.\ \app{softreview}.
This explains why Diagram~\ref{fig:virtdiags}c exactly vanishes upon zero-bin subtraction, see \app{virtualZBs}.
It can thus be effectively omitted in the calculation of $B_{g/g}^{(2,h)}$ as a whole (independent of whether or not an unnecessary rapidity regulator is applied).

\underline{Diagram~\ref{fig:virtdiags}e} represents the full (QCD) two-loop wave function correction from 1PI massive quark vacuum polarization diagrams.
One specific diagram included in this correction is diagram~\ref{fig:virtdiags}d.
It contains the complete $\kappa$ term of the two-loop wave function renormalization.
The complete two-loop 1PI wavefunction contribution from a heavy flavor can be obtained from diagram~\ref{fig:virtdiags}e by inserting the known expression (with $\kappa=1$) for the massive vacuum polarization function $\Pi^{(2)}(0,m^2)$ given in \rcite{Bierenbaum:2008yu}.%
\footnote{We recomputed $\Pi^{(2)}(0,m^2)$ and found the same result as \rcite{Bierenbaum:2008yu} up to a different overall sign. This might be a typo or due to a different (undocumented) convention. We fully agree with their $\Pi^{(1)}(0,m^2)$ though. For the calculation of diagram~\ref{fig:virtdiags}e we use our result for $\Pi^{(2)}(0,m^2)$, which passes the cross checks described in \sec{limits}.}
In order to verify our statement of \subsec{dispersionmethod} that dropping the $\kappa$ terms (and thus the zero-bins) also in the purely virtual contribution yields the correct result we must carefully extract the $\kappa$ term due to wavefunction renormalization.
To this end we evaluate
\begin{align}
  \textrm{diagram~\ref{fig:virtdiags}d} \;\propto\;
  \biggl[ \frac{\partial}{\partial {p^2}} \,  g_\perp^{\mu \nu} \Pi^{(2,\kappa)}_{\mu \nu} (p^2,m^2) \biggr]_{p^2 \to 0}\,,
  \label{eq:diag3d}
\end{align}
where $\Pi^{(2,\kappa)}$ denotes the $\kappa$ Term of the gluon self-energy subgraph in diagram~\ref{fig:virtdiags}d (suppressing color indices).
Note that the gluon propagator connecting the subgraph with the $\ord{g^0}$ vertex for the
$\cB_{n\perp}^{\sigma c} \cB_{n\perp\sigma}^{c}$ operator is
($\propto g_\perp^{\mu\nu}$ and) in fact replaced by the derivative $\partial/ \partial {p^2}$.
This is necessary, because $g_\perp^{\mu \nu} \Pi^{(2,\kappa)}_{\mu \nu}/p^2$ (unlike the full $g_\perp^{\mu \nu} \Pi^\two_{\mu \nu}/p^2$)  is not regular in the on-shell limit $p^2 \to 0$. The limit is equivalent to $p^+\to0$ with $p^2= p^-p^+$ and may conveniently be taken after the derivative but before the loop integration in \eq{diag3d}.
The contribution of diagram~\ref{fig:virtdiags}d is subtracted from that of diagram~\ref{fig:virtdiags}e (with $\kappa=1$) to obtain the two-loop 1PI wavefunction correction to $B_{g/g}^{(2,h)}$ without $\kappa$ term (and zero-bins).

\underline{Diagrams~\ref{fig:virtdiags}f and~\ref{fig:virtdiags}g} represent the one-particle reducible contributions from wavefunction renormalization.
They are straightforward to compute using \eqs{Pimunudef}{Pioneloop}.
As usual the external leg correction diagrams~\ref{fig:virtdiags}d-g (and their left-right mirror graphs) must be multiplied with the fractional numbers given in the caption of \fig{virtdiags} to obtain the correct wavefunction renormalization contribution.\\

The soft zero-bin contributions of the  virtual diagrams are necessary when the virtual $\kappa$ terms are included.
Their calculation is discussed in \app{virtualZBs}.
When expressing the bare result for $B_{g/g}^{(2,h)}$ in terms of the $\MS$ renormalized coupling $\alpha_s \equiv \alpha_s^{\{n_l+1\}}  (\mu)$ and on-shell renormalized mass $m$ we have to add the counterterm contributions
\begin{equation}
  Z_{\alpha_s}^{(1,h)} B_{g/g}^{(1,l)}
  + Z_{\alpha_s}^{(1)} B_{g/g}^{(1,h)}
  + B_{g/g}^{(1,h)}(\pT,m + \delta m^\one, x)\big|_{\alpha_s^2}\,.
  \label{eq:MBggCTcontribs}
\end{equation}
The massless one-loop result $B_{g/g}^{(1,l)}$ is given in \eq{Bgg1l}.
In the second term of \eq{MBggCTcontribs} the one-loop heavy-flavor contribution in \eq{MBgg1} is multiplied with the full one-loop coupling counterterm for $n_l+1$ active flavors,
\begin{equation}
   Z_{\alpha_s}^{(1)} = Z_{\alpha_s}^{(1,h)} + Z_{\alpha_s}^{(1,l)} =
   - \frac{\alpha_s}{4\pi }\, \frac{\beta_0^{\{n_l+1\}}}{\eps}  =
   - \frac{\alpha_s}{4\pi} \frac{1}{\eps} \biggl( \frac{11}{3} C_A - \frac{4}{3} (n_l+1) T_F \biggr)\,.
\end{equation}
The one-loop mass counterterm in the on-shell scheme is
\begin{equation}
  \delta m^\one = \frac{\alpha_s C_F}{4 \pi} m
  \biggl[
  -\frac{3}{\epsilon }
  +3 L_m -4
  -\biggl(\frac{3 L_m^2}{2}-4 L_m+\frac{\pi ^2}{4}+8\biggr) \eps
  +\ord{\eps} \biggr],
\end{equation}
and, as indicated, only the correction $\propto \alpha_s^2$ is to be kept in the third term of \eq{MBggCTcontribs}.
The counterterm contributions in \eq{MBggCTcontribs} are included in our final result for $B_{g/g}^{(2,h)}$.

\section{Two-loop TMD beam function results}
\label{sec:results}

We now have the full (real + virtual) two-loop results for the heavy flavor contributions $B_{g/g}^{(2,h)}$ and $B_{g/q}^{(2,h)}$ to the bare partonic beam functions.
The one-loop expression $B_{g/g}^{(1,h)}$ is given in \eq{MBgg1} to the required order in the $\eps$ expansion, while $B_{g/q}^{(1,h)}=0$.
From these bare results we determine in this section the $n$-loop heavy-flavor contributions $\mathcal{I}_{gi}^{(n,h)}$ to the renormalized TMD beam function matching kernels $\mathcal{I}_{gi}^{(n)} = \mathcal{I}_{gi}^{(n,h)}+\mathcal{I}_{gi}^{(n,l)}$ for $n=1,2$.
We also obtain the $n$-loop beam function anomalous dimensions $\gamma_{B_g}^{(n-1,h)}$ and $\gamma_{\nu,B_g}^{(n-1,h)}$ for $n=1,2$.
The latter are fixed by RG consistency, which relates them to known expressions for hard and soft anomalous dimensions~\cite{Pietrulewicz:2017gxc}.
Our beam function calculation provides an explicit confirmation of these results, which serves as an important cross check.

The renormalized matching kernels $\mathcal{I}_{gk}$ in \eq{BeamFctMatching} are related to the bare partonic beam functions via%
\footnote{In this section we conventionally use the variable $z$ (instead of $x$) for purely partonic longitudinal light-cone momentum fractions.}
\begin{align}
 B_{g/j}^{\{n_l+1\}} (\vec{p}_{T},m, z) & =
 Z_{B_g}^{\{n_l+1\}} \Bigl(\pT,m,\mu,\frac{\nu}{\omega}\Bigr)
 \otimes_\perp \mathcal{I}_{gi}^{\{n_l+1\}}\Bigl(\pT,m,z,\mu,\frac{\nu}{\omega}\Bigr)
 \otimes_z f_{i/j}^{\{n_l\}}(z,\mu)\,,
 \label{eq:renrel}
\end{align}
where $Z_{B_g}$ is the $\MS$ renormalization factor of the TMD gluon beam function operator and the sum over all massless partons $i$ is understood.
Throughout this paper we always express (any contributions to) $B_{g/j}$, $Z_{B_g}$, and $\mathcal{I}_{gj}$ in terms of $\alpha_s \equiv \alpha_s^{\{n_l+1\}} (\mu)$ as indicated explicitly in \eq{renrel} by the superscript $\{n_l+1\}$.
We stress that this also applies to the massless $n$-loop expressions $B_{g/j}^{(n,l)}$, $Z_{B_g}^{(n,l)}$, and $\mathcal{I}_{gi}^{(n,l)}$, which arise from $n_l$ light quark flavors and gluons only.
For compactness of notation we will often drop the $\{n_l+1\}$ superscript in the following.
The (ultra-) collinear PDFs live in $n_l$-flavor QCD, where the heavy flavor has been integrated out, and are therefore naturally expressed in terms of $\alpha_s^{\{n_l\}}(\mu)$.
In the $\MS$ scheme we have
\begin{align}
  f_{i/j}^{\{n_l\}}(z,\mu) = \delta(1-z)\, \delta_{ij}
  - \frac{1}{\eps} \frac{\alpha_s^{\{n_l\}}(\mu)}{2\pi} P_{ij}^{(0)}(z)
  + \ord{\alpha_s^2}\,
\end{align}
with the one-loop splitting functions $P_{jk}^{(0)}(z)$ given in  \app{splitting}.
In the course of extracting $\mathcal{I}_{gj}^{(2,h)}$ and $Z_{B_g}^{(2,h)}$ from \eq{renrel} we have to convert $\alpha_s^{\{n_l\}}(\mu)$ to $\alpha_s \equiv \alpha_s^{\{n_l+1\}}(\mu)$ via the threshold matching relation
\begin{align}
  \alpha_s^{\{n_l\}}(\mu)
 &=
 \alpha_s \biggl[1 - \Pi^{(1)}(0,m^2)  +Z_{\alpha_s}^{(1,h)}
 +\ord{\alpha_s^2}\biggr]
 \nn\\
 &=
  \alpha_s \biggl[1+ \frac{\alpha_s T_F}{4\pi} \biggl( \frac{4 L_m}{3}-\frac{6 L_m^2+\pi ^2 }{9}\, \eps \biggr)
  +\ord{\eps^2,\alpha_s^2}\biggr]\,,
  \label{eq:alphasthreshold}
\end{align}
with $L_m \equiv \ln(m^2/\mu^2)$.
Expanding \eq{renrel} to $\ord{\alpha_s}$ and using
\begin{equation}
  \mathcal{I}_{gi}^{(0)} = \delta(1-z)\, \delpT\, \delta_{gi}\,,
  \qquad
  Z_{B_g}^{(0)} = \delpT\,,
  \qquad
  B^{(0,h)}_{g/j}=\mathcal{I}_{gi}^{(0,h)} = Z_{B_g}^{(0,h)} = 0\,,
\end{equation}
we have
\begin{align}
  B^{(1,h)}_{g/j} &= Z_{B_g}^{(1,h)} \delta(1-z)\,\delta_{gj}
  + \mathcal{I}_{gj}^{(1,h)} \,.
  \label{eq:Bgj1h}
\end{align}
With $B_{g/g}^{(1,h)}$ in \eq{MBgg1} and $B_{g/q}^{(1,h)}=0$ we thus obtain
\begin{align}
  \mathcal{I}_{gg}^{(1,h)} & =
  \frac{\alpha_s T_F}{4\pi} \,\delta (1-z)\, \delpT
  \biggl[ \frac{4 L_m}{3}-\frac{6 L_m^2+\pi ^2 }{9}\, \eps +\ord{\eps^2} \biggr]
  \,,
  \nn\\
  \mathcal{I}_{gq}^{(1,h)} & = \mathcal{I}_{g\bar{q}}^{(1,h)} = 0\,,
  \nn\\
  Z_{B_g}^{(1,h)} &= -\frac{\alpha_s T_F}{4\pi}\, \frac{4}{3 \eps} \, \delpT \,.
  \label{eq:I1hZ1h}
\end{align}
At $\ord{\alpha_s^2}$ \eq{renrel} yields
\begin{align}
  Z_{B_g}^{(2,h)} \delta(1-z)\,\delta_{gj}  + \mathcal{I}_{gj}^{(2,h)}
  ={}& B^{(2,h)}_{g/j}
  +\frac{\alpha_s}{2\pi\eps}   B^{(1,h)}_{g/i}
   \otimes_z \! P_{ij}^\zero
   + \frac{1}{2\pi \eps} \Bigl[ \alpha_s^{\{n_l\}}(\mu) - \alpha_s\Bigr]_{\alpha_s^2} P_{gj}^\zero\,\delpT
    \nn\\
  &
  - Z_{B_g}^{(1,h)} \operp \mathcal{I}^{(1,h)}_{gj}
  - Z_{B_g}^{(1,h)} \operp \mathcal{I}^{(1,l)}_{gj}
  - Z_{B_g}^{(1,l)} \operp \mathcal{I}^{(1,h)}_{gj}  \,,
\end{align}
where we have used \eq{Bgj1h} for compactness.
Inserting the massless one-loop results in \app{masslessBFresults}, the two-loop expressions $B^{(2,h)}_{g/q}$ and $B^{(2,h)}_{g/g}$, as well as the one-loop heavy-flavor contributions in \eq{I1hZ1h}, we find%
\footnote{Recall the definitions of $\hat{m}$ and $c_y$ in \eq{mhatcdef} and  of the plus-distributions $\LpT{n}$ in \eq{LpTdef}.}
\begin{align}
  \mathcal{I}^{(2,h)}_{gg} 
  &=
  \frac{\alpha_s^2 C_A T_F }{6 \pi ^2} \,\theta(z)
  \Biggl\{
  \frac{\theta(1-z)}{\pi \pTsq}
  \Biggl[
  12 \hat{m}^2  z^2 \bigl(1 - 4 \hat{m}^2 \bigr) U \bigl(z,\hat{m}^2 \bigr)
  +6 \hat{m}^2 (z-1) z \,U \bigl(1-z,\hat{m}^2 \bigr)
  \nn\\
  &\qquad +\frac{1}{c_z}\Bigl[2 \hat{m}^4 (47 z+29) z^2+ \hat{m}^2 (17 z+5) z
  -2 (z+1)\Bigr] \ln \frac{c_z-1}{c_z+1}
  \nn\\
  &\qquad -\frac{2}{z c_1}
  \Bigl[2 \hat{m}^4 \bigl(65 z^3-33 z^2+24 z-8\bigr)
   +  \hat{m}^2 \bigl(25 z^3-12 z^2-6 z+5\bigr)
   \nn\\
   &\qquad\quad -  \bigl(z^3+3 z-1\bigr) \Bigr]  \ln \frac{c_1-1}{c_1+1}
  +12 \hat{m}^2 (z-1) z\, c_{1-z} \ln \frac{c_{1-z}-1}{c_{1-z}+1}
  \nn\\
  &\qquad -\frac{ \hat{m}^2}{z} \bigl(83 z^3-95 z^2+48 z-16\bigr)
  +\frac{1}{3 z} \bigl(23 z^3-19 z^2+29 z-23\bigr)
  \Biggr]
  \nn\\
  &\quad+   \frac{1}{\pi \pTsq}\biggl[ 2
   \bigl(2 \hat{m}^2-1 \bigr) c_1
  \ln \frac{c_1-1}{c_1+1}
  +8 \hat{m}^2-\frac{10}{3} \biggr]
  \Bigl[ \delta (1-z) \ln \frac{\omega }{\nu} +\cL_0(1-z) \Bigr]
  \nn\\
  &\quad + 2 \LpT{0} L_m \Bigl[ 2  \delta (1-z)
  \ln   \frac{\omega }{\nu } +  P_{gg}(z)
  \Bigr]
  \nn\\
  &\quad +  \delpT \, \delta (1-z) \biggl[ \frac{5}{12} +2 L_m
  -  \biggl( L_m^2 +\frac{10}{3}L_m +\frac{28}{9}\biggr)
  \ln \frac{\omega }{\nu }
   \biggr]
   \Biggr\}
  \nn\\[1 ex]
  &+\frac{\alpha_s^2 C_F T_F }{6 \pi ^2}\, \theta(z)
  \Biggl\{
  \frac{\theta(1-z)}{\pi  \pTsq}
  \Biggl[
  6 \Bigl( 16 \hat{m}^4 z^2+2 \hat{m}^2 (z+3) z- z-1 \Bigr) U \bigl(z,\hat{m}^2 \bigr)
  \nn\\
  &\qquad+\frac{3}{c_z}\Bigl[ 64
   \hat{m}^4 z^2-2 \hat{m}^2 (7 z-3) z-3 (z+1)\Bigr]
   \ln \frac{c_z-1}{c_z+1}
   \nn\\
  &\qquad -\frac{2}{zc_1}
   \Bigl[  96 \hat{m}^4 z^3  + 2 \hat{m}^2 \bigl(7 z^3-12 z^2-1\bigr)
   - (z+2) \bigl(2 z^2+2 z-1\bigr) \Bigr]
   \ln \frac{c_1-1}{c_1+1}
  \nn\\
  &\qquad-48 \hat{m}^2 (z-1) z + \frac{1}{z} (z-1) \bigl(4 z^2+19 z-2\bigr)
  \Biggr]
 + \delpT \, \delta (1-z)  \biggl( \frac{3}{2} L_m- \frac{45}{8} \biggr)
  \Biggr\}
\nn\\[1 ex]
& + \frac{\alpha_s^2 T_F^2 }{9 \pi ^2}\,  L_m^2  \, \delpT \, \delta (1-z)
+ \ord{\eps}\,,
  \label{eq:Igg2h}
\\[3 ex]
  \mathcal{I}^{(2,h)}_{gq}  
  &= \mathcal{I}^{(2,h)}_{g\bar{q}} =
  \frac{\alpha_s^2 C_F T_F}{3 \pi ^3\, \pTsq}  \,\theta(z) \, P_{gq}(z)
  \biggl[\Bigl(2 (1-z)\mhsq -1\Bigr) c_{1-z}
  \ln \frac{c_{1-z}-1}{c_{1-z}+1} +4  (1-z)\mhsq  -  \frac{5}{3}\biggr]
  \nn\\
  &
  \hspace{11 ex}
  +\frac{\alpha_s T_F}{4 \pi} \, \frac{8}{3} L_m\, \mathcal{I}^{(1)}_{gq}(\vec{p}_{T},z,\mu)
  + \ord{\eps}  \, ,
  \label{eq:Igq2h}
\\[3 ex]
Z_{B_g}^{(2,h)} & =   \frac{\alpha_s^2 C_A T_F }{6 \pi ^2}
  \Biggl\{
  \frac{1}{\eta }
  \Biggl[
  \frac{2}{\pi \pTsq} \biggl(
  -\bigl(2 \hat{m}^2 -1\bigr) c_1
  \ln \frac{c_1-1}{c_1+1}
  -4 \hat{m}^2+\frac{5}{3}  \biggr)
  -2 \LpT{1}
\nn\\
   &\qquad
    + 2 \LpT{0} \biggl(\frac{1}{\epsilon }- L_m\biggr)
    -\delpT \biggl(
   \frac{1}{\epsilon^2}
   +\frac{5}{3 \epsilon }
   -\frac{\pi ^2}{6}
   -\frac{28}{9}
   -L_m^2
   -\frac{10}{3}L_m
   \biggr)
   + \ord{\eps} \Biggr]
\nn\\
   &\quad
  +\delpT \biggl[
  \biggl(\frac{1}{\epsilon ^2}+\frac{5}{3 \epsilon }\biggr)
  \ln \frac{\omega }{\nu } -\frac{1}{\epsilon }
  \biggr]
  \Biggr\} - \frac{\alpha_s^2 C_F T_F }{8 \pi^2 \eps} \delpT
  \,.
  \label{eq:ZBg2h}
\end{align}
In order to write \eq{Igg2h} in a compact form, we have introduced the auxiliary function
\begin{align}
  U(z,\hat{m}^2) ={}&
    \Li_2\biggl(\frac{1-c_z}{c_1+1}\biggr)
  +\Li_2\biggl(z\,\frac{c_1+1}{1-c_z}\biggr)
  +\Li_2\biggl(z\,\frac{c_1+1}{c_z+1}\biggr)
  +\Li_2\biggl(\frac{c_z+1}{c_1+1}\biggr)
   \nn\\
  &
   -\Li_2\biggl(z\, \frac{ c_z+1}{c_{z^2}+1}\biggr)
   -\Li_2\biggl(z \,\frac{1 - c_z}{c_{z^2}+1}\biggr)
   -\Li_2\biggl(\frac{c_{z^2}+1}{1-c_z}\biggr)
   -\Li_2\biggl(\frac{c_{z^2}+1}{c_z+1}\biggr)\,.
\end{align}
The beam function anomalous dimensions are derived from the counterterm $Z_{B_g}$ as follows:
\begin{align}
  \label{eq:gamma_mu_def}
   \gamma_{B_g} \!\Bigl(\mu,\frac{\nu}{\omega} \Bigr) \delpT
  &=-\big(Z_{B_g}\big)^{-1} \otimes_\perp \Bigl( \mu\frac{\df}{\df\mu}Z_{B_g}  \Bigr)\,,\\
  \gamma_{\nu,B_g}(\pT,m,\mu)
  &=-\big(Z_{B_g}\big)^{-1} \otimes_\perp \Bigl( \nu\frac{\df}{\df\nu}Z_{B_g}  \Bigr)\,.
  \label{eq:gamma_nu_def}
\end{align}
At $\ord{\alpha_s}$ and $\ord{\alpha_s^2}$ we thus obtain the heavy-flavor contributions
\begin{align}
    \gamma_{B_g}^{(0,h)}  \delpT={}&
    - \mu\frac{\df}{\df\mu} Z_{B_g}^{(1,h)} \Big|_{\alpha_s}
    = - (-2\eps \alpha_s) \frac{\partial}{\partial \alpha_s} Z_{B_g}^{(1,h)}
    =  - \frac{\alpha_s  T_F}{4\pi} \frac{8}{3}\, \delpT \,,
    \label{eq:gammaB0h}
    \\
    \gamma_{B_g}^{(1,h)} \delpT ={}&
    - \mu\frac{\df}{\df\mu} Z_{B_g}^{(2,h)} \Big|_{\alpha_s^2}
    - \Bigl(-\frac{\alpha_s^2}{2\pi} \beta_0 \Bigr) \frac{\partial}{\partial \alpha_s} Z_{B_g}^{(1,h)}
    -  \frac{2 \alpha_s^2 T_F}{3 \pi} \frac{\partial}{\partial \alpha_s} Z_{B_g}^{(1,l)}
    \nn\\
    &- \Bigl(Z_{B_g}^{(1,l)} + Z_{B_g}^{(1,h)}\Bigr)  \gamma_{B_g}^{(0,h)}
    +  (-2\eps \alpha_s) Z_{B_g}^{(1,h)}  \otimes_\perp
    \biggl(\frac{\partial}{\partial \alpha_s} Z_{B_g}^{(1,l)} \biggr)
    \nn\\
    ={}& \frac{\alpha_s^2}{(4 \pi)^2}
    \biggl\{
    \frac{32}{9} C_A T_F \Bigl[5
    \ln \frac{\omega }{\nu } - 3\Bigr]-8 C_F T_F
    \biggr\}\,   \delpT\,,
    \label{eq:gammaB1h}
    \\[2ex]
    \gamma_{\nu,B_g}^{(0,h)} ={}& - \nu\frac{\df}{\df\nu} Z_{B_g}^{(1,h)} = 0\,,
    \\
    \gamma_{\nu,B_g}^{(1,h)} ={}&
    \Bigl[ \eta Z_{B_g}^{(2,h)} \Bigr]_{\eta \to 1}
    - \nu \frac{\partial}{\partial \nu} Z_{B_g}^{(2,h)}
    - Z_{B_g}^{(1,h)} \otimes_\perp  \Bigl[ \eta Z_{B_g}^{(1,l)} \Bigr]_{\eta \to 1}
    + Z_{B_g}^{(1,h)} \otimes_\perp \nu \frac{\partial}{\partial \nu} Z_{B_g}^{(1,l)}
    \nn\\
     ={}&
    \frac{\alpha_s^2 C_A T_F}{16 \pi ^2} \biggl\{
    -\frac{16}{9 \pi  \pTsq} \biggl[
    3 (2 \hat{m}^2 -1) c_1 \ln \frac{c_1-1}{c_1+1}  + 12 \hat{m}^2 -5 \biggr]
    \nn\\
    & +\frac{8}{27}\, \delpT \bigl(9 L_m^2+30 L_m+28\bigr)
    -\frac{16}{3} \LpT{0} L_m
   \biggr\}\,.
    \label{eq:gammanu1h}
\end{align}
In the derivation of anomalous dimensions within the RRG formalism using the $\eta$ regulator it is in general important to retain higher order $\eps$ terms in the $1/\eta$ poles of the corresponding counterterms.
In this particular case it is crucial to include the $\ord{\eps/\eta}$ term of $Z_{B_g}^{(1,l)}$, as given in \eq{ZB1l}, in the formulas for the two-loop anomalous dimensions.
Moreover, for \eq{gammaB1h} it is necessary to restore the full $\mu$ dependence for finite $\eps$ in the $1/\eta$ terms of $Z_{B_g}^{1,l}$ and $Z_{B_g}^{2,h}$, which are $\propto \mu^{2\eps}$ and $\mu^{4\eps}$, respectively.
The results in \eqs{gammaB0h}{gammaB1h} exactly equal the contributions of a single light flavor, see \rcite{Luebbert:2016itl}.
The heavy-flavor contribution to the two-loop rapidity anomalous dimension in \eq{gammanu1h} satisfies, according to \eq{nuconsistency}, the RG consistency relation
\begin{equation}
  2 \gamma_{\nu,B_g}^{(1,h)} +  \gamma_{\nu, S_g}^{(1,h)} = 0\,,
\end{equation}
where $\gamma_{\nu, S_g}^{(1,h)}$ is part of the anomalous dimension of the TMD soft function $S_{gg}$. It is determined via Casimir rescaling from the result in \rcite{Pietrulewicz:2017gxc} and given in \eq{gammaS1h}.

The results in this section represent the heavy-flavor contributions to the gluon TMD beam function and its anomalous dimensions through $\ord{\alpha_s^2}$.
The corresponding massless contributions $\mathcal{I}_{gi}^{(n,l)}$, $\gamma_{B_g}^{(n-1,l)}$, and $\gamma_{\nu,B_g}^{(n-1,l)}$ for $n=1,2$ are given in \rcite{Luebbert:2016itl} and must be added to obtain the complete expressions entering the factorized cross section in \eq{factXsec} and the (R)RGEs in \eqs{muRGE}{nuRGE}, respectively:
\begin{equation}
  \mathcal{I}_{gi}^{(n)} =
  \mathcal{I}_{gi}^{(n,h)} + \mathcal{I}_{gi}^{(n,l)}\,,
  \quad
  \gamma_{B_g}^{(n-1)} = \gamma_{B_g}^{(n-1,h)} + \gamma_{B_g}^{(n-1,l)}\,,
  \quad
  \gamma_{\nu,B_g}^{(n-1)} =
  \gamma_{\nu,B_g}^{(n-1,h)} + \gamma_{\nu,B_g}^{(n-1,l)} \,.
  \label{eq:hplusl}
\end{equation}
The massless part of the soft function $S_{gg}^{(n,l)}$ and its anomalous dimensions are also found in \rcite{Luebbert:2016itl}, while the soft heavy-flavor contributions are computed in \rcite{Pietrulewicz:2017gxc} and collected for convenience in \app{Softfctresults}.
We emphasize again that our explicit results imply that both massive and massless contributions must be evaluated with $\alpha_s \equiv \alpha_s^{\{n_l+1\}} (\mu)$. In the coefficients of the $\alpha_s$ expansion of the massless contributions from \rcite{Luebbert:2016itl} we must however consistently replace $n_f \to n_l$ (e.g.\ in the expression for $\beta_0$).

\section{Beam function asymptotics}
\label{sec:limits}

In this section we study the limiting behavior of the beam function matching coefficients in \eqs{Igg2h}{Igq2h} when  $m \ll p_T \ll Q$  (``small mass limit") and $p_T \ll  m \ll Q$  (``large mass limit").
Following \rcite{Pietrulewicz:2017gxc} we establish in this way the connection between the cross section in \eq{factXsec} and the corresponding factorization formulas in the two limits (with $p_T \equiv |\pT| \sim q_T$), where also
the matching coefficients $\mathcal{I}_{gi}$ themselves exhibit a factorized structure.
The associated factorization ingredients and thus the asymptotic expressions for the $\mathcal{I}_{gi}$ are either available in the literature or can be inferred by consistency from related factorization formulas for other processes.
Verifying the expected limiting behavior therefore represents a valuable and strong check of our  results for the $\mathcal{I}_{gi}$.
In particular the virtual contributions, which are proportional to $\delpT$ and thus unaffected by the small and large mass expansions, are directly cross-checked.

\subsection{Small mass limit}
\label{subsec:smallmasslimit}

In the small mass limit $m \ll q_T$ (for $\lqcd \ll m$, $q_T \ll Q$) the factorized cross section of gluon-fusion color-singlet production takes the form
(see \rcite{Pietrulewicz:2017gxc} for quark-initiated processes)
\begin{align}
 \frac{\df \sigma}{\df q^2_T \,\df Q^2 \,\df Y} &=  H^{\{n_l+1\}}_{gg}(Q,\mu)\, \int \df^2 p_{T,a}\,\df^2 p_{T,b}\, \df^2 p_{T,s} \, \delta(q_T^2-|\vec{p}_{T,a}+\vec{p}_{T,b}+\vec{p}_{T,s}|^2)
 \nn\\
 & \quad \times
 \Bigl[\sum_{i \in \{Q,\bar{Q},q,\bar{q},g\}} \sum_{k \in \{q,\bar{q},g\}} \mathcal{I}_{gi,\mu \nu}\Bigl(\vec{p}_{T,a},x_a,\mu,\frac{\nu}{\omega_a}\Bigr) \otimes \mathcal{M}_{ik}\bigl(m,x_a,\mu\bigr)\otimes f^{\{n_l\}}_k (x_a,\mu)\Bigr]
 \nn \\
 & \quad  \times \Bigl[\sum_{j \in \{Q,\bar{Q},q,\bar{q},g\}} \sum_{l \in \{q,\bar{q},g\}} \mathcal{I}^{\mu \nu}_{gj}\Bigl(\vec{p}_{T,b},x_b,\mu,\frac{\nu}{\omega_b}\Bigr) \otimes \mathcal{M}_{jl}\bigl(m,x_a,\mu\bigr)\otimes f^{\{n_l\}}_l (x_b,\mu)\Bigr]  \nn \\
 & \quad \times S_{gg}(\vec{p}_{T,s},\mu,\nu)  \, \Bigl[1+\mathcal{O}\Bigl(\frac{q_T}{Q},\frac{m^2}{q_T^2},\frac{\lqcd^2}{m^2}\Bigr)\Bigr]  \, .
 \label{eq:factXsecsmallmass}
\end{align}
Compared to \eq{factXsec} the mass-dependent beam function matching coefficients are factorized in \eq{factXsecsmallmass} into the corresponding coefficients for $n_l+1$ massless quarks and the known PDF (flavor-threshold) matching factors $\mathcal{M}_{ij}$:
\begin{align}
 \mathcal{I}_{gk,\mu \nu}\Bigl(\vec{p}_{T},m,z,\mu,\frac{\nu}{\omega}\Bigr) = \!\sum_{i \in \{Q,\bar{Q},q,\bar{q},g\}} \!\!\!\!\!\! \mathcal{I}^{\{n_l+1\}}_{gi,\mu \nu}\Bigl(\vec{p}_{T},x,\mu,\frac{\nu}{\omega}\Bigr) \otimes_z \mathcal{M}_{ik}\bigl(m,z,\mu\bigr)\, \Bigl[1+\mathcal{O}\Bigl(\frac{m^2}{\pTsq}\Bigr)\Bigr] \, .
 \label{eq:Igksmallmass}
\end{align}
Here the explicit superscript $\{n_l+1\}$, which has been suppressed in \eq{factXsecsmallmass}, indicates that the beam function matching coefficients must be evaluated using $\alpha_s \equiv \alpha_s^{\{n_l+1\}}$ in order to avoid large logarithms $\sim \ln^n(m^2/ \pTsq)$.
The relevant $\mathcal{M}_{ij}$ are collected up to $\ord{\alpha_s^2}$ for convenience in \app{PDFmatching}.

We now show explicitly that our mass-dependent two-loop results for the  unpolarized coefficients $\mathcal{I}_{gi}$ in \sec{results} indeed satisfy \eq{Igksmallmass}.
At one-loop we trivially have
\begin{equation}
  \mathcal{I}^{(1,h)}_{gg}(\vec{p}_{T},m,z,\mu)
  = \delpT\,\mathcal{M}^{(1)}_{gg}(m,z,\mu) \,,
\end{equation}
because the only contributing diagram is virtual, see \fig{diagslowestorder}~a, and thus survives the expansion in \eq{Igksmallmass} as a whole.
For the matching coefficient $\mathcal{I}^{(2,h)}_{gq}$ in \eq{Igq2h} we find in the small mass limit
\begin{align}
  \mathcal{I}^{(2,h)}_{gq}(\vec{p}_{T},m,z,\mu)
  \underset{m \ll p_T}{\longrightarrow}{}&
  \frac{\alpha_s^2 C_F T_F}{3\pi^2} \biggl\{
  P_{gq}(z)\,\mathcal{L}_1(\pT,\mu)
  - \mathcal{L}_0(\pT,\mu)\,P_{gq}(z) \biggl(\ln(1-z)+\frac{5}{3}\biggr)\nn \\
  &
  +\delpT \biggl[\frac{L_m^2 P_{gq}(z)}{2}+L_m P_{gq}(z) \left(\ln (1-z)+\frac{5}{3}\right)+L_m \theta(1-z)z
  \nn\\
  &\quad +P_{gq}(z) \left(\frac{1}{2} \ln^2(1-z)+\frac{5}{3} \ln
   (1-z)+\frac{28}{9}\right)
  \biggr]
  \biggr\}
  +\ord{\eps}
  \nn \\
  & = \mathcal{I}^{(2,l)}_{gq}(\vec{p}_{T},z,\mu)\Big|_{T_F}^{n_l=1}+\delta^{(2)}(\pT)\, \mathcal{M}^{(2)}_{gq}(m,z,\mu)
  \label{eq:I2hgqsmallmass}
\end{align}
in agreement with \eq{Igksmallmass}.
The first term in the last line represents the contribution to $\mathcal{I}^{(2)}_{gq}$ due to a single massless quark flavor.
The explicit expression is given in \eq{I2lgqnlglei1TF}.
Note that for $p_T>0$ the $m^2/\pTsq$ expansion leading to \eq{I2hgqsmallmass}  is straightforward.
However, to determine the correct distributional structure and in particular to fix the coefficient of $\delpT$ on the right-hand side of \eq{I2hgqsmallmass}
we also had to expand the cumulant, i.e.\ the  $\pT$-integral from $0$ to an arbitrary $\pT^{\,\mathrm{cut}}$ (with $m \ll |\pT^{\,\mathrm{cut}}|$), of the left-hand side.

In the same manner we verify%
\footnote{The (non-distributional) part of the $\delpT$ coefficient that is regular as $z\to1$  on the right hand side of \eqs{I2hggCFTFsmallmass}{I2hggCATFsmallmass} we checked numerically for convenience.}
that our result for the matching coefficient $\mathcal{I}^{(2,h)}_{gg}$  in \eq{Igg2h} is in accordance with \eq{Igksmallmass}:
\begin{align}
  \mathcal{I}^{(2,h)}_{gg}(\vec{p}_{T},m,z,\mu)\Big|_{T_F^2}  \quad={}\quad&
  \delpT\, \mathcal{M}^{(2)}_{gg}(m,z,\mu)\Big|_{T_F^2}
  \, ,
  \label{eq:I2hggTFTFsmallmass}
  \\[2 ex]
  \mathcal{I}^{(2,h)}_{gg}(\vec{p}_{T},m,z,\mu)\Big|_{C_F T_F}
  \,\underset{m \ll p_T}{\longrightarrow}\,{}& \mathcal{I}^{(2,l)}_{gg}\Bigl(\vec{p}_{T},z,\mu,\frac{\nu}{\omega}\Bigr)\Big|_{C_F T_F}^{n_l=1}
  + \delpT\,\mathcal{M}^{(2)}_{gg}(m,z,\mu) \Big|_{ C_F T_F}
  \nn \\
  & +2 \, \mathcal{I}^{(1)}_{gq}(\vec{p}_{T},z,\mu) \otimes_z \mathcal{M}^{(1)}_{Qg}(m,z,\mu) \, ,
  \label{eq:I2hggCFTFsmallmass}
  \\[2 ex]
  \mathcal{I}^{(2,h)}_{gg}\Bigl(\vec{p}_{T},m,z,\mu,\frac{\nu}{\omega}\Bigr)\Big|_{C_A T_F} \,\underset{m \ll p_T}{\longrightarrow}\,{}&  \mathcal{I}^{(2,l)}_{gg}\Bigl(\vec{p}_{T},z,\mu,\frac{\nu}{\omega}\Bigr)
  \Big|_{C_A T_F}^{n_l=1}
  +  \delpT\,\mathcal{M}^{(2)}_{gg}(m,z,\mu) \Big|_{ C_A T_F}
  \nn \\
  &+ \mathcal{I}^{(1,l)}_{gg}\Bigl(\vec{p}_{T},z,\mu,\frac{\nu}{\omega}\Bigr) \otimes_z \mathcal{M}^{(1)}_{gg}(m,z,\mu) \, .
  \label{eq:I2hggCATFsmallmass}
\end{align}
Note that we have used $\mathcal{I}^{(1)}_{gQ}(\vec{p}_{T},z,\mu) \equiv \mathcal{I}^{(1)}_{gq}(\vec{p}_{T},z,\mu)$ in the last term of \eq{I2hggCFTFsmallmass}, because here also the massive quark flavor $Q$ is to be treated as massless.
The factor of two in front of this term allows for the equal contribution due to the respective antiflavor $\bar{Q}$.
The explicit expressions for the required massless coefficients  $\mathcal{I}^{(n,l)}_{gi}$ are collected in \app{masslessBFresults}.

\subsection{Large mass limit}
\label{subsec:largemasslimit}

In the large mass (or ``decoupling") limit $q_T \ll m \ll Q$ the factorized cross section reads
\begin{align}\label{eq:factXseclargemass}
 \frac{\df \sigma}{\df q^2_T \,\df Q^2 \,\df Y} &= \sum_{i,j \in \{q,\bar{q}\}}  H^{\{n_l+1\}}_{gg}(Q,\mu)\, H^g_{c}\Bigl(m,\mu,\frac{\nu}{\omega_a}\Bigr) H^g_{c}\Bigl(m,\mu,\frac{\nu}{\omega_b}\Bigr)  H^g_{s}(m,\mu,\nu)
\\
 & \quad \times
 \int \df^2 p_{T,a}\,\df^2 p_{T,b}\, \df^2 p_{T,s} \, \delta(q_T^2-|\vec{p}_{T,a}+\vec{p}_{T,b}+\vec{p}_{T,s}|^2)\,
 S_{gg}^{\{n_l\}}(\vec{p}_{T,s},\mu,\nu)
 \nn \\
& \quad \times
 \Bigl[\sum_{k \in \{q,\bar{q},g\}} \mathcal{I}^{\{n_l\}}_{gk,\mu \nu}\Bigl(\vec{p}_{T,a},x_a,\mu,\frac{\nu}{\omega_a}\Bigr) \otimes f^{\{n_l\}}_k (x_a,\mu)\Bigr]
 \nn \\
 & \quad  \times \Bigl[\sum_{l \in \{q,\bar{q},g\}} \mathcal{I}^{\mu \nu \{n_l\}}_{gl}\Bigl(\vec{p}_{T,b},x_b,\mu,\frac{\nu}{\omega_b}\Bigr) \otimes f^{\{n_l\}}_l (x_b,\mu)\Bigr]  \biggl[1\!+\!\mathcal{O}\Bigl(\frac{q_T}{Q},\frac{m^2}{Q^2},\frac{q_T^2}{m^2},
 \frac{\lqcd^2}{q_T^2}
 \Bigr)\biggr].
 \nn
\end{align}
The mass-dependent beam function matching coefficients of \eq{BeamFctMatching} now factorize into a hard threshold matching factor $H^g_c$ and massless matching coefficients with $n_l$ active quark flavors:
\begin{align}
  \mathcal{I}_{gk,\mu \nu}\Bigl(\vec{p}_{T},m,x,\mu,\frac{\nu}{\omega}\Bigr) = H_c^g \Bigl(m,\mu,\frac{\nu}{\omega}\Bigr) \, \mathcal{I}^{\{n_l\}}_{gk,\mu \nu}\Bigl(\vec{p}_{T},x,\mu,\frac{\nu}{\omega}\Bigr) \, \Bigl[1+\mathcal{O}\Bigl(\frac{\pTsq}{m^2}\Bigr)\Bigr] \, .
  \label{eq:Igklargemass}
\end{align}
Similarly, the soft function in \eq{factXsec} factorizes into the hard matching factor $H^g_s$ and the massless soft function with $n_l$ flavors in \eq{factXseclargemass}.
While the matching function $H^g_s$ (arising from virtual soft mass modes) equals the one for a Drell-Yan--type process $H^q_s$ up to Casimir rescaling (i.e\ the replacement $C_F \to C_A$)~\cite{Pietrulewicz:2017gxc}, the matching functions  $H^g_c$ (arising from virtual collinear mass modes) differ from their quark-initiated counterparts  $H^q_c$ and were so far not given in the literature.
They can, however, be inferred from consistency relations between the different formulations of the factorization theorem in \rcite{Hoang:2015iva} for deep inelastic scattering in the ($x\to1$) endpoint region and a corresponding alternative factorization theorem based on the approach of \rcite{Pietrulewicz:2017gxc}.%
\footnote{The alternative ``mass-mode" and ``universal" factorization approaches of \rcite{Pietrulewicz:2017gxc} and \rcites{Pietrulewicz:2014qza,Hoang:2015iva}, respectively, as well as the consistency relations between their ingredients are discussed in detail in \rcite{Hoang:2019fze} using the double differential hemisphere mass distribution in the process $e^+ e^- \to Q\bar{Q}$ as an example.}
Concretely, one can show the relation
\begin{align}
H^g_c \Bigl(m,\mu,\frac{\nu}{\omega}\Bigr) \,H^g_s (m,\mu,\nu)\, \mathcal{S}^g_c (\omega(1-z),m,\mu,\nu) = \frac{1}{\omega}
\,\mathcal{M}^g(1-z,m,\mu) \, ,
\label{eq:DISconsistency}
\end{align}
where $\mathcal{S}_c^g$ is, up to Casimir rescaling, the csoft function in \rcite{Pietrulewicz:2017gxc} and $ \mathcal{M}^g(1-z,m,\mu) = \lim_{z \to 1} \mathcal{M}_{gg}(z,m,\mu)$ is the massive PDF matching coefficient  in the threshold limit.
The explicit expression for $H^g_c$ that we extracted up to $\ord{\alpha_s^2}$ from \eq{DISconsistency} is given in \eq{Hgc}.

We can now check our results for the heavy-flavor corrections to the gluon TMD beam function coefficients in the large mass limit against \eq{Igklargemass}.
At one loop we consistently have
\begin{equation}
  \mathcal{I}^{(1,h)}_{gg}(\vec{p}_{T},m,z,\mu)
  = H_c^{g(1)} (m,\mu) \,\delta(1-z) \, \delpT \,.
\end{equation}
The first term in the two-loop matching coefficient $\mathcal{I}^{(2,h)}_{gq}$ in \eq{Igq2h}  vanishes in the decoupling limit and we obtain
\begin{align}
\mathcal{I}^{(2,h)}_{gq}(\vec{p}_{T},m,z,\mu) \,\underset{m \gg p_T}{\longrightarrow}\,{}&
 \frac{\alpha_s T_F}{4 \pi} \, \frac{8}{3} L_m\, \mathcal{I}^{(1)}_{gq}(\vec{p}_{T},z,\mu)
\nn \\
  =\quad{}& \Bigl(H_c^{g(1)} (m,\mu) + \frac{\alpha_s T_F}{4 \pi} \,\frac{4}{3} L_m\Bigr)\,\mathcal{I}^{(1)}_{gq}(\vec{p}_{T},z,\mu) \,.
  \label{eq:Igq2hlargemass}
\end{align}
The last term in \eq{Igq2hlargemass} arises from the flavor threshold matching relation \eq{alphasthreshold}, which is used in \eq{Igklargemass} to switch from $\alpha_s^{\{n_l\}}$ to $\alpha_s \equiv \alpha_s^{\{n_l+1\}}$ in the massless one-loop coefficient $\mathcal{I}^{(1,l)}_{gk}$.
The large-mass expansion of the coefficient $\mathcal{I}^{(2,h)}_{gg}$ in \eq{Igg2h} yields
\begin{align}
\mathcal{I}^{(2,h)}_{gg}\Bigl(\vec{p}_{T},m,z,\mu,\frac{\nu}{\omega}\Bigr)
 \,\underset{m \gg p_T}{\longrightarrow}\,{}&
  \Bigl(H_c^{g(1)} (m,\mu) +\frac{\alpha_s T_F}{4 \pi} \,\frac{4}{3} L_m\Bigr)\,
  \mathcal{I}^{(1,l)}_{gg}\Bigl(\vec{p}_{T},z,\mu,\frac{\nu}{\omega}\Bigr) \nn \\
& +H_c^{g(2)} \Bigl(m,\mu,\frac{\nu}{\omega}\Bigr) \, \delta^{(2)}(\vec{p}_{T})\, \delta(1-z)\, .
  \label{eq:Igg2hlargemass}
\end{align}
The asymptotic behavior of the beam function coefficients in
\eqs{Igq2hlargemass}{Igg2hlargemass}  agrees with \eq{Igklargemass} and thus confirms our results.

\section{Numerical effect of bottom mass corrections}
\label{sec:numerics}

\begin{figure}[t]
  \centering
  \includegraphics[height=5.7cm]{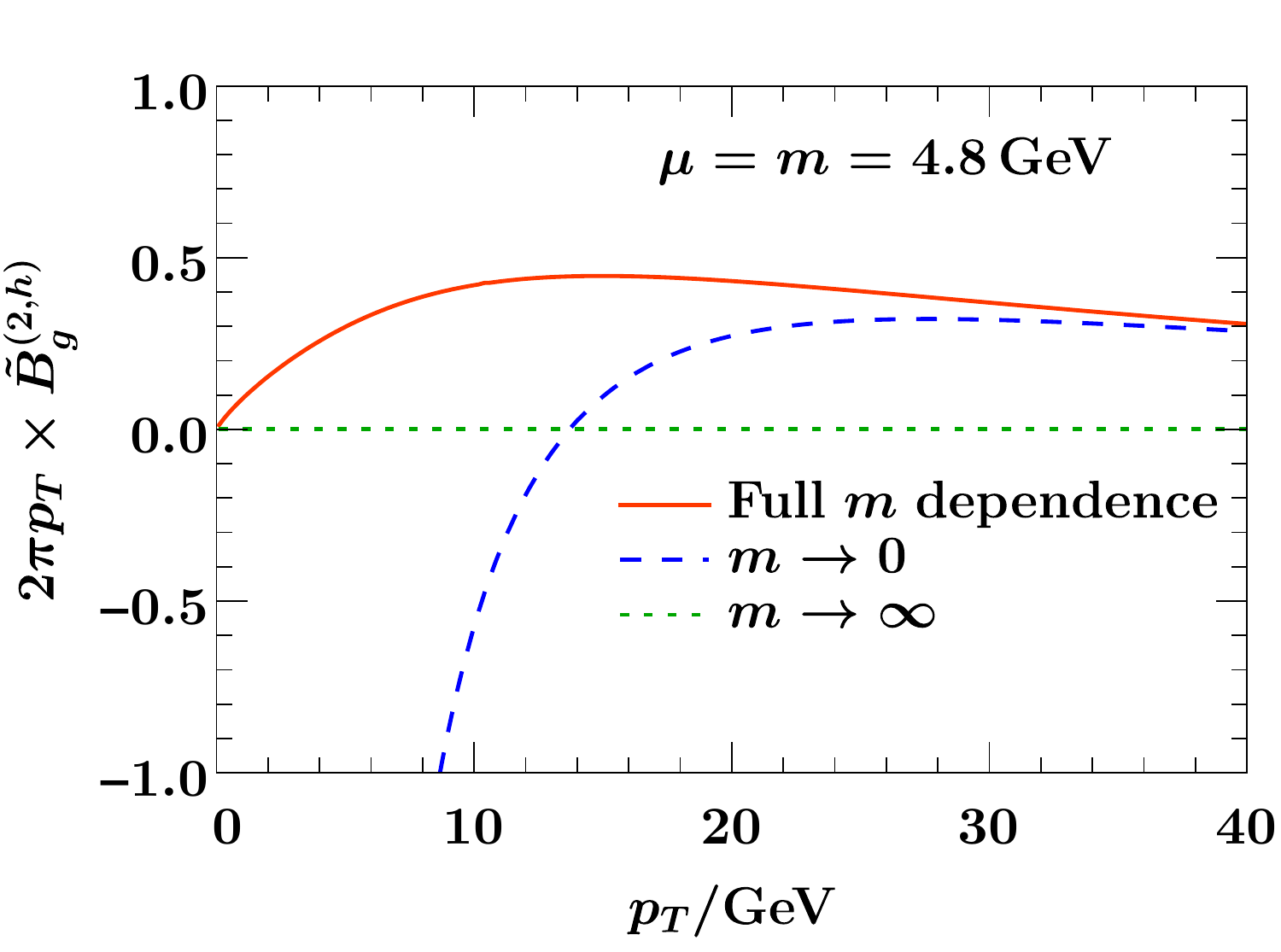} \hfill
  \includegraphics[height=5.7cm]{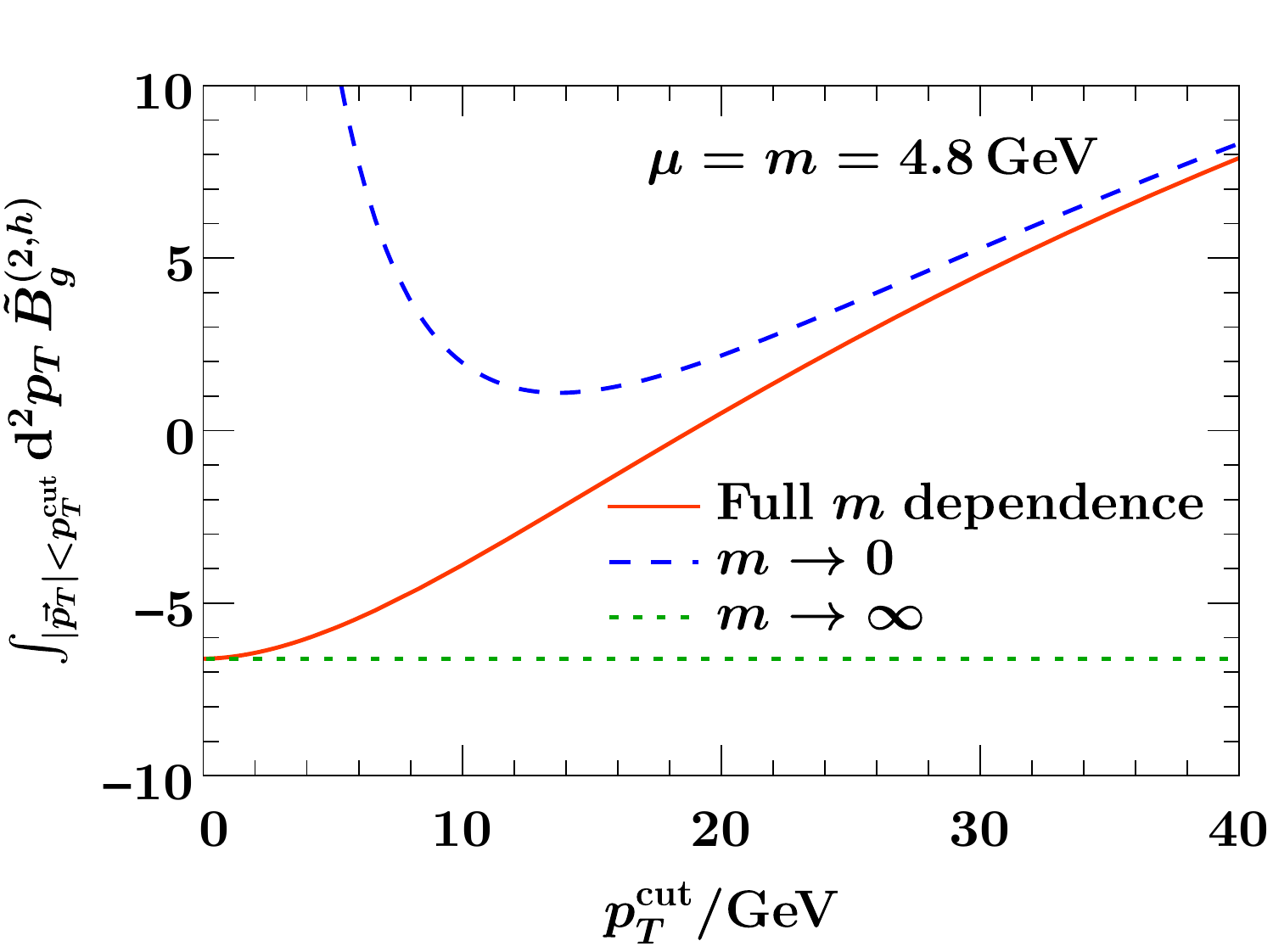}
  \caption{Massive quark corrections to the gluon TMD PDF $\tilde{B}_g$ defined in \eq{TMDPDF} (left panel) and its cumulant (right panel) at $\mathcal{O}(\alpha_s^2)$ as a function of $p_T$ and $p_T^\mathrm{cut}$, respectively.}
  \label{fig:Bg}
\end{figure}

As a first step to quantify the effect of the quark mass corrections obtained in \sec{results} on physical observables, in particular the Higgs transverse momentum spectrum, we assess here their numerical size at fixed order in $\alpha_s$.
A full-fledged analysis including resummation and the appropriate renormalization scale variations as well as the matching to the massless full QCD fixed-order result relevant at large transverse momentum is left for future work.

To remove the dependence on the rapidity renormalization scale $\nu$ we consider the symmetrized combination
\begin{align}
  \tilde{B}_g(\vec{p}_{T},m,\omega,x,\mu) =  \int \! \df^2 p_T' \,  B_g\Bigl(\vec{p}_{T}-\vec{p}^{\,\prime}_{T},m,x,\mu,\frac{\nu}{\omega}\Bigr) \, \sqrt{S_{gg}(\vec{p}^{\,\prime}_{T},m,\mu,\nu)}  \,,
  \label{eq:TMDPDF}
\end{align}
of TMD gluon beam and soft function, often referred to as TMD PDF.
We are interested in the $\ord{\alpha_s^2}$ correction due to the massive quark flavor, i.e.\ $\tilde{B}_g^{(2,h)}$.
For simplicity we set $\mu = m$ to evaluate $\tilde{B}_g^{(2,h)}$ in the following.
This choice eliminates the one-loop correction $\tilde{B}_g^{(1,h)} \propto L_m \equiv \ln (m^2/\mu^2)$ and thus yields%
\footnote{Recall that $q$ and $\bar{q}$ stand for the $n_l$ massless quark and antiquark flavors, respectively.}
\begin{align}
  \tilde{B}_g^{(2,h)}(\vec{p}_{T},m,\omega,x,m) ={}& \biggl[
  \mathcal{I}^{(2,h)}_{gg}\Bigl(\vec{p}_{T},m,x,m,\frac{\nu}{\omega}\Bigr)
  + \frac12 S_{gg}^{(2,h)}(\pT,m,m,\nu)\,\delta(1-x) \biggr]\!
  \otimes_x f^{\{n_l\}}_g(x,m)
  \nn\\
  &+ \sum_{k \in \{q,\bar{q}\}} \mathcal{I}^{(2,h)}_{gk}(\vec{p}_{T},m,x,m)
   \otimes_x f^{\{n_l\}}_k(x,m)
   \,,
\end{align}
with $S_{gg}^{(2,h)}$ as given in \eq{Sgg2h}.

In \fig{Bg}  we show plots of ($2 \pi p_T$ times)
  $\tilde{B}_g^{(2,h)}$ and its cumulant, i.e.\ the integral
\begin{equation}
  \int_{|\vec{p}_T| < p_T^\mathrm{cut}}
  \! \df^2 p_T \,
  \tilde{B}_g^{(2,h)}(\vec{p}_{T},m,\omega,x,\mu)
  = 2 \pi   \int_0^{p_T^\mathrm{cut}} \!\! \df p_T \,p_T\,
  \tilde{B}_g^{(2,h)}(\vec{p}_{T},m,\omega,x,\mu)\,,
\end{equation}
as a function of $p_T$ and $p_T^\mathrm{cut}$, respectively.
For the plots we set $\mu=m = m_b = 4.8$ GeV and $x=\omega/E_\mathrm{cm}$ with $\omega=Q=m_H =125$ GeV, $E_\mathrm{cm} = 13$ TeV, and we used MMHT2014 NNLO PDFs~\cite{Harland-Lang:2014zoa}.
The quark mass corrections can be expressed as an infinite series of the subleading terms $\sim (m/p_T)^{2n}$ in the small mass expansion with $n \ge 1$.
Note that fixing $\mu$ does not affect the $(m/p_T)^{2n}$ corrections we want to visualize here: The difference between the result with the full mass dependence (red curve) and its small mass limit (blue dashed curve) is (unlike the individual curves) $\mu$-independent, because the beam and soft function (and equivalently the hard function) $\mu$ anomalous dimension is mass-independent.
Note also that this difference is of $\ord{\alpha_s^2}$, i.e.\ the full and the small mass results are equal at $\ord{\alpha_s}$.

We observe that for $p_T \gg m$ the deviation between the full result and the small mass limit performed in \subsec{smallmasslimit} is indeed small, while for $p_T \sim 10$ GeV $\sim 2 m_b$ the deviations are of $\mathcal{O}(100 \%)$ and the small mass result does not provide a sensible approximation of the $\mathcal{O}(\alpha_s^2 T_F)$ contributions.
In the large mass limit and for $\mu=m$ the correction $\tilde{B}_g^{(2,h)}$ is proportional to $\delpT$ as can be verified from the results in \subsec{largemasslimit} and \rcite{Pietrulewicz:2017gxc}.
In the plots of \fig{Bg} the large mass limit is therefore illustrated by (dotted green)
horizontal lines at zero (left panel) and a nonzero value (right panel), which are touched by the curves of the full result at $p_T=0$ and $p_T^\mathrm{cut}=0$, respectively.

\begin{figure}[t]
  \centering
  \includegraphics[width=0.6\textwidth]{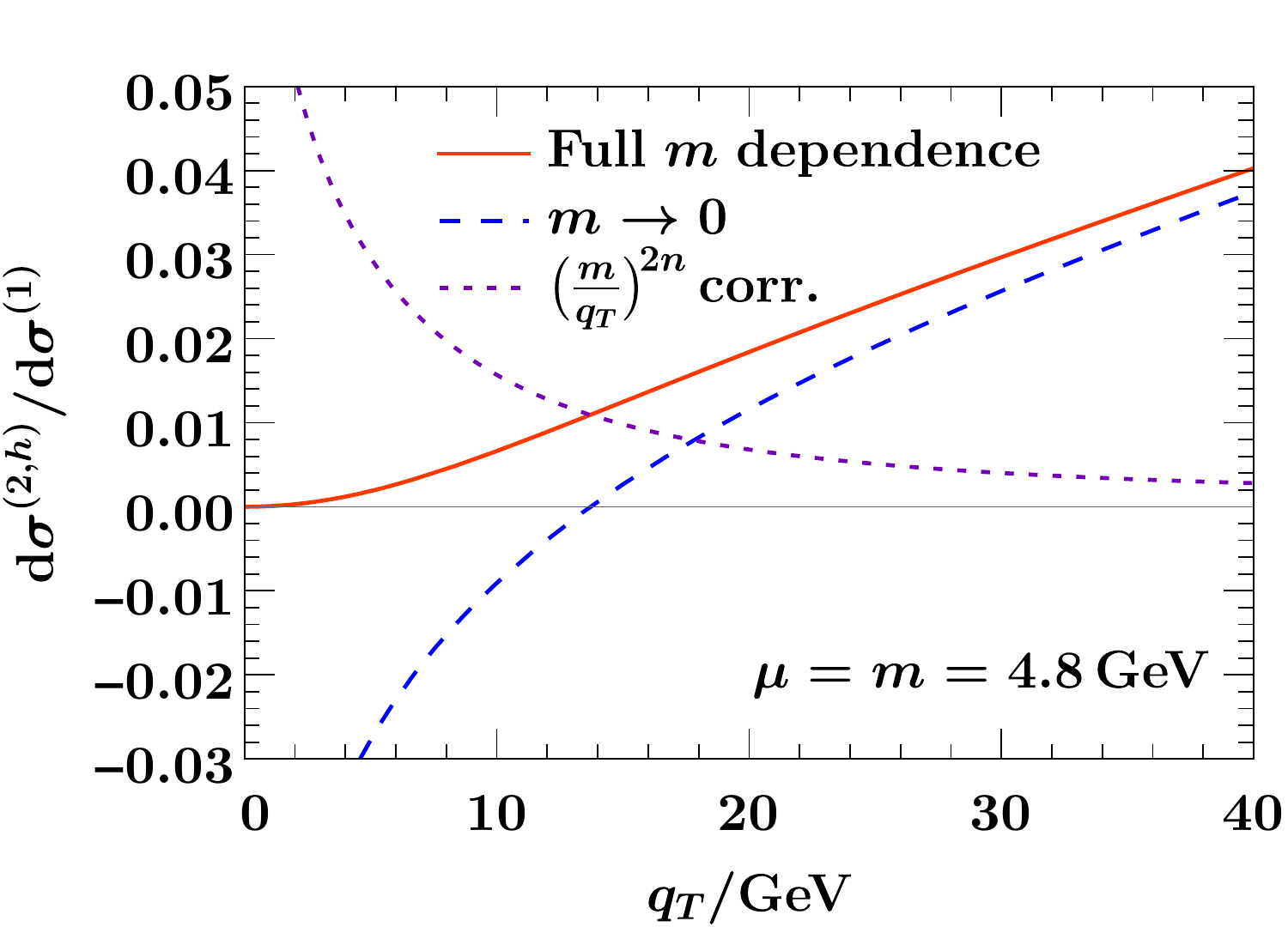}
  \caption{Relative size of the massive quark corrections to the gluon fusion Higgs transverse momentum distribution at NNLO (= NLO$_1$) with respect to the NLO (= LO$_1$) spectrum.}
  \label{fig:XsecRatio}
\end{figure}

Finally, we assess the NNLO (=NLO$_1$) quark mass corrections $\sim (m/q_T)^{2n}$  to the Higgs $q_T$ distribution in top-induced gluon fusion at the LHC.
In \fig{XsecRatio} we show their effect relative to the full NLO (= LO$_1$) result, i.e.\ relative to the leading order spectrum for $q_T>0$.
Concretely, we plot the cross section ratio%
\footnote{Note that in accordance with \eq{factXsec} the $q_T$-independent hard function factors do not contribute for $\mu=m$ and $q_T>0$ at the order of interest.}
\begin{align}
  \frac{\df\sigma^{(2,h)}}{\df\sigma^\one} \equiv
  \frac{\df\sigma^{(2,h)}}{\df q_T \,\df Y} \bigg/
  \frac{\df\sigma^\one}{\df q_T \,\df Y} \bigg|_{Y=0}^{\mu=m}
  = \frac{\tilde{B}_g^{(2,h)}(\vec{q}_{T},m,\omega,x,m)}{\tilde{B}_g^{(1)}(\vec{q}_{T},m,\omega,x,m)}
  =
  \frac{\tilde{B}_g^{(2,h)}(\vec{q}_{T},m,\omega,x,m)}{\tilde{B}_g^{(1,l)}(\vec{q}_{T},\omega,x,\mu=m)}
  \label{eq:XsecRatio}
\end{align}
as a function of $q_T>0$ with the same input as for \fig{Bg}.
The newly computed quark mass power corrections are nonsingular in the $m\to 0$ limit and composed of terms
$\propto (m/q_T)^{2n}$ with $n \in \mathbb{N}$, which can also contain positive powers of $\ln(m/q_T)$.
The total mass-nonsingular contribution to \eq{XsecRatio} is depicted in \fig{XsecRatio} by the dotted violet curve. It corresponds to the difference of the full result (solid red curve) and the small mass limit (blue dashed curve), i.e.\ the mass-singular contribution, which includes massive PDF matching factors, as described in \subsec{smallmasslimit}.
Similar to the plots in \fig{Bg}, this difference (the dotted violet curve) is $\mu$-independent, while  the full and small mass results for $\df\sigma^{(2,h)} / \df\sigma^\one$
individually depend on $\mu$, as $\df\sigma^{(1,h)}$ is nonzero for $\mu \neq m$.

Although the obligatory resummation (with mass-dependent RRG evolution kernels) in the peak region (i.e.\ for $q_T \sim 10$ GeV) may quantitatively change the result, \fig{XsecRatio} should nevertheless give a reasonable estimate of the potential size of the  bottom mass corrections to the Higgs $q_T$ distribution in gluon fusion mediated by a top loop.
As expected, the corrections (represented by the dotted violet curve) become negligibly small for $q_T \gg 10$ GeV, i.e\ away from the peak region.
Around the peak they amount to $\sim 1-2\%$ and increase for smaller $q_T$.
Despite their apparent small size, these corrections will likely matter for  precision analyses of the Higgs $q_T$ spectrum at N$^3$LL$^\prime$ accuracy (and beyond), which already has reached the few percent level~\cite{Billis:2021ecs}.

\section{Conclusion}
\label{sec:conclusion}

We have calculated the TMD gluon beam functions at NNLO in SCET with one massive and $n_l$ massless quark flavors.
Our results for the massive quark contributions to the renormalized beam function matching kernels are presented in \sec{results}.
The relevant two-loop diagrams associated with collinear real emissions are shown in \fig{diagslowestorder}b and \fig{realdiags2loopgg}. Their calculation is rather straightforward.
The purely virtual two-loop diagrams in \fig{virtdiags}, however, require a careful subtraction of non-trivial zero-bin contributions and involve terms that, upon rapidity regularization, exhibit a discontinuous behavior in the limit of vanishing rapidity regulator even though they are rapidity-finite.
Both types of contributions arise from two-loop diagrams with a massive quark bubble subdiagram on a virtual gluon line.
More precisely, they are generated by the part proportional to $p^\mu p^\nu$ of the virtual gluon propagator dressed with a massive quark bubble, where $p^\mu$ is the off-shell four-momentum flowing through the gluon line, see \subsec{dispersionmethod}.
The corresponding parts of the two-loop diagrams are referred to as $\kappa$ terms in this work.
We explicitly show that the zero-bin subtraction completely removes all $\kappa$ terms from the final beam function result.
In other words, one obtains the correct NNLO expressions by omitting the $\kappa$ terms in all diagrams with a massive quark bubble and thus also the non-vanishing zero-bin contributions.

An important application of our beam function results is the computation of bottom mass effects on the gluon-fusion production of Higgs bosons with small transverse momenta ($q_T \ll m_H$).
We have derived the full $(m_b/q_T)$-dependence of the Higgs $q_T$ distribution at leading order in the QCD coupling, i.e.\ at relative $\ord{\alpha_s}$, and at leading power in $1/m_H$, i.e.\ at $\ord{y_b^0}$ with $y_b$ the bottom Yukawa coupling.
We have also confirmed the anomalous dimensions relevant for the resummation of logarithms $\sim \ln(m_b^2/m_H^2) \sim \ln(q_T^2/m_H^2)$ at NNLL$^\prime$.
For N$^3$LL resummation only the quark mass corrections to the three-loop rapidity anomalous dimension is yet unknown.
Apart from the process-dependent hard function, i.e.\ an overall factor, our results directly carry over to the transverse momentum distribution of any other color singlet final state produced by gluon fusion.
We have performed a first numerical analysis at fixed order and found a few-percent level effect of the new bottom mass corrections on the Higgs $q_T$ spectrum in the peak region, where $q_T \sim m_b$.
A more sophisticated analysis including the resummation of large logarithms based on the factorization approach of \rcite{Pietrulewicz:2017gxc} for the different hierarchies between $m_b$, $m_H$, and $q_T \ll m_H$ is left for future work.

\begin{acknowledgments}
MS thanks Andr\'e Hoang for enlightening discussions and Frank Tackmann for comments on the manuscript.
This research was supported by the Munich Institute for Astro-, Particle and BioPhysics (MIAPbP) which is funded by the Deutsche Forschungsgemeinschaft (DFG, German Research Foundation) under Germany's Excellence Strategy – EXC-2094 – 390783311,
by the DFG through the Emmy-Noether Grant No.~TA 867/1-1, and the Collaborative Research Center (SFB) 676 Particles, Strings and the Early Universe.
\end{acknowledgments}

\appendix
\addtocontents{toc}{\protect\setcounter{tocdepth}{1}}

\section{Review of soft function calculation}
\label{app:softreview}

\begin{figure}[t]
  \begin{center}
    \includegraphics[height=0.21 \textwidth]{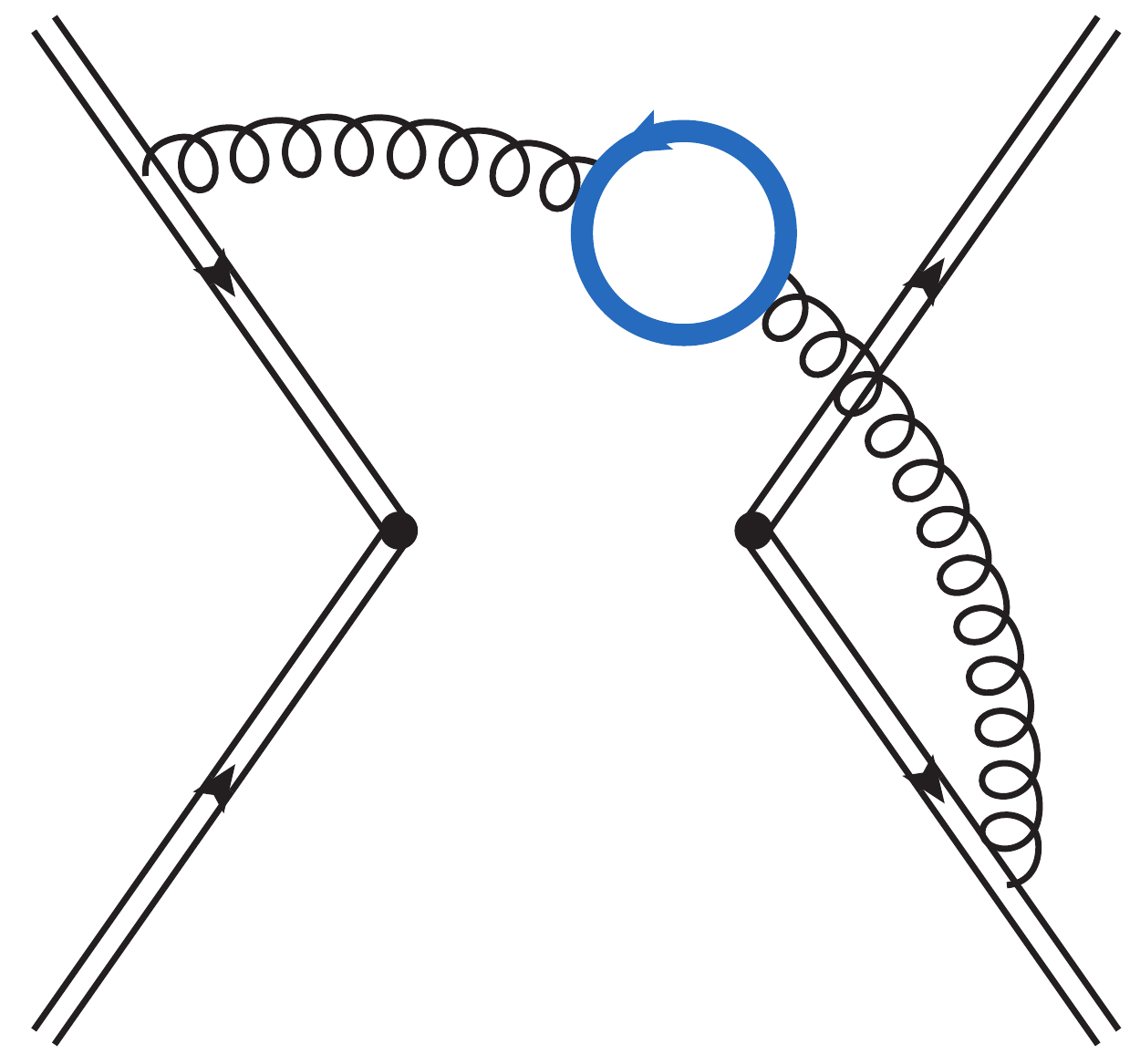}%
    \put(-57,0){(a)}
    \quad
    \includegraphics[height=0.21 \textwidth]{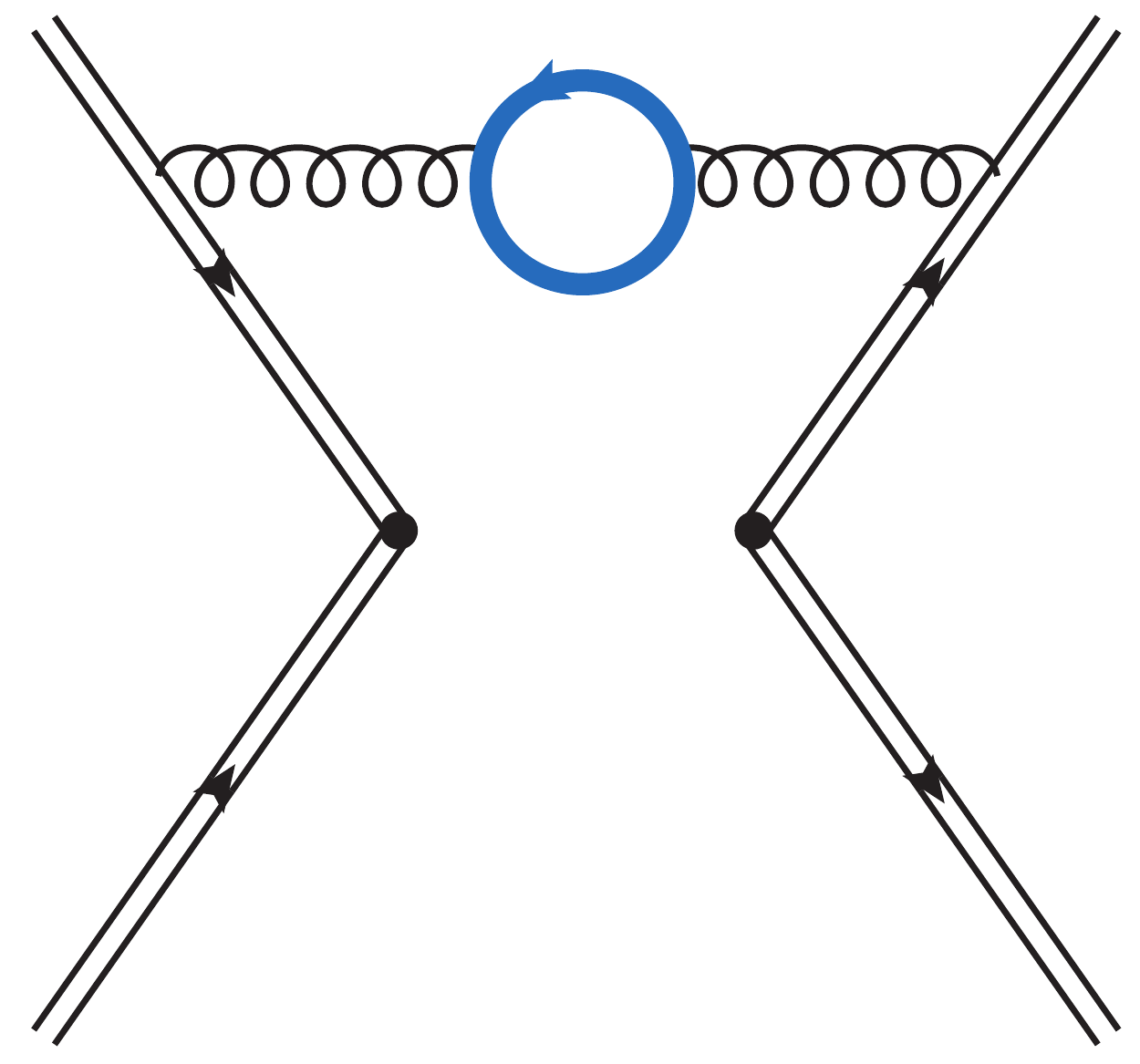}%
    \put(-57,0){(b)}
    \quad
    \includegraphics[height=0.21 \textwidth]{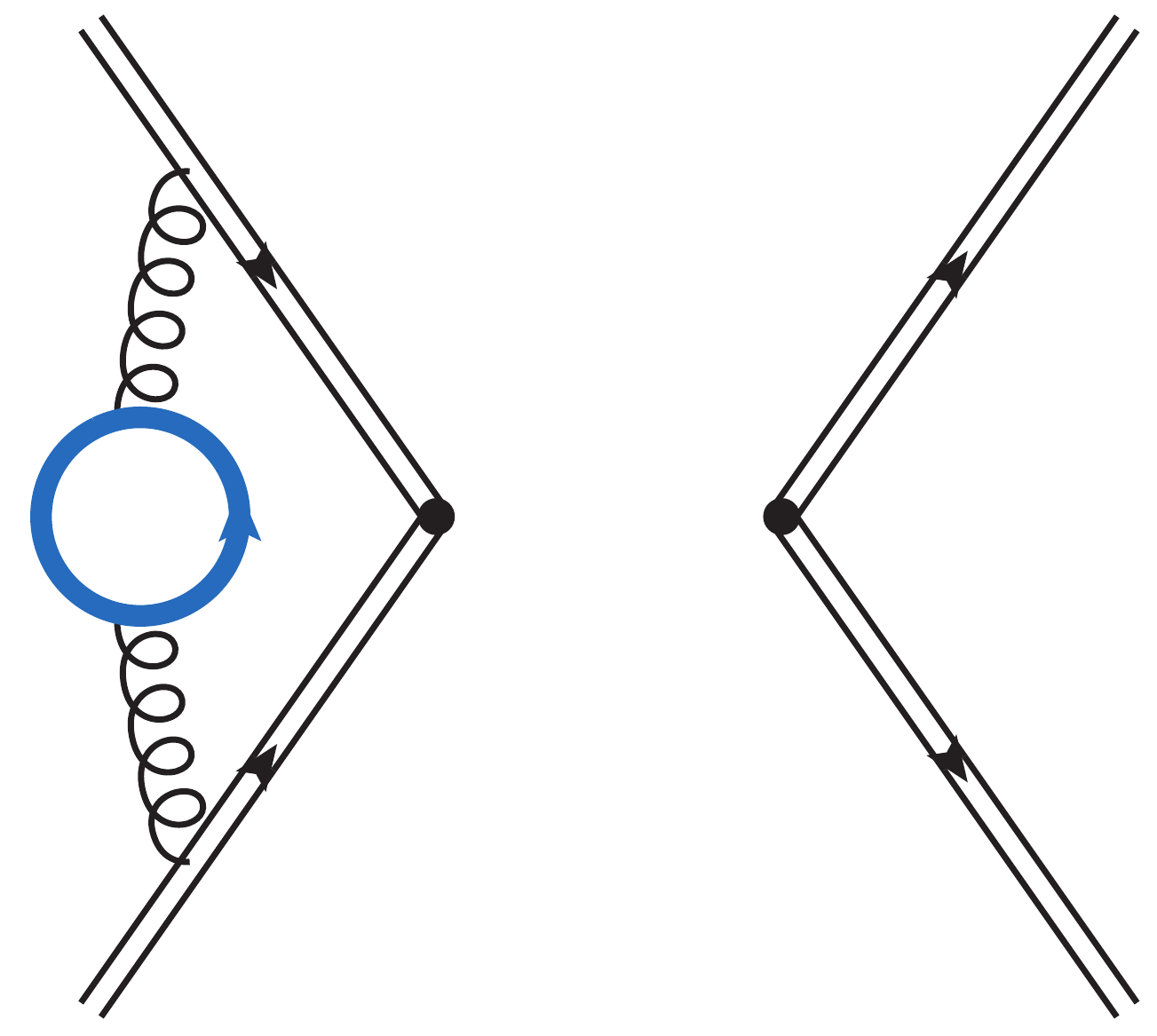}%
    \put(-57,0){(c)}
    \quad
    \includegraphics[height=0.21 \textwidth]{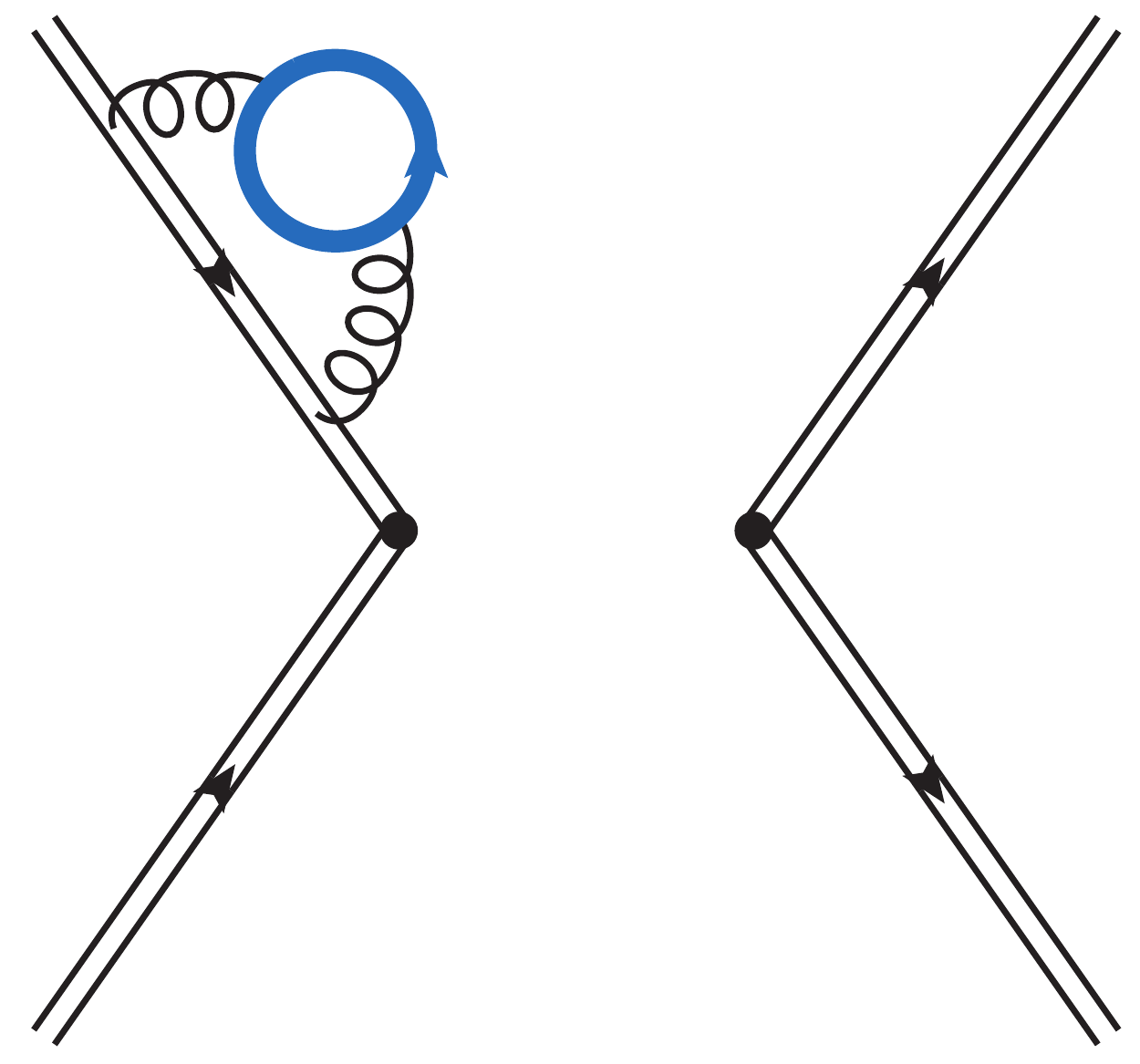}%
    \put(-57,0){(d)}
  \end{center}
  \caption{Diagrams with a massive quark loop (thick blue line) contributing to the soft function.
    The little arrows on the Wilson (double) lines indicate the original parton (here gluon) flow.
    All possible final-state cuts and mirror graphs are understood.
    Diagram~c corresponds to the purely virtual vertex correction. Diagram~d corresponds to a soft Wilson line self energy correction. It must be treated in analogy to a soft wave function correction, i.e.\ carefully regulated with an offshellness ($\delta$) and multiplied with a factor $1/2$, see \eq{softwavefct}.
    \label{fig:softdiags}
  }
\end{figure}

The two-loop heavy-flavor contribution $S_{gg}^{(2,h)}$ to the TMD soft function $S_{gg}$ in the factorization formula \eq{factXsec} was derived in \rcite{Pietrulewicz:2017gxc}.%
\footnote{The calculation in \rcite{Pietrulewicz:2017gxc} was done for incoming quark and antiquark, i.e.\ with Wilson lines in the fundamental color representation. Their two-loop result can however be directly translated to $S_{gg}^{(2,h)}$ due to Casimir scaling by replacing the overall color factor $C_F T_F \to C_A T_F$.}
The relevant soft function diagrams are displayed in \fig{softdiags}.
The calculation was performed using the dispersion relation in \eq{dispersion} with the $\kappa$ terms set to zero as justified by a gauge invariance argument following \rcite{Chiu:2009yx}.
In this appendix we explicitly demonstrate the cancellation of $\kappa$ terms among the diagrams in \fig{softdiags}. This requires a careful evaluation of diagram~\ref{fig:softdiags}d, which is closely related to the zero-bin contribution of the collinear diagram~\ref{fig:virtdiags}d, see \app{virtualZBs}.

By means of \eq{dispersion} the two-loop calculation of $S_{gg}^{(2,h)}$ is split into two one-loop calculations with a massive and a massless gluon exchanged between the soft Wilson lines, respectively.
The massless calculation yields the known result~\cite{Chiu:2012ir,Luebbert:2016itl} with an overall factor $\propto \Pi^{(1)}(0,m^2)$.
Here we focus on the one-loop diagrams corresponding to the graphs in \fig{softdiags} where the gluon line with a massive quark bubble is replaced by a gluon with mass $M$ over which we eventually integrate according to \eq{dispersion}.
For the real corrections from diagrams~\ref{fig:softdiags}a and~\ref{fig:softdiags}b we have to cut the massive gluon propagator, i.e.\ effectively replace it with a $\delta$-function that puts the gluon momentum on the mass shell (with positive energy).
The corresponding one-loop integrands read up to a common overall factor
\begin{align}
  \mathrm{[\ref{fig:softdiags}a]} & =
  \delta^{(d-2)}(\vec{k}_\perp-\vec{p}_{T,s})\,
  \theta(k_0) \, \delta(k^2-M^2)\,
  \biggl(
  \frac{2}{k^+ k^-}-\frac{\kappa}{M^2}
  \biggr) ,
  \nn\\
  \mathrm{[\ref{fig:softdiags}b]} & =
  \delta^{(d-2)}(\vec{k}_\perp-\vec{p}_{T,s})\,
  \theta(k_0) \, \delta(k^2-M^2)\,
  \frac{\kappa}{M^2}
  \,.
  \label{eq:realsoftints}
\end{align}
Applying the $\eta$-regulator~\cite{Chiu:2012ir} both integrands in \eq{realsoftints} are multiplied with the same factor $\nu^\eta |k^- - k^+|^{-\eta}$.
We see that the $\kappa$ terms of the two real-emission integrands cancel each other.

The (normalized) integrand of the virtual diagram~\ref{fig:softdiags}c is given by (suppressing the $\ri0$-prescription)
\begin{align}
  \mathrm{[\ref{fig:softdiags}c]} & =
  \frac{2}{k^+ k^- (k^2 - M^2)}-\frac{\kappa}{k^2 (k^2-M^2)}\,.
  \label{eq:softvirtdiagram}
\end{align}
Again the $\eta$-regulator adds a factor $\nu^\eta |k^- - k^+|^{-\eta}$.
Note, however, that the $\kappa$ term is rapidity-finite and the result for this term is independent of whether the $\eta \to 0$ limit is taken before or after the integration (in contrast to the collinear integral in \eq{Ieta}).
This $\kappa$ term is exactly canceled by the soft wave function renormalization represented by diagram~\ref{fig:softdiags}d.

The integrand $\mathrm{[\ref{fig:softdiags}d]}$, similar to that of diagram~\ref{fig:virtdiags}d, requires an offshellness $\delta$ of the external (Wilson) line with the self energy insertion, and the $\eta$-regulator must not be applied.
In SCET, a consistent implementation of the offshellness in soft diagrams requires the parametric scaling $\delta \sim m^2/Q \ll m$, because $\delta$ is related to the offshellness $p^2 \sim Q \delta\sim m^2$ of the underlying full QCD (and  corresponding collinear) diagrams, like in \eq{diag3d}.
To comply with the EFT power counting we therefore must expand the integrand of diagram~\ref{fig:softdiags}d in $\delta/k \sim \delta/M \sim m/Q$, see also \rcite{Chiu:2009yx}.
With the same normalization as for \eq{softvirtdiagram} and taking into account the factor $1/2$ for wave function renormalization  we have
(with $k^- \leftrightarrow k^+$ for two of the mirror diagrams)
\begin{align}
  \mathrm{[\ref{fig:softdiags}d]} & = \frac12\,
  \frac{\kappa}{k^2 (k^2-M^2)}
  \biggl[ - \frac{k^-}{\delta} + 1 +\ord{\delta}
  \biggr]
  \,.
  \label{eq:softwavefct}
\end{align}
The $1/\delta$ pole in \eq{softwavefct} vanishes upon integration over the soft loop momentum $k^\mu$ as it is antisymmetric under $k^\mu \leftrightarrow -k^\mu$.%
\footnote{Equation~\eqref{eq:softwavefct} is therefore equivalent to $\bigl[\frac{\partial}{\partial \delta}\, \delta\, \mathrm{[\ref{fig:softdiags}d]}\bigr]_{\delta \to 0}$, where the integrand of diagram~\ref{fig:softdiags}d is evaluated in analogy to \eq{diag3d}. The $\delta \to 0$ limit must then consistently be taken before the loop integration.}
So, (before integration over $M$) the total $\eta$-regulated contribution of diagrams~\ref{fig:softdiags}c,~\ref{fig:softdiags}d, and their mirror graphs is proportional to
\begin{align}
  \lim_{\delta \to 0}\,
  \int\!\! \frac{\df^d k}{(2 \pi)^d}
  \biggl(
  2\times \mathrm{[\ref{fig:softdiags}c]}\, \nu^\eta |k^- - k^+|^{-\eta}
  + 4\times \mathrm{[\ref{fig:softdiags}d]}
  \biggr)
  =
  \int\!\! \frac{\df^d k}{(2 \pi)^d} \,
  \frac{ 2\nu^\eta |k^- - k^+|^{-\eta}}{k^+ k^- (k^2 - M^2)}
\end{align}
and thus $\kappa$-independent.
The final $\MS$ renormalized result for $S_{gg}^{(2,h)}$ is given in \eq{Sgg2h}.


\section{Soft zero-bin contributions}
\label{app:ZBs}

In this appendix we give details on the calculation of the soft zero-bin contributions. These must be subtracted from the sum of the diagrams in \figs{realdiags2loopgg}{virtdiags} in order to avoid double counting soft contributions already contained in the diagrams of \fig{softdiags}.
The zero-bin contributions are obtained by expanding the integrand (including the measurement $\delta$-functions) of the corresponding collinear diagrams assuming a loop-momentum scaling such that the momentum passing through at least one of the propagators is soft, while the momenta of the other propagators may be soft or collinear~\cite{Manohar:2006nz}.
In this way, for each two-loop diagram in \figs{realdiags2loopgg}{virtdiags} different zero-bins associated with soft-collinear and soft-soft loop-momentum regions arise.
Except for some zero-bins from diagrams with a massive quark bubble, however, we find that all of these are power-suppressed (w.r.t.\ $1/Q$) and/or individually scaleless and thus do not contribute to the (leading-power) beam function kernel $\mathcal{I}_{gg}$.
In particular, there is no non-trivial zero-bin contribution from diagrams with massive quark triangle or box subgraphs.
This is expected, because the zero-bin subtractions are supposed to remove the overlap with the contributions from the soft graphs in \fig{virtdiags}, which all contain a massive quark bubble.

For the relevant zero-bins the scaling of the loop momentum running inside the massive quark bubble is the same than the momentum scaling of the gluons attached to it.
It is convenient to evaluate these contributions
by means of the dispersion relation in \eq{dispersion}.
The relevant two-loop zero-bins can thus be expressed as one-parameter integrals of
soft  zero-bins of the corresponding one-loop diagrams with a massive gluon propagator. The second term in the second line of \eq{dispersion} gives rise to a contribution proportional to the total massless one-loop zero-bin, which is scaleless and vanishes~\cite{Chiu:2012ir}.

\subsection{Zero-bin contributions of real-emission graphs}
\label{app:realZBs}

In this section we show that, like in the massless case, the total zero-bin contribution from the real-emission diagrams in \fig{realdiags2loopgg} vanishes.
Employing the dispersion relation in \eq{dispersion} we find the following relevant zero-bin integrands from real emission diagrams up to a common prefactor
$\propto \delta(1-x)\, \mathrm{Im}\big[\Pi^{(1)}(p^2,m^2)\big]/M$ (and suppressing the $\ri0$ prescription):
\begin{align}
  2\times \mathrm{[\ref{fig:realdiags2loopgg}e]}_{\textrm{0-bin}} & =
  \frac{2 \xi }{(\ell^- \ell^+\!-\pTsq)^2}\,,
  \nn\\
  2\times \mathrm{[\ref{fig:realdiags2loopgg}f]}_{\textrm{0-bin}} & =
  \frac{4}{\ell^- \ell^+ (\ell^- \ell^+\!-\pTsq)}
  -\frac{2 \xi }{(\ell^- \ell^+-\pTsq)^2}\,,
  \nn\\
  \mathrm{[\ref{fig:realdiags2loopgg}g]}_{\textrm{0-bin}} & =
  \frac{\kappa}{(\ell^- \ell^+\!-\pTsq)
    (\ell^- \ell^+-M^2-\pTsq)}\,,
  \nn\\
  2\times\mathrm{[\ref{fig:realdiags2loopgg}h]}_{\textrm{0-bin}} & =
  \frac{4}{\ell^- \ell^+ (\ell^- \ell^+-M^2-\pTsq)}
  -\frac{2 \kappa}{(\ell^- \ell^+\!-\pTsq)
    (\ell^- \ell^+-M^2-\pTsq)}\,,
  \nn\\
  \mathrm{[\ref{fig:realdiags2loopgg}i]}_{\textrm{0-bin}} & =
  \frac{\kappa}{(\ell^- \ell^+\!-\pTsq) (\ell^- \ell^+-M^2-\pTsq)}
  \,.
  \label{eq:realZBints}
\end{align}
The factors of 2 on the right account for left-right mirror graphs.
The integration variables are $\ell^+$, $\ell^-$, and $M$.
The integration over $\vec{\ell}_\perp$ has already been performed exploiting the TMD beam function measurement $\propto \delta^{(d-2)}(\pT -\vec{\ell}_T)$.
The zero-bin loop-momenta scale as follows: $\ell^+ \sim \ell^- \sim |\pT| \sim M \sim m$.

In \eq{realZBints} no rapidity regulator has been implemented yet.
Here, the naive implementation of the $\eta$-regulator according to the prescription in \eq{etaregWL} fails, because it violates gauge symmetry (see also footnote \ref{footnote:rapreg}).
This concerns not only the $\xi$-dependent terms, but indirectly also the $\kappa$-dependent terms,
since the latter are tied to the $\xi$-dependent terms of the corresponding massless one-loop zero-bin integrands by the replacement $\kappa \to \xi$ and
$M \to 0$.
The naive  prescription in \eq{etaregWL} would assign a factor $|\ell^-|^{-\eta}$ to
$\mathrm{[\ref{fig:realdiags2loopgg}f]}_{\textrm{0-bin}}$ and $\mathrm{[\ref{fig:realdiags2loopgg}h]}_{\textrm{0-bin}}$, and a factor
$|\ell^-|^{-2\eta}$ to $\mathrm{[\ref{fig:realdiags2loopgg}i]}_{\textrm{0-bin}}$.
These terms would thus vanish upon $\ell$ integration and leave a non-zero gauge-dependent zero-bin contribution from $\mathrm{[\ref{fig:realdiags2loopgg}e]}_{\textrm{0-bin}}$ and
$\mathrm{[\ref{fig:realdiags2loopgg}g]}_{\textrm{0-bin}}$.
The $\eta$-regulator must therefore be implemented \textit{after} the cancellation of the $\xi$ and $\kappa$ terms in the sum of the zero-bin integrands.
This treatment is consistent with the cancellation of the $\kappa$ terms in the real-emission soft function diagrams in \fig{softdiags}a,b, which are both regulated by the same factor $\propto |\ell^- - \ell^+|^{-\eta}$, cf.\ \eq{realsoftints}.
The $\kappa$- and $\xi$-independent terms in \eq{realZBints} integrate to zero (with or without $\eta$-regulator). Hence the total zero-bin contribution from real emission graphs vanishes.

\subsection{Virtual zero-bin contributions}
\label{app:virtualZBs}

The relevant (unregulated) zero-bin integrands  of the virtual diagrams in \fig{virtdiags} are
\begin{align}
  \mathrm{[\ref{fig:virtdiags}b]}_{\textrm{0-bin}} & =
  \frac{2}{\ell^+\ell^- (\ell^2-M^2)}
  -\frac{\kappa}{\ell^2 (\ell^2-M^2)}
  \,,\nn\\
  \mathrm{[\ref{fig:virtdiags}c]}_{\textrm{0-bin}} & =
  \frac{\kappa}{\ell^2 (\ell^2-M^2)}
  \,,\nn\\
  \mathrm{[\ref{fig:virtdiags}e]}_{\textrm{0-bin}} = \mathrm{[\ref{fig:virtdiags}d]}_{\textrm{0-bin}} & =
  \frac12 \frac{\kappa}{\ell^2 (\ell^2-M^2)}
  \,,
\end{align}
where $\ell^\mu \sim M \sim m$ and we suppressed a common factor $\propto \delta^\two(\pT)\,  \delta(1-x)\, \mathrm{Im}\big[\Pi^{(1)}(p^2,m^2)\big]/M$ (as well as the $\ri0$ prescription).

Implementing the $\eta$-regulator adds a factor $\nu^\eta|\ell^-|^{-\eta}$ to the integrand $\mathrm{[\ref{fig:virtdiags}b]}_{\textrm{0-bin}}$, so that the integrated zero-bin contribution of diagram~\ref{fig:virtdiags}b vanishes (for $\eta \neq 0$).
As discussed in \subsec{rapreg}, the $\kappa$ term of $\mathrm{[\ref{fig:virtdiags}b]}_{\textrm{0-bin}}$ exactly cancels the term in the integrand of diagram~\ref{fig:virtdiags}b that gives rise to the discontinuous $\eta \to 0$ limit due to the integral in \eq{Ieta}.
The rapidity-finite part (including the $\kappa$ term) of the zero-bin subtracted diagram~\ref{fig:virtdiags}b therefore does not need to be rapidity-regulated.
The zero-bin integrand $\mathrm{[\ref{fig:virtdiags}c]}_{\textrm{0-bin}}$ exactly equals that of the unsubtracted diagram~\ref{fig:virtdiags}c. The zero-bin subtracted diagram~\ref{fig:virtdiags}c therefore vanishes (regardless of rapidity regularization).

The zero-bin integrand $\mathrm{[\ref{fig:virtdiags}e]}_{\textrm{0-bin}}$ is derived from the unintegrated \eq{diag3d} respecting the scaling $p^2 \sim m^2$ and thus $p^+ \sim m^2/Q$.
As a consequence of SCET power counting we must take the $p^+ \to 0$ limit at the level of the zero-bin integrand. This is consistent with the evaluation of the soft diagram~\ref{fig:softdiags}d in \app{softreview}.
The factor $1/2$ in $\mathrm{[\ref{fig:virtdiags}e]}_{\textrm{0-bin}}$ is due to wavefunction renormalization.
The zero-bin contribution of diagram~\ref{fig:virtdiags}e effectively cancels the soft wave function contribution represented by diagram~\ref{fig:softdiags}d once the corrections from all four external legs (of the soft function and the two partonic beam functions, respectively) are combined in the factorized cross section in \eq{factXsec}.
This is expected because the unsubtracted collinear wavefunction renormalization exactly equals that of full QCD~\cite{Bauer:2000yr}.

After zero-bin subtraction the $\kappa$ terms of diagrams~\ref{fig:virtdiags}b and~\ref{fig:virtdiags}d cancel each other exactly.
This resembles the cancellation of the $\kappa$ dependence  from diagrams~\ref{fig:softdiags}c and~\ref{fig:softdiags}d
within the virtual contribution to the soft function as shown in \app{softreview}.

\section{Perturbative ingredients}
\label{app:known}

\subsection{Splitting functions}
\label{app:splitting}

The leading-order (one-loop) PDF anomalous dimensions $\gamma_{f,ij}^\zero = \alpha_s P_{ij}^\zero/\pi$ are given by
\begin{align}
  P_{q_i q_j}^\zero(z) &= C_F\, \theta(z)\, \delta_{ij} P_{qq}(z)
  \,,\nn\\
  P_{q_ig}^\zero(z) = P_{\bar q_ig}^\zero(z) &= T_F\, \theta(z) P_{qg}(z)
  \,,\nn\\
  P_{gg}^\zero(z) &= C_A\, \theta(z) P_{gg}(z) + \frac{\beta_0}{2}\,\delta(1-z)
  \,,\nn\\
  P_{gq_i}^\zero(z) = P_{g\bar q_i}^\zero(z) &= C_F\, \theta(z) P_{gq}(z)
  \,,
\end{align}
with $q_i$ explicitly denoting here the different massless quark flavors, and the usual one-loop (LO) quark and gluon splitting functions
\begin{align}
  P_{qq}(z)
  &= \cL_0(1-z)(1+z^2) + \frac{3}{2}\,\delta(1-z)
  \equiv \biggl[\theta(1-z)\,\frac{1+z^2}{1-z}\biggr]_+
  \,,\nn\\
  P_{qg}(z) &= \theta(1-z)\bigl[(1-z)^2+ z^2\bigr]
  \,,\nn\\
  P_{gg}(z)
  &= 2 \cL_0(1-z) \frac{(1 - z + z^2)^2}{z}
  \,,\nn\\
  P_{gq}(z) &= \theta(1-z)\, \frac{1+(1-z)^2}{z}
  \,.
  \label{eq:Pij}
\end{align}

\subsection{Beam function results for massless quarks}
\label{app:masslessBFresults}

The bare massless one-loop partonic TMD beam functions are given by
\begin{align}
B_{g/q}^{(1)}(\pT,z,\mu) &=
 \mathcal{I}^{(1)}_{gq}(\vec{p}_{T},z,\mu)
 - \frac{\alpha_s}{2\pi \eps} P_{gq}^{(0)}(z)\, \delpT \,,
  \label{eq:Bgq1}
\\
B_{g/g}^{(1,l)}\!\Bigl(\pT,z,\mu,\frac{\nu}{\omega}\Bigr) &=
Z_{B_g}^{(1,l)} \delta(1-z) + \mathcal{I}^{(1,l)}_{gq}(\vec{p}_{T},z,\mu)
 - \frac{\alpha_s}{2\pi \eps} P_{gg}^{(0)}(z)\, \delpT \,,
\label{eq:Bgg1l}
\end{align}
with the (renormalized) massless one-loop matching coefficients
\begin{align}
\mathcal{I}^{(1,l)}_{gg}\Bigl(\pT,z,\mu,\frac{\nu}{\omega}\Bigr) ={}&
\frac{\alpha_s C_A}{4\pi}
\biggl[
2\delta(1-z) \ln\frac{\omega}{\nu} +  \theta(z) P_{gg}(z)
\biggr]
\nn\\
&\times
\biggl[
2 \LpT{0} + \frac{1}{6} \pi ^2 \delpT\, \eps -2 \LpT{1}\, \eps
+\ord{\eps^2}
\biggr]\, ,
\\
\mathcal{I}^{(1)}_{gq}(\vec{p}_{T},z,\mu) ={}&
\mathcal{I}^{(1,l)}_{gq}(\vec{p}_{T},z,\mu) =
\frac{\alpha_s C_F}{4\pi}\, \theta(z)\,
\biggl\{
2\theta(1-z)z\,\delpT
+2 P_{gq}(z)\,\mathcal{L}_0(\pT,\mu)
\nn\\
&\hspace{-12 ex}
+\biggl[\frac{\pi ^2}{6} P_{gq}(z)  \delpT -2\theta(1 - z)z\, \LpT{0} - 2  P_{gq}(z)  \LpT{1} \biggr] \epsilon
+\ord{\eps^2}
\biggr\}\,,
\end{align}
and the gluon beam function counterterm
\begin{align}
  Z_{B_g}^{(1,l)} ={}& \frac{\alpha_s}{4\pi} \biggl\{
  \frac{1}{\epsilon } \biggl(\beta_0
  - 4 C_A \ln \frac{\omega }{\nu }\biggr) \delpT
  \nn\\
  &
  +\frac{C_A}{\eta }\biggl[
  \frac{4}{\eps} \delpT
  -4 \LpT{0}
  - \biggl(\frac{\pi^2}{3} \delpT- 4 \LpT{1} \biggr) \eps
  +\ord{\eps^2}
  \biggr]
  \biggr\}\,
  \label{eq:ZB1l}\,.
\end{align}
The plus distribution $\mathcal{L}_n(\pT,\mu)$ is defined in \eq{LpTdef}.

At two loops the contributions due to a single massless quark flavor read~\cite{Luebbert:2016itl,Gehrmann:2014yya}
\begin{align}
  \mathcal{I}^{(2,l)}_{gg}
  \Big|_{T_F}^{n_l=1} &= \frac{\alpha_s^2 T_F}{16\pi^2} \theta(z) \,
  \Biggl\{C_F\,  \theta(1-z) \biggl[\mathcal{L}_1(\pT,\mu)
  \biggl(16(1+z)\ln z+\frac{8(4+3z-3z^2-4z^3)}{3z}\biggr)
  \nn \\
  &\quad +\mathcal{L}_0(\pT,\mu)
  \biggl(-8(1+z)\ln^2z -24(1+z)\ln z + \frac{8(2-21z+15z^2+4z^3)}{3z}\biggr)
  \nn \\
  &\quad +\frac{8}{3} \delpT  \biggl( \frac{1+z}{2} \ln^3 z+ \frac{9+3z}{4}\ln^2 z
  +9(1+z)\ln z
  -\frac{1-24z+24z^2-z^3}{z}\biggr)\biggr]
  \nn \\
  & + C_A \biggl[\mathcal{L}_1(\pT,\mu)\biggl(
  \frac{16}{3} \delta(1-z) \ln \frac{\omega}{\nu}
  +\frac{8}{3} \,P_{gg}(z)\biggr)
  \nn\\
  &\quad +\mathcal{L}_0(\pT,\mu)
  \biggl(-\frac{80}{9}\delta(1-z) \ln \frac{\omega}{\nu}
  -\frac{80}{9}\mathcal{L}_0(1-z)
  -\frac{16(1+z)}{3} \theta(1-z) \ln z
  \nn\\
  &\qquad -\frac{8(23-29z+19z^2-23z^3)}{9z}\, \theta(1-z)  \biggr)
  \nn\\
  &\quad
  +\delpT \biggl(\frac{224}{27} \delta(1-z)  \ln \frac{\omega}{\nu}
  +\frac{224}{27} \mathcal{L}_0(1-z)
  +\theta(1-z) \biggl\{
  \frac{4(1+z)}{3} \ln^2 z
  \nn\\
  &\qquad +\frac{4(13+10z)}{9} \ln z-\frac{4z}{3}\ln(1-z)
  +\frac{4(121-166z+110z^2-139z^3)}{27z}
  \biggr\}
  \biggr)\biggr]\Biggr\}\, ,
  \\
  \mathcal{I}^{(2,l)}_{gq}
  \Big|_{T_F}^{n_l=1} &= \frac{\alpha_s^2 C_F T_F}{16\pi^2} \,\theta(z)\,\Biggl\{\frac{16}{3} P_{gq}(z)\, \LpT{1}
  \nn \\
  &\quad - \frac{16}{3} \LpT{0}\,P_{gq}(z) \biggl[\ln(1-z)+\frac{5}{3}\biggr] +\frac{4}{3} \delta^{(2)}(\pT)\biggl[P_{gq}(z) \ln^2(1-z)
  \nn \\
  &\qquad + 2\biggl(\frac{5}{3} P_{gq}(z)- \theta(1-z) z \biggr)  \ln(1-z)+\frac{56}{9}P_{gq}(z)-\frac{10}{3} \theta(1-z) z\biggr]\Biggr\}\, .
  \label{eq:I2lgqnlglei1TF}
\end{align}

\subsection{Massive PDF matching factors}
\label{app:PDFmatching}

The massive PDF matching coefficients were calculated up to two loops in \rcite{Buza:1996wv}. At one loop the relevant expressions read
\begin{align}
\mathcal{M}^{(1)}_{gg}(m,z,\mu) & = \frac{\alpha_s T_F}{4\pi} \,\frac{4}{3} L_m  \, \delta(1-z)\, , \nn \\
\mathcal{M}^{(1)}_{Qg}(m,z,\mu) & = -\frac{\alpha_s T_F}{4\pi}\,2 L_m \,\theta(z)\,P_{qg}(z)\,   ,
\end{align}
where $L_m \equiv \ln(m^2/\mu^2)$.
At two loops we need
\begin{align}
  \mathcal{M}^{(2)}_{gg}
  & = \frac{\alpha_s^2 T_F}{16\pi^2}\,\theta(z)
  \biggl\{
  C_F \biggl[
  \theta(1-z) \biggl(8(1+z)\ln z\, L_m^2 + \frac{4(4+3z-3z^2-4z^3)}{3z}L_m^2
  \nn \\
  & \quad + 8(1+z) \ln^2z\, L_m +8(3+5z)\ln z\, L_m
  -\frac{8(2-24z+12z^2+10z^3)}{3z} L_m
  \nn \\
  & \quad +  \frac{4(1+z)}{3}\ln^3 z+2(3+5z)\ln^2 z +16(2+3z)\ln z-\frac{8(1-10z+6z^2+3z^3)}{z} \biggr) \nn \\
  & \quad + \bigl( 4 L_m - 15 \bigr)\, \delta(1-z) \biggr]
  \nn\\
  & +C_A \biggl[
  \biggl(\frac{8}{3}L_m^2+\frac{80}{9}L_m+\frac{224}{27}\biggr)
  \mathcal{L}_0(1-z)
  +\biggl(\frac{16}{3}L_m +\frac{10}{9}\biggr)\delta(1-z)
  \nn \\
  & \quad + \frac{8}{3}\, \theta(1-z)\biggl(\frac{1-2z+z^2-z^3}{z} L_m^2
  + 2(1+z) \ln z\,L_m  +  \frac{23-29z+19z^2-23z^3}{3z} L_m
  \nn \\
  & \quad +\frac{1+z}{2} \ln^2 z+\frac{13+22z}{6} \ln z -\frac{z}{2} \ln(1-z)+\frac{139-157z+137z^2-175z^3}{18z}
  \biggr)
  \biggr]
  \nn \\
  & +\frac{16}{9} L_m^2T_F \,\delta(1-z)
  \biggr\} \, ,
\\
\mathcal{M}^{(2)}_{gq}
  & = \frac{\alpha_s^2 C_F T_F}{16\pi^2}\,\theta(1-z)\,\theta(z)\,\Biggl\{\frac{8(2-2z+z^2)}{3z}  L_m^2
  + \frac{16(2-2z+z^2)}{3z}\ln(1-z)L_m\nn \\
  & \quad + \frac{32(5-5z+4z^2)}{9z}L_m +\frac{4(2-2z+z^2)}{3z}\ln^2(1-z)
  +\frac{16(5-5z+4z^2)}{9z} \ln(1-z)
  \nn \\
  & \quad +\frac{8(56-56z +43 z^2)}{27z} \Biggr\} \, .
\end{align}

\subsection{Hard massive threshold correction}

The matching correction $H_c^g$ due to collinear mass modes arising in the limit $Q \gg m \gg q_T$ can be inferred from the literature via \eq{Igq2hlargemass}.
 Up to two-loop order we find
\begin{align}
H^g_{c}\Bigl(m,\mu,\frac{\nu}{\omega}\Bigr) & = 1+ \frac{\alpha_s T_F}{4\pi} \,\frac{4}{3} L_m + \frac{\alpha_s^2  T_F}{16\pi^2} \biggl\{C_A\biggl[\biggl(\frac{8}{3}L_m^2+\frac{80}{9}L_m+\frac{224}{27}\biggr)
\ln\frac{\nu}{\omega} +\frac{16}{3}L_m+\frac{10}{9}\biggr] \nn \\
& \qquad +C_F\bigl(4L_m-15\bigr)+\frac{16}{9} T_F L_m^2\biggr\}
 +\ord{\alpha_s^3}
 \, .
 \label{eq:Hgc}
\end{align}

\subsection{Soft function and rapidity anomalous dimension}
\label{app:Softfctresults}

The two-loop soft function correction due to the heavy quark flavor reads~\cite{Pietrulewicz:2017gxc}
\begin{align}
S_{gg}^{(2,h)}(\pT,m,\mu,\nu)
&= \frac{\alpha_s^2 C_A T_F}{16\pi^2}\,\biggl\{ \delta^{(2)}(\pT)\biggl[\biggl(-\frac{16}{3}L_m^2-\frac{160}{9}L_m-\frac{448}{27}\biggr)\ln\frac{\nu}{\mu} +\frac{8}{9}L_m^3+\frac{40}{9}L_m^2
\nn \\ & \qquad
+ \biggl(\frac{448}{27}-\frac{4\pi^2}{9}\biggr)L_m +\frac{656}{27}-\frac{10\pi^2}{27}-\frac{56\zeta_3}{9}\biggr]
\nn\\& \quad
+\frac{16}{9\pi\pTsq}
 \biggl[2\biggl(-5+12\hat{m}^2+3c_1(1-2\hat{m}^2)\ln\frac{c_1+1}{c_1-1} \biggr)\ln\frac{\nu}{m}
\nn \\ & \qquad
+3c_1(1-2\hat{m}^2)\biggl(\Li_2\biggl(\frac{(c_1-1)^2}{(c_1+1)^2}\biggr)
+\ln\frac{c_1+1}{c_1-1} \ln\frac{\mhsq (c_1+1)^2}{4c_1^{\,2}} -
\frac{\pi^2}{6}\biggr)
\nn \\ & \qquad
+c_1(5-16\hat{m}^2)\ln\frac{c_1+1}{c_1-1} + 8\hat{m}^2 \biggr]\biggr\}
+\frac{\alpha_s T_F}{4\pi }\,\frac{4}{3}L_m\,S_{gg}^{(1)}(\pT,\mu,\nu)
\,,
\label{eq:Sgg2h}
\end{align}
where $\alpha_s\equiv\alpha_s^{\{n_l\}}(\mu)$, $\hat{m}\equiv m/|\pT|$, $c_1=\sqrt{1+4\hat{m}^2}$, and the one-loop contribution is~\cite{Chiu:2012ir,Luebbert:2016itl}
\begin{align}
S_{gg}^{(1)}(\pT,\mu,\nu)
&= \frac{\alpha_s C_A}{4\pi}\,\biggl[
-4\mathcal{L}_1(\pT,\mu)
+8\ln\frac{\nu}{\mu}\, \mathcal{L}_0(\pT,\mu)
-\frac{\pi^2}{3}\delta^{(2)}(\pT)
\biggr] \, .
\end{align}
The massless $\eta$-regulated two-loop soft function $S_{gg}^{(2,l)}$ can be found in \rcite{Luebbert:2016itl}.
The two-loop massive quark correction to the soft rapidity anomalous dimension is given by~\cite{Pietrulewicz:2017gxc}
 \begin{align}\label{eq:gammaS1h}
 \gamma_{\nu, S}^{(1,h)}(\pT,m, \mu)
 &= \frac{ \alpha_s^2 C_A T_F}{16\pi^2}\,\biggl\{
 \frac{32}{3}L_m \,\LpT{0}
 - \delpT \biggl[ \frac{16}{3}L_m^2 +\frac{160}{9}L_m +\frac{448}{27}\biggr]
 \nn \\ & \quad
 +\frac{32}{9\pi \pTsq} \biggl[-5+12\hat{m}^2+3c_1(1-2\hat{m}^2)\ln\frac{c_1+1}{c_1-1} \biggr]\biggr\}
 \,.\end{align}


\phantomsection
\addcontentsline{toc}{section}{References}
\bibliographystyle{jhep}
\bibliography{../bmassggF}

\providecommand{\href}[2]{#2}\begingroup\raggedright\begin{thebibliography}{10}

\bibitem{Cepeda:2019klc}
M.~Cepeda {\em et~al.}, {\it {Report from Working Group 2}: {Higgs Physics at
  the HL-LHC and HE-LHC}},  {\em CERN Yellow Rep. Monogr.} {\bf 7} (2019)
  221--584, [\href{http://arXiv.org/abs/1902.00134}{{\tt arXiv:1902.00134}}].

\bibitem{Grazzini:2016paz}
M.~Grazzini, A.~Ilnicka, M.~Spira, and M.~Wiesemann, {\it {Modeling BSM effects
  on the Higgs transverse-momentum spectrum in an EFT approach}},  {\em JHEP}
  {\bf 03} (2017) 115, [\href{http://arXiv.org/abs/1612.00283}{{\tt
  arXiv:1612.00283}}].

\bibitem{Bishara:2016jga}
F.~Bishara, U.~Haisch, P.~F. Monni, and E.~Re, {\it {Constraining Light-Quark
  Yukawa Couplings from Higgs Distributions}},  {\em Phys. Rev. Lett.} {\bf
  118} (2017), no.~12 121801, [\href{http://arXiv.org/abs/1606.09253}{{\tt
  arXiv:1606.09253}}].

\bibitem{Soreq:2016rae}
Y.~Soreq, H.~X. Zhu, and J.~Zupan, {\it {Light quark Yukawa couplings from
  Higgs kinematics}},  {\em JHEP} {\bf 12} (2016) 045,
  [\href{http://arXiv.org/abs/1606.09621}{{\tt arXiv:1606.09621}}].

\bibitem{Bonner:2016sdg}
G.~Bonner and H.~E. Logan, {\it {Constraining the Higgs couplings to up and
  down quarks using production kinematics at the CERN Large Hadron Collider}},
  \href{http://arXiv.org/abs/1608.04376}{{\tt arXiv:1608.04376}}.

\bibitem{Billis:2021ecs}
G.~Billis, B.~Dehnadi, M.~A. Ebert, J.~K.~L. Michel, and F.~J. Tackmann, {\it
  {Higgs pT Spectrum and Total Cross Section with Fiducial Cuts at Third
  Resummed and Fixed Order in QCD}},  {\em Phys. Rev. Lett.} {\bf 127} (2021),
  no.~7 072001, [\href{http://arXiv.org/abs/2102.08039}{{\tt
  arXiv:2102.08039}}].

\bibitem{Re:2021con}
E.~Re, L.~Rottoli, and P.~Torrielli, {\it {Fiducial Higgs and Drell-Yan
  distributions at N$^3$LL$^\prime$+NNLO with RadISH}},
  \href{http://arXiv.org/abs/2104.07509}{{\tt arXiv:2104.07509}}.

\bibitem{Becher:2020ugp}
T.~Becher and T.~Neumann, {\it {Fiducial $q_T$ resummation of color-singlet
  processes at N$^3$LL+NNLO}},  {\em JHEP} {\bf 03} (2021) 199,
  [\href{http://arXiv.org/abs/2009.11437}{{\tt arXiv:2009.11437}}].

\bibitem{Bizon:2018foh}
W.~Bizo\'n, X.~Chen, A.~Gehrmann-De~Ridder, T.~Gehrmann, N.~Glover, A.~Huss,
  P.~F. Monni, E.~Re, L.~Rottoli, and P.~Torrielli, {\it {Fiducial
  distributions in Higgs and Drell-Yan production at N$^{3}$LL+NNLO}},  {\em
  JHEP} {\bf 12} (2018) 132, [\href{http://arXiv.org/abs/1805.05916}{{\tt
  arXiv:1805.05916}}].

\bibitem{Chen:2018pzu}
X.~Chen, T.~Gehrmann, E.~W.~N. Glover, A.~Huss, Y.~Li, D.~Neill, M.~Schulze,
  I.~W. Stewart, and H.~X. Zhu, {\it {Precise QCD Description of the Higgs
  Boson Transverse Momentum Spectrum}},  {\em Phys. Lett. B} {\bf 788} (2019)
  425--430, [\href{http://arXiv.org/abs/1805.00736}{{\tt arXiv:1805.00736}}].

\bibitem{Bizon:2017rah}
W.~Bizon, P.~F. Monni, E.~Re, L.~Rottoli, and P.~Torrielli, {\it
  {Momentum-space resummation for transverse observables and the Higgs
  p$_{\perp}$ at N$^{3}$LL+NNLO}},  {\em JHEP} {\bf 02} (2018) 108,
  [\href{http://arXiv.org/abs/1705.09127}{{\tt arXiv:1705.09127}}].

\bibitem{Duhr:2022yyp}
C.~Duhr, B.~Mistlberger, and G.~Vita, {\it {Four-Loop Rapidity Anomalous
  Dimension and Event Shapes to Fourth Logarithmic Order}},  {\em Phys. Rev.
  Lett.} {\bf 129} (2022), no.~16 162001,
  [\href{http://arXiv.org/abs/2205.02242}{{\tt arXiv:2205.02242}}].

\bibitem{Moult:2022xzt}
I.~Moult, H.~X. Zhu, and Y.~J. Zhu, {\it {The four loop QCD rapidity anomalous
  dimension}},  {\em JHEP} {\bf 08} (2022) 280,
  [\href{http://arXiv.org/abs/2205.02249}{{\tt arXiv:2205.02249}}].

\bibitem{Agarwal:2021zft}
B.~Agarwal, A.~von Manteuffel, E.~Panzer, and R.~M. Schabinger, {\it {Four-loop
  collinear anomalous dimensions in QCD and N=4 super Yang-Mills}},  {\em Phys.
  Lett. B} {\bf 820} (2021) 136503,
  [\href{http://arXiv.org/abs/2102.09725}{{\tt arXiv:2102.09725}}].

\bibitem{Jones:2018hbb}
S.~P. Jones, M.~Kerner, and G.~Luisoni, {\it {Next-to-Leading-Order QCD
  Corrections to Higgs Boson Plus Jet Production with Full Top-Quark Mass
  Dependence}},  {\em Phys. Rev. Lett.} {\bf 120} (2018), no.~16 162001,
  [\href{http://arXiv.org/abs/1802.00349}{{\tt arXiv:1802.00349}}]. [Erratum:
  Phys.Rev.Lett. 128, 059901 (2022)].

\bibitem{Bonciani:2022jmb}
R.~Bonciani, V.~Del~Duca, H.~Frellesvig, M.~Hidding, V.~Hirschi, F.~Moriello,
  G.~Salvatori, G.~Somogyi, and F.~Tramontano, {\it {Next-to-leading-order QCD
  Corrections to Higgs Production in association with a Jet}},
  \href{http://arXiv.org/abs/2206.10490}{{\tt arXiv:2206.10490}}.

\bibitem{Caola:2018zye}
F.~Caola, J.~M. Lindert, K.~Melnikov, P.~F. Monni, L.~Tancredi, and C.~Wever,
  {\it {Bottom-quark effects in Higgs production at intermediate transverse
  momentum}},  {\em JHEP} {\bf 09} (2018) 035,
  [\href{http://arXiv.org/abs/1804.07632}{{\tt arXiv:1804.07632}}].

\bibitem{Lindert:2017pky}
J.~M. Lindert, K.~Melnikov, L.~Tancredi, and C.~Wever, {\it {Top-bottom
  interference effects in Higgs plus jet production at the LHC}},  {\em Phys.
  Rev. Lett.} {\bf 118} (2017), no.~25 252002,
  [\href{http://arXiv.org/abs/1703.03886}{{\tt arXiv:1703.03886}}].

\bibitem{Grazzini:2013mca}
M.~Grazzini and H.~Sargsyan, {\it {Heavy-quark mass effects in Higgs boson
  production at the LHC}},  {\em JHEP} {\bf 09} (2013) 129,
  [\href{http://arXiv.org/abs/1306.4581}{{\tt arXiv:1306.4581}}].

\bibitem{Melnikov:2016emg}
K.~Melnikov and A.~Penin, {\it {On the light quark mass effects in Higgs boson
  production in gluon fusion}},  {\em JHEP} {\bf 05} (2016) 172,
  [\href{http://arXiv.org/abs/1602.09020}{{\tt arXiv:1602.09020}}].

\bibitem{Caola:2016upw}
F.~Caola, S.~Forte, S.~Marzani, C.~Muselli, and G.~Vita, {\it {The Higgs
  transverse momentum spectrum with finite quark masses beyond leading order}},
   {\em JHEP} {\bf 08} (2016) 150, [\href{http://arXiv.org/abs/1606.04100}{{\tt
  arXiv:1606.04100}}].

\bibitem{Greiner:2016awe}
N.~Greiner, S.~H\"oche, G.~Luisoni, M.~Sch\"onherr, and J.-C. Winter, {\it
  {Full mass dependence in Higgs boson production in association with jets at
  the LHC and FCC}},  {\em JHEP} {\bf 01} (2017) 091,
  [\href{http://arXiv.org/abs/1608.01195}{{\tt arXiv:1608.01195}}].

\bibitem{Bagnaschi:2015bop}
E.~Bagnaschi, R.~V. Harlander, H.~Mantler, A.~Vicini, and M.~Wiesemann, {\it
  {Resummation ambiguities in the Higgs transverse-momentum spectrum in the
  Standard Model and beyond}},  {\em JHEP} {\bf 01} (2016) 090,
  [\href{http://arXiv.org/abs/1510.08850}{{\tt arXiv:1510.08850}}].

\bibitem{Banfi:2013eda}
A.~Banfi, P.~F. Monni, and G.~Zanderighi, {\it {Quark masses in Higgs
  production with a jet veto}},  {\em JHEP} {\bf 01} (2014) 097,
  [\href{http://arXiv.org/abs/1308.4634}{{\tt arXiv:1308.4634}}].

\bibitem{Mantler:2012bj}
H.~Mantler and M.~Wiesemann, {\it {Top- and bottom-mass effects in hadronic
  Higgs production at small transverse momenta through LO+NLL}},  {\em Eur.
  Phys. J. C} {\bf 73} (2013), no.~6 2467,
  [\href{http://arXiv.org/abs/1210.8263}{{\tt arXiv:1210.8263}}].

\bibitem{Keung:2009bs}
W.-Y. Keung and F.~J. Petriello, {\it {Electroweak and finite quark-mass
  effects on the Higgs boson transverse momentum distribution}},  {\em Phys.
  Rev. D} {\bf 80} (2009) 013007, [\href{http://arXiv.org/abs/0905.2775}{{\tt
  arXiv:0905.2775}}].

\bibitem{Liu:2022ajh}
Z.~L. Liu, M.~Neubert, M.~Schnubel, and X.~Wang, {\it {Factorization at
  Next-to-Leading Power and Endpoint Divergences in $gg\to h$ Production}},
  \href{http://arXiv.org/abs/2212.10447}{{\tt arXiv:2212.10447}}.

\bibitem{Liu:2021chn}
T.~Liu, S.~Modi, and A.~A. Penin, {\it {Higgs boson production and quark
  scattering amplitudes at high energy through the next-to-next-to-leading
  power in quark mass}},  {\em JHEP} {\bf 02} (2022) 170,
  [\href{http://arXiv.org/abs/2111.01820}{{\tt arXiv:2111.01820}}].

\bibitem{Melnikov:2016qoc}
K.~Melnikov, L.~Tancredi, and C.~Wever, {\it {Two-loop $gg \to Hg$ amplitude
  mediated by a nearly massless quark}},  {\em JHEP} {\bf 11} (2016) 104,
  [\href{http://arXiv.org/abs/1610.03747}{{\tt arXiv:1610.03747}}].

\bibitem{Harlander:2014hya}
R.~V. Harlander, A.~Tripathi, and M.~Wiesemann, {\it {Higgs production in
  bottom quark annihilation: Transverse momentum distribution at NNLO$+$NNLL}},
   {\em Phys. Rev. D} {\bf 90} (2014), no.~1 015017,
  [\href{http://arXiv.org/abs/1403.7196}{{\tt arXiv:1403.7196}}].

\bibitem{Pietrulewicz:2017gxc}
P.~Pietrulewicz, D.~Samitz, A.~Spiering, and F.~J. Tackmann, {\it
  {Factorization and Resummation for Massive Quark Effects in Exclusive
  Drell-Yan}},  \href{http://arXiv.org/abs/1703.09702}{{\tt arXiv:1703.09702}}.

\bibitem{Bauer:2000ew}
C.~W. Bauer, S.~Fleming, and M.~E. Luke, {\it {Summing Sudakov logarithms in
  $B\to X_s \gamma$ in effective field theory}},  {\em Phys.~Rev.} {\bf D63}
  (2000) 014006, [\href{http://arXiv.org/abs/hep-ph/0005275}{{\tt
  hep-ph/0005275}}].

\bibitem{Bauer:2000yr}
C.~W. Bauer, S.~Fleming, D.~Pirjol, and I.~W. Stewart, {\it {An Effective field
  theory for collinear and soft gluons: Heavy to light decays}},  {\em
  Phys.~Rev.} {\bf D63} (2001) 114020,
  [\href{http://arXiv.org/abs/hep-ph/0011336}{{\tt hep-ph/0011336}}].

\bibitem{Bauer:2001ct}
C.~W. Bauer and I.~W. Stewart, {\it {Invariant operators in collinear effective
  theory}},  {\em Phys.Lett.} {\bf B516} (2001) 134--142,
  [\href{http://arXiv.org/abs/hep-ph/0107001}{{\tt hep-ph/0107001}}].

\bibitem{Bauer:2001yt}
C.~W. Bauer, D.~Pirjol, and I.~W. Stewart, {\it {Soft collinear factorization
  in effective field theory}},  {\em Phys.~Rev.} {\bf D65} (2002) 054022,
  [\href{http://arXiv.org/abs/hep-ph/0109045}{{\tt hep-ph/0109045}}].

\bibitem{Bauer:2002nz}
C.~W. Bauer, S.~Fleming, D.~Pirjol, I.~Z. Rothstein, and I.~W. Stewart, {\it
  {Hard scattering factorization from effective field theory}},  {\em
  Phys.Rev.} {\bf D66} (2002) 014017,
  [\href{http://arXiv.org/abs/hep-ph/0202088}{{\tt hep-ph/0202088}}].

\bibitem{Beneke:2002ph}
M.~Beneke, A.~Chapovsky, M.~Diehl, and T.~Feldmann, {\it {Soft collinear
  effective theory and heavy to light currents beyond leading power}},  {\em
  Nucl.Phys.} {\bf B643} (2002) 431--476,
  [\href{http://arXiv.org/abs/hep-ph/0206152}{{\tt hep-ph/0206152}}].

\bibitem{Collins:1984kg}
J.~C. Collins, D.~E. Soper, and G.~F. Sterman, {\it {Transverse Momentum
  Distribution in Drell-Yan Pair and W and Z Boson Production}},  {\em Nucl.
  Phys.} {\bf B250} (1985) 199.

\bibitem{Becher:2010tm}
T.~Becher and M.~Neubert, {\it {Drell-Yan production at small $q_T$, transverse
  parton distributions and the collinear anomaly}},  {\em Eur.Phys.J.} {\bf
  C71} (2011) 1665, [\href{http://arXiv.org/abs/1007.4005}{{\tt
  arXiv:1007.4005}}].

\bibitem{Chiu:2012ir}
J.-Y. Chiu, A.~Jain, D.~Neill, and I.~Z. Rothstein, {\it {A Formalism for the
  Systematic Treatment of Rapidity Logarithms in Quantum Field Theory}},  {\em
  JHEP} {\bf 1205} (2012) 084, [\href{http://arXiv.org/abs/1202.0814}{{\tt
  arXiv:1202.0814}}].

\bibitem{GarciaEchevarria:2011rb}
M.~G. Echevarria, A.~Idilbi, and I.~Scimemi, {\it {Factorization Theorem For
  Drell-Yan At Low $q_T$ And Transverse Momentum Distributions
  On-The-Light-Cone}},  {\em JHEP} {\bf 07} (2012) 002,
  [\href{http://arXiv.org/abs/1111.4996}{{\tt arXiv:1111.4996}}].

\bibitem{Ebert:2020yqt}
M.~A. Ebert, B.~Mistlberger, and G.~Vita, {\it {Transverse momentum dependent
  PDFs at N$^3$LO}},  {\em JHEP} {\bf 09} (2020) 146,
  [\href{http://arXiv.org/abs/2006.05329}{{\tt arXiv:2006.05329}}].

\bibitem{Luo:2020epw}
M.-x. Luo, T.-Z. Yang, H.~X. Zhu, and Y.~J. Zhu, {\it {Unpolarized quark and
  gluon TMD PDFs and FFs at N$^{3}$LO}},  {\em JHEP} {\bf 06} (2021) 115,
  [\href{http://arXiv.org/abs/2012.03256}{{\tt arXiv:2012.03256}}].

\bibitem{Chiu:2011qc}
J.-Y. Chiu, A.~Jain, D.~Neill, and I.~Z. Rothstein, {\it {The Rapidity
  Renormalization Group}},  {\em Phys.Rev.Lett.} {\bf 108} (2012) 151601,
  [\href{http://arXiv.org/abs/1104.0881}{{\tt arXiv:1104.0881}}].

\bibitem{Berger:2010xi}
C.~F. Berger, C.~Marcantonini, I.~W. Stewart, F.~J. Tackmann, and W.~J.
  Waalewijn, {\it {Higgs Production with a Central Jet Veto at NNLL+NNLO}},
  {\em JHEP} {\bf 04} (2011) 092, [\href{http://arXiv.org/abs/1012.4480}{{\tt
  arXiv:1012.4480}}].

\bibitem{Gehrmann:2010ue}
T.~Gehrmann, E.~W.~N. Glover, T.~Huber, N.~Ikizlerli, and C.~Studerus, {\it
  {Calculation of the quark and gluon form factors to three loops in QCD}},
  {\em JHEP} {\bf 06} (2010) 094, [\href{http://arXiv.org/abs/1004.3653}{{\tt
  arXiv:1004.3653}}].

\bibitem{Collins:1981uw}
J.~C. Collins and D.~E. Soper, {\it {Parton Distribution and Decay Functions}},
   {\em Nucl. Phys.} {\bf B194} (1982) 445--492.

\bibitem{Fleming:2006cd}
S.~Fleming, A.~K. Leibovich, and T.~Mehen, {\it {Resummation of Large Endpoint
  Corrections to Color-Octet $J/\psi$ Photoproduction}},  {\em Phys. Rev.} {\bf
  D74} (2006) 114004, [\href{http://arXiv.org/abs/hep-ph/0607121}{{\tt
  hep-ph/0607121}}].

\bibitem{Stewart:2009yx}
I.~W. Stewart, F.~J. Tackmann, and W.~J. Waalewijn, {\it {Factorization at the
  LHC: From PDFs to Initial State Jets}},  {\em Phys.Rev.} {\bf D81} (2010)
  094035, [\href{http://arXiv.org/abs/0910.0467}{{\tt arXiv:0910.0467}}].

\bibitem{Luebbert:2016itl}
T.~L{\"u}bbert, J.~Oredsson, and M.~Stahlhofen, {\it {Rapidity renormalized TMD
  soft and beam functions at two loops}},  {\em JHEP} {\bf 03} (2016) 168,
  [\href{http://arXiv.org/abs/1602.01829}{{\tt arXiv:1602.01829}}].

\bibitem{Stewart:2010qs}
I.~W. Stewart, F.~J. Tackmann, and W.~J. Waalewijn, {\it {The Quark Beam
  Function at NNLL}},  {\em JHEP} {\bf 1009} (2010) 005,
  [\href{http://arXiv.org/abs/1002.2213}{{\tt arXiv:1002.2213}}].

\bibitem{Gaunt:2014xga}
J.~R. Gaunt, M.~Stahlhofen, and F.~J. Tackmann, {\it {The Quark Beam Function
  at Two Loops}},  {\em JHEP} {\bf 04} (2014) 113,
  [\href{http://arXiv.org/abs/1401.5478}{{\tt arXiv:1401.5478}}].

\bibitem{Gaunt:2014cfa}
J.~Gaunt, M.~Stahlhofen, and F.~J. Tackmann, {\it {The Gluon Beam Function at
  Two Loops}},  {\em JHEP} {\bf 08} (2014) 020,
  [\href{http://arXiv.org/abs/1405.1044}{{\tt arXiv:1405.1044}}].

\bibitem{Gaunt:2020xlc}
J.~R. Gaunt and M.~Stahlhofen, {\it {The fully-differential gluon beam function
  at NNLO}},  {\em JHEP} {\bf 07} (2020), no.~07 234,
  [\href{http://arXiv.org/abs/2004.11915}{{\tt arXiv:2004.11915}}].

\bibitem{Catani:2022sgr}
S.~Catani and P.~K. Dhani, {\it {Collinear functions for QCD resummations}},
  \href{http://arXiv.org/abs/2208.05840}{{\tt arXiv:2208.05840}}.

\bibitem{Luo:2019bmw}
M.-X. Luo, T.-Z. Yang, H.~X. Zhu, and Y.~J. Zhu, {\it {Transverse Parton
  Distribution and Fragmentation Functions at NNLO: the Gluon Case}},  {\em
  JHEP} {\bf 01} (2020) 040, [\href{http://arXiv.org/abs/1909.13820}{{\tt
  arXiv:1909.13820}}].

\bibitem{Echevarria:2016scs}
M.~G. Echevarria, I.~Scimemi, and A.~Vladimirov, {\it {Unpolarized Transverse
  Momentum Dependent Parton Distribution and Fragmentation Functions at
  next-to-next-to-leading order}},  {\em JHEP} {\bf 09} (2016) 004,
  [\href{http://arXiv.org/abs/1604.07869}{{\tt arXiv:1604.07869}}].

\bibitem{Gehrmann:2014yya}
T.~Gehrmann, T.~Luebbert, and L.~L. Yang, {\it {Calculation of the transverse
  parton distribution functions at next-to-next-to-leading order}},  {\em JHEP}
  {\bf 06} (2014) 155, [\href{http://arXiv.org/abs/1403.6451}{{\tt
  arXiv:1403.6451}}].

\bibitem{Catani:2011kr}
S.~Catani and M.~Grazzini, {\it {Higgs Boson Production at Hadron Colliders:
  Hard-Collinear Coefficients at the NNLO}},  {\em Eur. Phys. J. C} {\bf 72}
  (2012) 2013, [\href{http://arXiv.org/abs/1106.4652}{{\tt arXiv:1106.4652}}].
  [Erratum: Eur.Phys.J.C 72, 2132 (2012)].

\bibitem{Gangal:2016kuo}
S.~Gangal, J.~R. Gaunt, M.~Stahlhofen, and F.~J. Tackmann, {\it {Two-Loop Beam
  and Soft Functions for Rapidity-Dependent Jet Vetoes}},  {\em JHEP} {\bf 02}
  (2017) 026, [\href{http://arXiv.org/abs/1608.01999}{{\tt arXiv:1608.01999}}].

\bibitem{Bell:2022nrj}
G.~Bell, K.~Brune, G.~Das, and M.~Wald, {\it {The NNLO quark beam function for
  jet-veto resummation}},  \href{http://arXiv.org/abs/2207.05578}{{\tt
  arXiv:2207.05578}}.

\bibitem{Abreu:2022zgo}
S.~Abreu, J.~R. Gaunt, P.~F. Monni, L.~Rottoli, and R.~Szafron, {\it {Quark and
  gluon two-loop beam functions for leading-jet $p_T$ and slicing at NNLO}},
  \href{http://arXiv.org/abs/2207.07037}{{\tt arXiv:2207.07037}}.

\bibitem{Chiu:2009yx}
J.-y. Chiu, A.~Fuhrer, A.~H. Hoang, R.~Kelley, and A.~V. Manohar, {\it
  {Soft-Collinear Factorization and Zero-Bin Subtractions}},  {\em Phys. Rev.}
  {\bf D79} (2009) 053007, [\href{http://arXiv.org/abs/0901.1332}{{\tt
  arXiv:0901.1332}}].

\bibitem{Manohar:2006nz}
A.~V. Manohar and I.~W. Stewart, {\it {The Zero-Bin and Mode Factorization in
  Quantum Field Theory}},  {\em Phys.~Rev.} {\bf D76} (2007) 074002,
  [\href{http://arXiv.org/abs/hep-ph/0605001}{{\tt hep-ph/0605001}}].

\bibitem{Gritschacher:2013pha}
S.~Gritschacher, A.~H. Hoang, I.~Jemos, and P.~Pietrulewicz, {\it {Secondary
  Heavy Quark Production in Jets through Mass Modes}},  {\em Phys.Rev.} {\bf
  D88} (2013) 034021, [\href{http://arXiv.org/abs/1302.4743}{{\tt
  arXiv:1302.4743}}].

\bibitem{Pietrulewicz:2014qza}
P.~Pietrulewicz, S.~Gritschacher, A.~H. Hoang, I.~Jemos, and V.~Mateu, {\it
  {Variable Flavor Number Scheme for Final State Jets in Thrust}},  {\em
  Phys.Rev.} {\bf D90} (2014), no.~11 114001,
  [\href{http://arXiv.org/abs/1405.4860}{{\tt arXiv:1405.4860}}].

\bibitem{Hoang:2019fze}
A.~H. Hoang, C.~Lepenik, and M.~Stahlhofen, {\it {Two-Loop Massive Quark Jet
  Functions in SCET}},  {\em JHEP} {\bf 08} (2019) 112,
  [\href{http://arXiv.org/abs/1904.12839}{{\tt arXiv:1904.12839}}].

\bibitem{Ebert:2016gcn}
M.~A. Ebert and F.~J. Tackmann, {\it {Resummation of Transverse Momentum
  Distributions in Distribution Space}},  {\em JHEP} {\bf 02} (2017) 110,
  [\href{http://arXiv.org/abs/1611.08610}{{\tt arXiv:1611.08610}}].

\bibitem{Buza:1996wv}
M.~Buza, Y.~Matiounine, J.~Smith, and W.~L. van Neerven, {\it {Charm
  electroproduction viewed in the variable flavor number scheme versus fixed
  order perturbation theory}},  {\em Eur. Phys. J.} {\bf C1} (1998) 301--320,
  [\href{http://arXiv.org/abs/hep-ph/9612398}{{\tt hep-ph/9612398}}].

\bibitem{Smirnov:2014hma}
A.~V. Smirnov, {\it {FIRE5: a C++ implementation of Feynman Integral
  REduction}},  {\em Comput. Phys. Commun.} {\bf 189} (2015) 182--191,
  [\href{http://arXiv.org/abs/1408.2372}{{\tt arXiv:1408.2372}}].

\bibitem{Bierenbaum:2008yu}
I.~Bierenbaum, J.~Blumlein, S.~Klein, and C.~Schneider, {\it {Two-Loop Massive
  Operator Matrix Elements for Unpolarized Heavy Flavor Production to
  O(epsilon)}},  {\em Nucl. Phys. B} {\bf 803} (2008) 1--41,
  [\href{http://arXiv.org/abs/0803.0273}{{\tt arXiv:0803.0273}}].

\bibitem{Hoang:2015iva}
A.~H. Hoang, P.~Pietrulewicz, and D.~Samitz, {\it {Variable Flavor Number
  Scheme for Final State Jets in DIS}},  {\em Phys. Rev.} {\bf D93} (2016),
  no.~3 034034, [\href{http://arXiv.org/abs/1508.04323}{{\tt
  arXiv:1508.04323}}].

\bibitem{Harland-Lang:2014zoa}
L.~A. Harland-Lang, A.~D. Martin, P.~Motylinski, and R.~S. Thorne, {\it {Parton
  distributions in the LHC era: MMHT 2014 PDFs}},  {\em Eur. Phys. J. C} {\bf
  75} (2015), no.~5 204, [\href{http://arXiv.org/abs/1412.3989}{{\tt
  arXiv:1412.3989}}].

\end{thebibliography}\endgroup

\end{document}